\documentclass[aps,prx,reprint,twocolumn,floatfix,superscriptaddress,eqsecnum]{revtex4-2}

\usepackage{xcolor}
\usepackage{amsmath}
\usepackage{amstext}
\usepackage{amssymb}
\usepackage{amsfonts}
\usepackage{amsbsy}
\usepackage{url}
\usepackage{bm}
\usepackage{graphicx}
\usepackage{mathrsfs}
\usepackage{tikz}
\usetikzlibrary{arrows.meta,calc}
\definecolor{pastelgreen}{RGB}{106,168,138}
\definecolor{paleblue}{RGB}{223,233,246}
\definecolor{softblue}{RGB}{110,140,180}
\definecolor{rose}{RGB}{196,124,132}

\usepackage{multirow}

\newcommand{\ket}[1]{{\left | #1 \right\rangle}}
\newcommand{\bra}[1]{{\left \langle #1 \right|}}

\usepackage[colorlinks=true, urlcolor=blue, linkcolor=blue, citecolor=blue]{hyperref}
\hypersetup{
  pdftitle={Chiral Color Ice: Exact Local Handedness Constraints and
    M\"obius Zero Modes in Frustrated Magnets},
  pdfauthor={P\'eter Kr\'anitz, Yasir Iqbal, Karlo Penc},
  pdfkeywords={frustrated magnetism, chiral color ice, parent
    Hamiltonian, scalar spin chirality, M\"obius transformation,
    frustration-free model, residual entropy, Anderson tower}
}
\usepackage{orcidlink}

\allowdisplaybreaks

\begin{document}

\author{P\'eter Kr\'anitz\,\orcidlink{0009-0002-9703-5367}}
\email{kranitz.peter@wigner.hun-ren.hu}
\affiliation{Department of Theoretical Physics, Institute of Physics, Budapest University of Technology and Economics, M\H{u}egyetem rakpart 3, H-1111 Budapest, Hungary}
\affiliation{Institute for Solid State Physics and Optics, HUN-REN Wigner Research Centre for Physics, H-1525 Budapest, P.O. Box 49, Hungary}
\affiliation{Department of Physics, Indian Institute of Technology Madras, Chennai 600036, India}
\author{Yasir Iqbal\,\orcidlink{0000-0003-1614-6920}}
\email{yiqbal@physics.iitm.ac.in}
\affiliation{Department of Physics, Indian Institute of Technology Madras, Chennai 600036, India}
\author{Karlo Penc\,\orcidlink{0000-0002-2197-1370}}
\email{penc.karlo@wigner.hun-ren.hu}
\affiliation{Institute for Solid State Physics and Optics, HUN-REN Wigner Research Centre for Physics, H-1525 Budapest, P.O. Box 49, Hungary}
\affiliation{Department of Physics, Indian Institute of Technology Madras, Chennai 600036, India}

\title{Chiral Color Ice: Exact Local Handedness Constraints and M\"obius Zero Modes in Frustrated Magnets}

\begin{abstract}
Local constraints govern the low-energy physics of frustrated matter,
but familiar ice-type rules constrain flux-like quantities and are
insensitive to handedness. Here we show that handedness itself can be
imposed as an exact local quantum constraint without selecting an axis in
spin space. We construct positive-semidefinite, SU(2)-invariant parent
Hamiltonians whose complete zero-energy space on a tetrahedron has a
prescribed chirality sign, rather than selecting a particular chiral wave
function. For spin $1/2$ the local term is a rank-one projector onto a
chiral tetrahedral singlet, while for arbitrary spin it factorizes as
$\mathcal B^\dagger \mathcal B$ through a singlet-annihilation operator,
with a completely characterized kernel given by the span of the globally
rotated chiral color-ice states. For coherent states, the same zero-energy
condition becomes an $S$-independent nonlinear constraint in which three
spin directions determine the fourth through a M\"obius transformation;
compositions of these maps define constraint holonomies on extended
lattices. Connecting the same local constraint in different ways produces
qualitatively different collective regimes: corner-sharing lattices
retain exponentially large quantum ground-state kernels, with rigorous
lower bounds already exceeding conventional ice benchmarks; edge-sharing
lattices support subdimensional plane or line zero modes; while triangular
constructions suppress nonuniform coherent deformations and contain the
complete Anderson tower of tetrahedral magnetic order at exactly zero
energy. Two inequivalent triangular coverings further show that harmonic
zero-mode counting does not determine the size of the quantum kernel.
These results establish a tractable setting in which local handedness,
nonlinear constraint geometry, and quantum degeneracy can be disentangled
and related directly to the connectivity of the constraint network.
\end{abstract}

\maketitle

\tableofcontents

\section{Introduction}
\label{sec:intro}

Frustrated magnets show how simple local rules can generate collective
behavior far richer than conventional magnetic order.  In these systems
the low-energy physics is organized not around a unique ordered
configuration but around an extensive manifold of states selected by a
local constraint.  Spin ice is the paradigm: the two-in--two-out rule on
each tetrahedron of the pyrochlore lattice produces a macroscopically
degenerate manifold with power-law correlations, an emergent gauge
field, and effective magnetic monopole excitations
\cite{Bramwell2001,Castelnovo2008}.  Related constraint counting
underlies much of the modern understanding of kagome and pyrochlore
antiferromagnets, where corner-sharing motifs impose local conditions
that remain visible in the long-distance physics
\cite{Moessner1998,Balents2010}.  Conventional ice-type rules,
however, are achiral: the constraint fixes a flux-like,
time-reversal-even quantity, so the constrained manifold is invariant
under time reversal and handedness plays no role.  Chiral ice-type
constructions have so far relied on Ising-like variables, external
fields, or purely classical interactions
\cite{Rahmani2013,Chern2014,Lozano2024}, and chiral correlations can
also emerge spontaneously on a quantum ice background
\cite{Onoda2010}.
Whether an intrinsically \emph{chiral} local constraint---one that
selects a handedness on every tetrahedron---can be formulated exactly
and enforced by a local, spin-rotation-invariant quantum Hamiltonian
has remained an open
question.

Chirality is the natural route to time-reversal symmetry breaking
without magnetic order.  Its local quantum measure is the scalar spin
chirality
\begin{equation}
\hat{\chi}_{ijk}
=
\hat{\mathbf S}_i\cdot
\left(
\hat{\mathbf S}_j\times \hat{\mathbf S}_k
\right),
\label{eq:SpinScalarChirality}
\end{equation}
defined on an ordered triple $(i,j,k)$ of sites.  This operator is
invariant under global spin rotations but changes sign under time
reversal and under odd permutations of the three sites: it detects the
handedness of a noncoplanar spin triad without singling out any
direction in spin space.  Scalar chirality lies at the heart of the
theory of chiral spin states and chiral spin liquids, from the
Kalmeyer--Laughlin wave function to the Wen--Wilczek--Zee order
parameter and its realizations in frustrated and itinerant magnets
\cite{Kalmeyer1987,WWZ1989,Baskaran1989,Martin2008,Batista2016}.
Exact parent Hamiltonians are known for individual chiral
wave functions---for the Kalmeyer--Laughlin state and its lattice
descendants \cite{Schroeter2007,Thomale2009,Nielsen2013}---and
explicit scalar-chirality interactions are known to stabilize chiral
spin-liquid phases in kagome Mott insulators \cite{Bauer2014}.  In
all of these settings, however, chirality characterizes a
\emph{single} correlated ground state, and the parent Hamiltonians
that pin it exactly require long-ranged or finely structured
multispin couplings \cite{Momoi2003}.  Here we pursue a complementary question: can
scalar chirality be built into the very \emph{constraint} that
defines an extensively degenerate manifold---in the way the ice rule
builds in a local flux condition---using only a local interaction on
a single tetrahedron?

\begin{figure}[tbp]
    \centering
    \includegraphics[width=0.5\columnwidth]{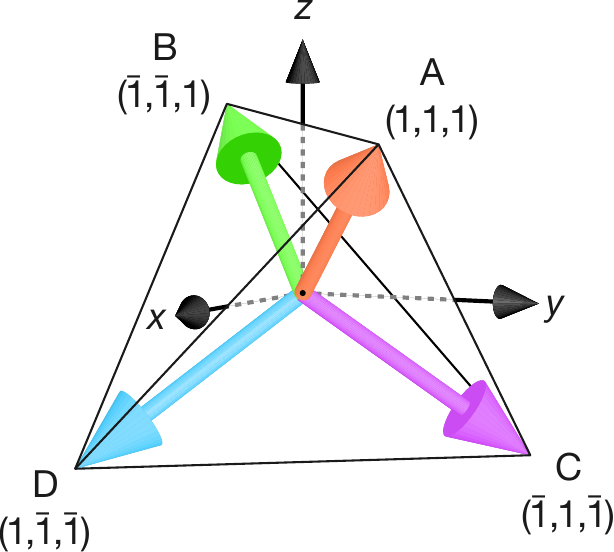}
    \caption{The elementary motif of chiral color ice: four spins
    pointing from the center to the vertices of a regular tetrahedron
    in spin space.  A color is associated with each direction, red for
    $\mathbf{n}^{A}=\frac{1}{\sqrt{3}}(1,1,1)$, green for
    $\mathbf{n}^{B}$, magenta for $\mathbf{n}^{C}$, and cyan for
    $\mathbf{n}^{D}$ of Eq.~(\ref{eq:color_directions}).  The
    configuration carries no net magnetization and selects no axis in
    spin space, but each oriented face of the tetrahedron carries a
    scalar spin chirality of magnitude $4/(3\sqrt{3})$, so the frame
    has a definite handedness: the two enantiomeric assignments are
    related by time reversal and cannot be connected by any proper
    spin rotation.}
    \label{fig:tetrahedral}
\end{figure}

The minimal object that carries chirality without magnetization is a
regular tetrahedral frame in spin space.  As illustrated in
Fig.~\ref{fig:tetrahedral}, four spins pointing from the center to the
vertices of a regular tetrahedron sum to zero and single out no axis;
instead they define a genuinely three-dimensional frame whose four
oriented faces all carry a nonzero scalar chirality.  The chirality
changes sign under time reversal and under any improper transformation
that reverses the orientation of the frame, so the two handednesses of
the tetrahedral configuration are related by time reversal but cannot
be connected by any proper global spin rotation.  Tetrahedral
(``all-in--all-out'' in spin space) configurations of this kind appear
as regular magnetic orders on triangular and pyrochlore geometries
\cite{Momoi1997,Messio2011,Tsunetsugu2001}, as the cyclic phase of
spin-2 condensates \cite{Barnett2006}, and, most importantly for our
purposes, as local building blocks of the \emph{color ice} manifold:
assigning a color $A$, $B$, $C$, $D$ to each of the four tetrahedral
spin directions, the classical ground states of the bilinear-biquadratic
pyrochlore antiferromagnet are exactly the four-colorings of the
lattice in which every tetrahedron contains each color once
\cite{Wan2016}.  Recently it was shown that a purely chiral classical
interaction on the pyrochlore lattice selects the chiral subset of
these colorings---those with a fixed handedness on every
tetrahedron---realizing a classical chiral spin liquid
\cite{Lozano2024}.  We refer to this constrained manifold as
\emph{chiral color ice} (CCI).  It combines a discrete color
constraint, extensive degeneracy, and local time-reversal symmetry
breaking, and it differs in an essential way from spin ice: the colors
are not Ising labels but the legs of a chiral frame in spin space.

These observations raise a sharp question.  Can one construct a
\emph{local quantum} spin Hamiltonian for which the chiral color-ice
states are exact zero-energy ground states?  The problem is more subtle
than it may appear.  In the absence of spin anisotropy, the constraint
must be enforced covariantly: the Hamiltonian cannot pin a particular
orientation of the tetrahedral frame, so its local kernel must contain
the twelve product states of one chiral color sector \emph{together
with all their global SU(2) rotations}.  The task is thus to build a
local, positive-semidefinite, spin-rotation-invariant operator that
distinguishes a handedness---a time-reversal-odd property---without
introducing any spin-space axis.

Parent-Hamiltonian constructions are the natural tool for this purpose.
They have produced some of the sharpest exact statements in quantum
many-body physics, from the Majumdar--Ghosh and Klein points to the
AKLT models and the pseudopotential Hamiltonians of the fractional
quantum Hall effect, and they continue to organize the search for
exactly solvable points in frustrated and topological matter
\cite{Majumdar1969,Klein1982,Affleck1987,Haldane1983,Parameswaran2009,Saito2024,Gioia2025,Raja2026}.
Particularly close in spirit are the coloring-based constructions on
kagome and related triangular-motif lattices, where special spin-$1/2$
Hamiltonians possess macroscopically degenerate ground-state manifolds
spanned by $120^\circ$ three-coloring states
\cite{Changlani2018,Changlani2019,Lee2020,Palle2021,Pal2021}, with an
extensive entropy famously computed by Baxter \cite{Baxter1970}; in
parallel work, fully packed valence-bond loop manifolds have been
obtained as the exact kernels of local spin-1 models on the same
checkerboard and pyrochlore geometries considered here
\cite{Hari2026}.  Those manifolds, however, are achiral---and the
coloring ones are built from coplanar states and exist only at
points of XXZ \emph{anisotropy}, which singles out the plane of the
colors in spin space.  A
parent-Hamiltonian framework for finite-color, noncoplanar,
time-reversal-breaking constraints---enforced without any spin-space
anisotropy---has not been available.  Supplying it is the purpose of
this paper.

Our construction starts from the chiral color-ice configurations on a
single tetrahedron, promotes the classical directions to spin-$S$
coherent states, and identifies the local subspace spanned by the
allowed product states and their global SU(2) rotations.  The
orthogonal complement of this subspace defines a local
positive-semidefinite parent term, and summing it over the tetrahedra
of a lattice yields a frustration-free Hamiltonian whose zero-energy
kernel contains every chiral color-ice state.  The spin-$1/2$ case
exposes the mechanism with particular clarity: on one tetrahedron the
SU(2)-rotated chiral four-coloring states span fifteen of the sixteen
dimensions of the Hilbert space, and the single excluded state is one
of the two time-reversal-conjugate \emph{chiral tetrahedral singlets},
distinguished only by the sign of its scalar chirality.  The local
parent Hamiltonian is the projector onto this singlet: it selects a
handedness, yet no direction in spin space.  For arbitrary $S$ the same
structure is encoded algebraically.  We construct a Schwinger-boson
operator $\mathcal{B}$ that annihilates precisely one chiral singlet
shared among the four spins, and show that the local parent Hamiltonian
takes the manifestly positive-semidefinite form
$\mathcal{B}^{\dagger}\mathcal{B}$.  In this representation each
spin $S$ is composed of $2S$ Schwinger bosons, and $\mathcal{B}$
removes one boson from each of the four sites, combined into the
four-site spin-$1/2$ chiral singlet of the forbidden handedness.  The
zero-energy condition $\mathcal{B}\ket{\Psi}=0$ therefore states that
no four spin-$1/2$ constituents, one drawn from each site, are found
in that chiral singlet---a singlet-annihilation constraint that
generalizes the ice rule to a chirality-selecting rule, uniformly in
$S$.  The selection is exact at the quantum level, not merely for the
classical solutions: every state of the local zero-energy
kernel---entangled superpositions included---carries tetrahedral
chirality of the selected sign, in the operator sense that the
projected chirality is sign-semidefinite on the kernel, and the same
inequality descends tetrahedron by tetrahedron to the many-body
common kernel of every lattice model built from these terms.

The kernel obtained in this way is larger than the discrete set of
colorings from which the construction departs: on a single tetrahedron
its dimension is exactly $(2S+1)^4-(2S)^4$, to be compared with the
twelve colorings of one chiral sector.  A useful way to see why is that the constraint fixes the
\emph{sign} of the tetrahedral chirality but not its magnitude: the
regular tetrahedral configuration and the fully polarized,
zero-chirality configurations belong to one and the same family of
coherent zero modes.  What makes this enlargement
tractable---indeed, what organizes the entire many-body problem---is
a rigid geometric structure hidden in the constraint.  In
stereographic coordinates $z_i$ on the Bloch sphere, the
coherent-state zero-energy condition---a quadratic polynomial
equation $p_-(z_1,z_2,z_3,z_4)=0$ derived in
Sec.~\ref{sec:mobius}---is affine in each $z_i$
separately, so fixing three spins determines the fourth uniquely on
the Riemann sphere: the local constraint acts as a \emph{M\"obius
completion rule}.  The completion map $z_3\mapsto z_4$ at fixed
$z_1,z_2$ is an elliptic M\"obius transformation of order six whose
two fixed points are $z_1$ and $z_2$ themselves; equivalently, four
distinct spins solve $p_-=0$ exactly when their coordinates form an
equianharmonic quadruple on the sphere.  Propagating the rule through
corner-sharing lattices converts ground-state counting into the
combinatorics of composite M\"obius transformations: around any closed
loop, consistency selects the fixed points of the accumulated map,
producing loop zero modes whose count we obtain in closed form for all
$S$.  Together with an exact mapping of the polarized sector onto the
monomer--dimer problem of the dual (square or diamond) lattice, these
modes yield rigorous exponential lower bounds on the ground-state
degeneracy of the checkerboard and pyrochlore models---bounds that
already exceed the Pauling and Lieb residual entropies of conventional
ice.  Chiral color ice is thus \emph{more} degenerate than ice, even
though its constraint is stronger than a naive count would suggest:
the extra entropy is the price exacted by spin-rotation invariance,
which forces the ferromagnet, its magnon descendants, and, more
generally, every product state with at most one deviated spin per
tetrahedron into the kernel, and places the parent Hamiltonians on
the boundary of the ferromagnetic phase.

The same local term can be assembled on lattices whose tetrahedra share
edges or faces rather than corners, and there the physics changes
qualitatively: the overlapping constraints become restrictive enough
to single out chiral four-sublattice states.  This is the organizing
principle of the
paper.  One local rule, propagated through simplex networks of
increasing connectivity, produces
\begin{equation}
\begin{aligned}
\text{corner sharing}\ &\longrightarrow\ \text{extensive degeneracy},\\
\text{edge sharing}\ &\longrightarrow\ \text{subdimensional zero modes},\\
\text{face sharing}\ &\longrightarrow\ \text{rigid chiral order},
\end{aligned}
\label{eq:connectivity_hierarchy}
\end{equation}
so that the lattices treated below are not a catalogue of examples but
successive stages of a single mechanism.  The relative chirality of the two
inversion-related families of tetrahedra now matters decisively.
With an \emph{alternating} assignment---opposite chirality on the two
families, the pattern inherited from pyrochlore color ice---edge
sharing on the fcc lattice eliminates local zero modes and singles
out the chiral four-sublattice coloring, up to a residual
subextensive family
of planar deformations; flattening fcc bilayers produces square- and
honeycomb-lattice models whose residual deformations are line-like;
and on the triangular lattice, where the flattened tetrahedra share
faces, the ordered state is rigid.  With a \emph{uniform} assignment,
by contrast, the coloring rule is frustrated---no four-coloring can
satisfy all tetrahedra simultaneously---and no four-sublattice
coloring survives, as we verify by exhaustive enumeration on the fcc,
square,
and honeycomb lattices.  In the quantum spin-$1/2$ models on the
ordered branches of these lattices,
exact diagonalization shows that the zero-energy kernel contains
the complete set of multiplets carrying the quantum numbers of the
Anderson tower of states of the chiral four-sublattice order.  At the
frustration-free point these multiplets are all exactly degenerate;
a degeneracy-lifting perturbation that favors the tetrahedrally
ordered component would resolve them into the Anderson
tower from which the symmetry-broken chiral order emerges
\cite{Sotnikov2023}, although which part of the larger kernel a
given perturbation actually selects is left open here.  The two
triangular coverings---edge-sharing and face-sharing assemblies of
the same flattened tetrahedra---provide the sharpest control on this
hierarchy: they support identical harmonic zero modes around the
ordered state, yet exact diagonalization finds markedly different
zero-energy kernels, so harmonic zero-mode counting and quantum
degeneracy are independent characteristics of the constraint
network.

The models constructed here therefore serve a double purpose.  In both
roles they are genuinely quantum, SU(2)-symmetric spin models rather
than classical vertex systems: the classical language we use
throughout refers to their coherent-state (product-state) solutions,
which form only part of the kernel.  On corner-sharing
lattices the models define exactly solvable, extensively degenerate
points---chiral analogs of the ice manifold---whose product-state
degeneracy is organized by M\"obius geometry, in contrast with the
divergence-free condition that organizes ice.  On edge- and
face-sharing lattices they provide exact parent Hamiltonians whose
zero-energy kernels contain the chiral four-sublattice states
together with the full Anderson-tower content of the corresponding
order.  In both cases the constraint is enforced by a local,
SU(2)-invariant, time-reversal-breaking term built solely from scalar
chiralities and products of Heisenberg exchanges. Its
time-reversal-even multispin part is of the kind generated by ring
exchange in Mott insulators \cite{Momoi1997}, whereas the signed
scalar-chirality part requires an explicitly time-reversal-breaking
ingredient, such as orbital magnetic flux, an applied magnetic
field, or circular driving
\cite{WWZ1989,Sen1995,Motrunich2006,Kitamura2017};
interactions of this multispin type are also the ones that inverse
quantum simulation methods are beginning to target
\cite{Kokail2026}.

The paper is organized as follows. Section~\ref{sec:cci} defines
chiral color ice at the classical level and its quantum
coherent-state counterpart.
Section~\ref{sec:parent} constructs the local parent Hamiltonian on a
tetrahedron, first for spin $1/2$ and then for arbitrary $S$, and
proves the exact chirality-sign constraint throughout its complete
local kernel. Section~\ref{sec:schwinger} develops the Schwinger-boson
singlet-annihilation formulation and derives the resulting local
kernel dimensions. Section~\ref{sec:mobius} analyzes the coherent
zero-mode variety of a single tetrahedron, derives the M\"obius
completion rule, and establishes the corresponding bounds on the
tetrahedral chirality. Section~\ref{sec:lattice_construction}
assembles these local constraints into lattice Hamiltonians, defines
the quantum common kernel and its handedness constraint, and
introduces M\"obius propagation and loop holonomy.
Section~\ref{sec:degeneracy} establishes the extensive ground-state
degeneracy on the checkerboard and pyrochlore lattices and compares it
with ice. Section~\ref{sec:otherlattices} treats the fcc, square,
honeycomb, and triangular lattices, whose zero-energy kernels carry
the full Anderson-tower content of chiral four-sublattice order, and
closes by examining the chirality spectrum of the quantum common
kernel; a linearized coherent-state analysis
(Appendix~\ref{app:rigidity}) makes the connectivity hierarchy of
these constructions quantitative.
Section~\ref{sec:summary_results} collects the principal results in
three reference tables, one for each level of the kernel hierarchy,
and Section~\ref{sec:discussion} discusses
the emerging picture and the avenues it opens. Technical derivations, special cases,
and counting arguments are collected in
Appendixes~\ref{app:chirality}--\ref{app:tower}.

\begin{figure}[tbp]
    \centering
    \includegraphics[width=0.8\columnwidth]{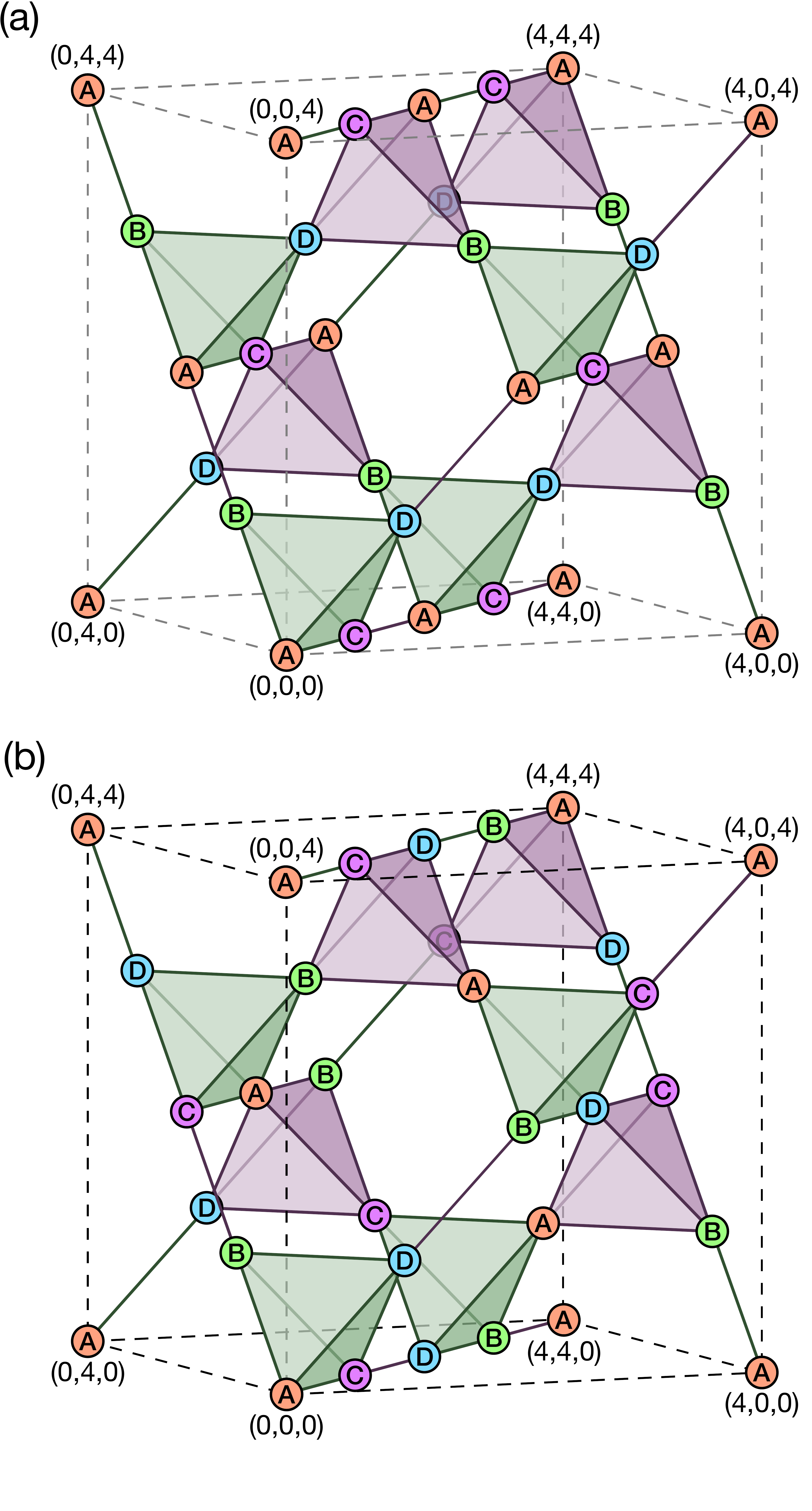}
 \caption{Chiral color-ice states compatible with the 16-site cubic
unit cell of the pyrochlore lattice.
(a)~The uniform state with a 4-site unit cell, in which the color is
constant on each fcc sublattice.
(b)~An analog of the $R$ state of
Ref.~\cite{Bergman_PRL.96.097207_2006}, generated from (a) by
combining the translation $(2,2,0)$ with the color permutation
$(AD)(BC)$.
Both states satisfy the same fixed-ordering CCI constraint on every
tetrahedron and hence belong to the same chiral sector.
In the geometric convention of Sec.~\ref{sec:cci_classical}, the
tetrahedral
chirality nevertheless has opposite signs on
the two orientation classes of tetrahedra (green and purple), which
are interchanged by spatial inversion.}
\label{fig:bloch_sphere} 
\end{figure}

\section{Chiral color ice}
\label{sec:cci}

We begin by characterizing the local spin configurations that will form the building blocks of chiral color ice.
This section fixes the local motif---four spins pointing along the four tetrahedral directions---shows that such configurations split into two chiral sectors exchanged by time reversal, and defines chiral color ice: the colorings obtained by prescribing one sector on every tetrahedron of a lattice.

\subsection{Color ice and its chiral sectors}
\label{sec:cci_classical}

A local color-ice configuration on a tetrahedron consists of four
classical spins pointing from the center toward the four vertices of a
regular tetrahedron in spin space (a tetrahedral state
\cite{Momoi1997,Messio2011}),
\begin{align}
    \mathbf{n}^{A} &= \frac{1}{\sqrt{3}}(1,1,1),
    \quad
    \mathbf{n}^{B} = \frac{1}{\sqrt{3}}(\bar{1},\bar{1},1),
    \nonumber\\
    \mathbf{n}^{C} &= \frac{1}{\sqrt{3}}(\bar{1},1,\bar{1}),
    \quad
    \mathbf{n}^{D} = \frac{1}{\sqrt{3}}(1,\bar{1},\bar{1}),
    \label{eq:color_directions}
\end{align}
as shown in Fig.~\ref{fig:tetrahedral}.
Assigning a distinct color to each tetrahedral direction, a global
color-ice (CI) state on a lattice of tetrahedra is a four-coloring of
the sites such that every tetrahedron contains each of the four colors
exactly once.
Throughout the paper, the labels $A$, $B$, $C$, $D$ denote
the four colors, with reference directions given by
Eq.~(\ref{eq:color_directions}), and generic spin directions are
written $\mathbf n_i$ or, in stereographic coordinates, $z_i$. As
shown in Sec.~\ref{sec:coloring_loops} and
Appendix~\ref{app:loops}, parts of the analysis depend only on the
coloring pattern and not on the reference directions themselves; the
color labels then stand for any four distinct spin directions
compatible with the local constraint.
On the pyrochlore lattice, this manifold is the classical ground-state
manifold of the bilinear-biquadratic Heisenberg model
\cite{Wan2016,Moessner1998}.
In the absence of anisotropies, global SO(3) rotations generate
equivalent ground states, while time reversal reverses the scalar
chirality.

Because any three of the four directions in
Eq.~(\ref{eq:color_directions}) are noncoplanar, every triangular face
of a colored tetrahedron carries a nonzero scalar chirality,
\begin{equation}
    \chi_{ijk}
    =
    \mathbf{n}_i\cdot
    \bigl(\mathbf{n}_j\times\mathbf{n}_k\bigr),
\end{equation}
with magnitude $|\chi_{ijk}|=4/(3\sqrt{3})$ for the normalized
directions of Eq.~(\ref{eq:color_directions}).
A convenient measure of the chirality of a tetrahedron $(i,j,k,l)$ is
the tetrahedral chirality, defined as the oriented sum of the four face
chiralities,
\begin{equation}
    \chi_{\mathrm{tet}}
    =
    \chi_{ijk} + \chi_{ilj} + \chi_{ikl} + \chi_{jlk} .
    \label{eq:tetrahedronchirality}
\end{equation}

Let $S_4$ denote the group of permutations of the four sites of a
tetrahedron.  Starting from the reference coloring $(A,B,C,D)$ on the
ordered sites $(i,j,k,l)$, the $4!=24$ site permutations generate all
possible color assignments.  We denote the parity of
$\sigma\in S_4$ by $\operatorname{sgn}(\sigma)=+1$ for even
permutations and $\operatorname{sgn}(\sigma)=-1$ for odd
permutations.  The twelve even permutations form the alternating
subgroup $A_4\subset S_4$.

For a coloring obtained from the reference assignment by the site
permutation $\sigma$, the tetrahedral chirality is
\begin{equation}
    \chi_{\mathrm{tet}}
    =
    -\operatorname{sgn}(\sigma)\,
    \frac{16}{3\sqrt{3}},
\end{equation}
for the conventions of Eqs.~(\ref{eq:color_directions}) and
(\ref{eq:tetrahedronchirality}).  Thus, even and odd site
permutations generate the two chiral sectors. Different
tetrahedra of a generic CI configuration may belong to opposite chiral
sectors.

We define the \emph{chiral color-ice} (CCI) \emph{coloring set} by
fixing the chiral sector on every tetrahedron.  We choose an ordering
of the four sites of each tetrahedron $t$ once and for all and assign
a sign $\varepsilon_t=\pm1$.  The coloring set
CCI($\{\varepsilon_t\}$) consists
of all colorings for which the site permutation $\sigma_t\in S_4$
relating the local coloring to the reference assignment satisfies
\begin{equation}
    \operatorname{sgn}(\sigma_t)=\varepsilon_t .
\end{equation}
In this fixed-ordering convention, the tetrahedral chirality is
therefore
\begin{equation}
    \chi_{\mathrm{tet},t}
    =
    -\varepsilon_t\,\frac{16}{3\sqrt{3}} .
\end{equation}
The chirality constraint thus retains twelve of the $4!=24$ local
color arrangements on each tetrahedron.
On the pyrochlore lattice, natural choices include the two uniform patterns, \(\varepsilon_t=+1\) or \(\varepsilon_t=-1\) for all \(t\), as well as alternating patterns in which \(\varepsilon_t\) changes sign between the two orientations of tetrahedra.

Throughout the paper we quote tetrahedral chiralities in a single
\emph{geometric} convention. The permutation parity defined above
refers to a fixed \emph{site ordering}; on the pyrochlore lattice, we
use the four fcc sublattice labels for this purpose. For the values
of $\chi_{\mathrm{tet}}$, by contrast, the sites of every
tetrahedron are ordered such that each face circuit entering
Eq.~(\ref{eq:tetrahedronchirality}), viewed from the center of that
tetrahedron, runs clockwise. Since $\chi_{\mathrm{tet}}$ is
invariant under even permutations of the four sites, this
prescription fixes its sign unambiguously, and it does so with one
and the same rule for every tetrahedron. On a single tetrahedron,
where no spatial embedding is specified, we use the reference site
ordering.

The tetrahedra of the pyrochlore lattice fall into two orientation
classes exchanged by spatial inversion---the up and down
tetrahedra---which we denote $\mathcal O_1$ and $\mathcal O_2$,
taking $\mathcal O_1$ to be the up class; the corresponding classes
of the other lattices are identified below. The fixed sublattice
ordering agrees with the geometric ordering on $\mathcal O_1$ but
differs from it by an odd permutation on $\mathcal O_2$: inversion
interchanges the two classes and reverses the view from the center,
while leaving the spin directions unchanged, because spins are axial
vectors. For a CCI coloring, the geometric tetrahedral chirality is
therefore $-\varepsilon_t\,16/(3\sqrt3)$ for $t\in\mathcal O_1$ and
$+\varepsilon_t\,16/(3\sqrt3)$ for $t\in\mathcal O_2$.

For example, in the uniform-parity pyrochlore state shown in
Fig.~\ref{fig:bloch_sphere}(a), every tetrahedron carries the same
sublattice coloring and therefore the same permutation parity.
Nevertheless, the geometric tetrahedral chirality is
$-16/(3\sqrt{3})$ on the up tetrahedra and $+16/(3\sqrt{3})$ on the
down tetrahedra.
Hence, a CCI state with uniform parity in the fixed-sublattice
convention exhibits an alternating pattern of geometric chirality
between the two tetrahedral orientations.  This is the convention
underlying the classical ground-state manifold selected by the uniform
chiral interaction of Ref.~\cite{Lozano2024}.
The resulting pattern of geometric tetrahedral chiralities is reversed by spatial inversion \(\mathcal{I}\) and by time reversal \(\mathcal{T}\) separately, but is invariant under their product \(\mathcal{IT}\). Figure~\ref{fig:bloch_sphere} shows two CCI states with this chirality pattern that are compatible with the 16-site cubic unit cell.

\subsection{Quantum chiral color-ice states}
\label{sec:cci_quantum}

We represent the classical CCI configurations quantum mechanically by
site-factorized spin coherent states.  On each site $j$, we associate
a classical direction $\mathbf n$ with the spin-$S$ coherent state
$\ket{\mathbf n_j}$, defined, up to an overall phase, as the
maximal-weight eigenstate
\begin{equation}
   \bigl(\mathbf n\cdot\hat{\mathbf S}_j\bigr)
   \ket{\mathbf n_j}
   =
   S\ket{\mathbf n_j} .
   \label{eq:coherent_def}
\end{equation}
Then, for a single tetrahedron with ordered sites $(1,2,3,4)$, the 
reference CCI product state is
\begin{equation}
  \ket{\Psi^{ABCD}}
  =
  \ket{\mathbf n^A_1}
  \otimes
  \ket{\mathbf n^B_2}
  \otimes
  \ket{\mathbf n^C_3}
  \otimes
  \ket{\mathbf n^D_4} ,
  \label{eq:Psi_CCI}
\end{equation}
which belongs to the $\varepsilon=+1$ chiral sector in the
convention of Sec.~\ref{sec:cci_classical}.  
The twelve even site permutations $\sigma\in A_4$ generate the twelve
local CCI product states in this sector, including the reference state
(\ref{eq:Psi_CCI}).

For a global CCI coloring $\{\alpha_j\}$, we define the corresponding
quantum product state as
\begin{equation}
    \ket{\Psi^{\{\alpha_j\}}}
    =
    \bigotimes_j \ket{\mathbf n^{\alpha_j}_j} ,
    \qquad
    \alpha_j\in\{A,B,C,D\},
    \label{eq:global_CCI}
\end{equation}
where the coloring $\{\alpha_j\}$ satisfies the prescribed CCI
constraint on every tetrahedron.

If the parent Hamiltonian is SU(2) invariant, every global spin
rotation of a CCI product state must also be a ground state.  A global
SU(2) transformation rotates all spins simultaneously and induces
the corresponding rigid SO(3) rotation of the tetrahedral frame
$\{\mathbf n^A,\mathbf n^B,\mathbf n^C,\mathbf n^D\}$, while leaving
the coloring pattern $\{\alpha_j\}$ unchanged.  We therefore require
the parent Hamiltonian to annihilate the CCI product states
(\ref{eq:global_CCI}) together with all their global SU(2)
rotations.

\section{The parent Hamiltonian on a tetrahedron}
\label{sec:parent}

We now construct a local operator whose kernel contains the chiral
color-ice states of a chosen handedness.  Spin-rotation invariance
strongly constrains this operator: for spin $1/2$, it fixes the local
term essentially uniquely and, as we show below, suggests its natural
generalization to arbitrary spin $S$.

\subsection{Construction principle}
\label{sec:principle}

The site-factorized structure of the CCI states allows us to impose
the parent constraint locally on a single tetrahedron.  For fixed spin
length $S$, we take the state $\ket{\Psi^{ABCD}}$ of
Eq.~(\ref{eq:Psi_CCI}) as a convenient reference for one of the two
chiral sectors and define
\begin{equation}
  \mathscr{V}^{S}
  =
  \mathrm{span}
  \left\{
  \bigl[\hat U^{(S)}(R)\bigr]^{\otimes 4}
  \ket{\Psi^{ABCD}}
  \,\middle|\,
  R\in SO(3)
  \right\}.
  \label{eq:VS_def}
\end{equation}
Here $\hat U^{(S)}(R)$ denotes the spin-$S$ rotation associated with
$R$, acting identically on all four spins.  For half-integer $S$, the
two SU(2) rotations corresponding to the same $R\in SO(3)$ differ
by a sign in the single-spin representation.  This sign disappears
in the fourfold tensor product, so the operator in
Eq.~(\ref{eq:VS_def}) is unambiguously defined.

The rotational \emph{orbit} in Eq.~(\ref{eq:VS_def})---the set of
states obtained by acting with all global spin rotations on the
reference state---already contains all
twelve local CCI product states in the same chiral sector.  Let
$\hat P_\sigma$ denote the operator that transports the spin on site
$i$ to site $\sigma(i)$.  For every even site permutation
$\sigma\in A_4$, there exists a proper rotation
$R_\sigma\in SO(3)$ of the tetrahedral frame such that
\begin{equation}
  \hat P_\sigma\ket{\Psi^{ABCD}}
  =
  e^{i\phi_\sigma}
  \bigl[\hat U^{(S)}(R_\sigma)\bigr]^{\otimes4}
  \ket{\Psi^{ABCD}},
\end{equation}
where $e^{i\phi_\sigma}$ is an irrelevant overall phase.  Thus the
twelve states generated by even site permutations already belong to
the rotational orbit in Eq.~(\ref{eq:VS_def}).  Odd site permutations
reverse the handedness and generate the opposite chiral sector.

We then seek a positive-semidefinite local parent term
$\mathcal H_{\mathrm{CCI}}^{(S)}$ whose kernel contains the subspace
$\mathscr V^S$ associated with the chosen chiral sector.  The parent
term for the opposite chirality is obtained by time reversal.  We
therefore define
\begin{equation}
  \mathcal H_{\mathrm{CCI}}^{(S),\varepsilon}
  =
  \begin{cases}
    \mathcal H_{\mathrm{CCI}}^{(S)},
      & \varepsilon=+1, \\[1mm]
    \mathcal T\,\mathcal H_{\mathrm{CCI}}^{(S)}\mathcal T^{-1},
      & \varepsilon=-1 .
  \end{cases}
  \label{eq:local_CCI_pm}
\end{equation}

In practice, we determine $\mathscr{V}^{S}$ numerically by sampling
global spin rotations of the reference state $\ket{\Psi^{ABCD}}$,
including the twelve even site permutations, and orthogonalizing the
resulting set. The rank rapidly saturates as the number of sampled
rotations increases, yielding $\dim\mathscr{V}^{S}$.

\begin{table*}[tp]
\caption{
Total-spin decomposition of three subspaces on a single tetrahedron of
four spins $S$: the fixed-chirality subspace $\mathscr V^S$ spanned by
the globally rotated CCI product states [Eq.~(\ref{eq:VS_def})], 
the full color-ice subspace $\mathscr W^S$ spanned by the
globally rotated CCI product states of both chiral sectors, and the
full Hilbert space $\mathscr H^S$.
Each entry gives the multiplicity of the corresponding
SU(2) multiplet with total spin $S_{\mathrm{tet}}$; weighting each
entry by $2S_{\mathrm{tet}}+1$ gives the total dimensions in the last
row.
For $S_{\mathrm{tet}}<2S$, the multiplicities are
$2S_{\mathrm{tet}}+1$ in $\mathscr V^S$ and
$2(2S_{\mathrm{tet}}+1)$ in $\mathscr W^S$, in agreement with the
Anderson towers of tetrahedral order derived in
Appendix~\ref{app:tower}.
}
\label{tab:Stot_vs_S}
\begin{ruledtabular}
\begin{tabular}{c ccc ccc ccc ccc ccc}
$S_{\mathrm{tet}}$
& \multicolumn{3}{c}{$S=\frac{1}{2}$}
& \multicolumn{3}{c}{$S=1$}
& \multicolumn{3}{c}{$S=\frac{3}{2}$}
& \multicolumn{3}{c}{$S=2$}
& \multicolumn{3}{c}{$S=\frac{5}{2}$} \\
& $\mathscr{V}^{1/2}_{S_{\mathrm{tet}}}$ & $\mathscr{W}^{1/2}_{S_{\mathrm{tet}}}$ & $\mathscr{H}^{1/2}_{S_{\mathrm{tet}}}$
& $\mathscr{V}^{1}_{S_{\mathrm{tet}}}$ & $\mathscr{W}^{1}_{S_{\mathrm{tet}}}$ & $\mathscr{H}^{1}_{S_{\mathrm{tet}}}$
& $\mathscr{V}^{3/2}_{S_{\mathrm{tet}}}$ & $\mathscr{W}^{3/2}_{S_{\mathrm{tet}}}$ & $\mathscr{H}^{3/2}_{S_{\mathrm{tet}}}$
& $\mathscr{V}^{2}_{S_{\mathrm{tet}}}$ & $\mathscr{W}^{2}_{S_{\mathrm{tet}}}$ & $\mathscr{H}^{2}_{S_{\mathrm{tet}}}$
& $\mathscr{V}^{5/2}_{S_{\mathrm{tet}}}$ & $\mathscr{W}^{5/2}_{S_{\mathrm{tet}}}$ & $\mathscr{H}^{5/2}_{S_{\mathrm{tet}}}$
\\
\hline
$0$
& $1$ & $2$ & $2$
& $1$ & $2$ & $3$
& $1$ & $2$ & $4$
& $1$ & $2$ & $5$
& $1$ & $2$ & $6$
\\
$1$
& $3$ & $3$ & $3$
& $3$ & $6$ & $6$
& $3$ & $6$ & $9$
& $3$ & $6$ & $12$
& $3$ & $6$ & $15$
\\
$2$
& $1$ & $1$ & $1$
& $5$ & $6$ & $6$
& $5$ & $10$ & $11$
& $5$ & $10$ & $16$
& $5$ & $10$ & $21$
\\
$3$
&     &     &
& $3$ & $3$ & $3$
& $7$ & $10$ & $10$
& $7$ & $14$ & $17$
& $7$ & $14$ & $24$
\\
$4$
&     &     &
& $1$ & $1$ & $1$
& $5$ & $6$ & $6$
& $9$ & $14$ & $15$
& $9$ & $18$ & $24$
\\
$5$
&     &     &
&     &     &
& $3$ & $3$ & $3$
& $7$ & $10$ & $10$
& $11$ & $18$ & $21$
\\
$6$
&     &     &
&     &     &
& $1$ & $1$ & $1$
& $5$ & $6$ & $6$
& $9$ & $14$ & $15$
\\
$7$
&     &     &
&     &     &
&     &     &
& $3$ & $3$ & $3$
& $7$ & $10$ & $10$
\\
$8$
&     &     &
&     &     &
&     &     &
& $1$ & $1$ & $1$
& $5$ & $6$ & $6$
\\
$9$
&     &     &
&     &     &
&     &     &
&     &     &
& $3$ & $3$ & $3$
\\
$10$
&     &     &
&     &     &
&     &     &
&     &     &
& $1$ & $1$ & $1$
\\
\hline
Total dim.
& $15$  & $16$   & $16$
& $65$  & $80$   & $81$
& $175$ & $240$  & $256$
& $369$ & $544$  & $625$
& $671$ & $1040$ & $1296$
\end{tabular}
\end{ruledtabular}
\end{table*}

\subsection{Spin one-half: chiral tetrahedral singlets}
\label{sec:spinhalf}

\begin{figure}[tbp]
    \centering
    \includegraphics[width=0.7\columnwidth]{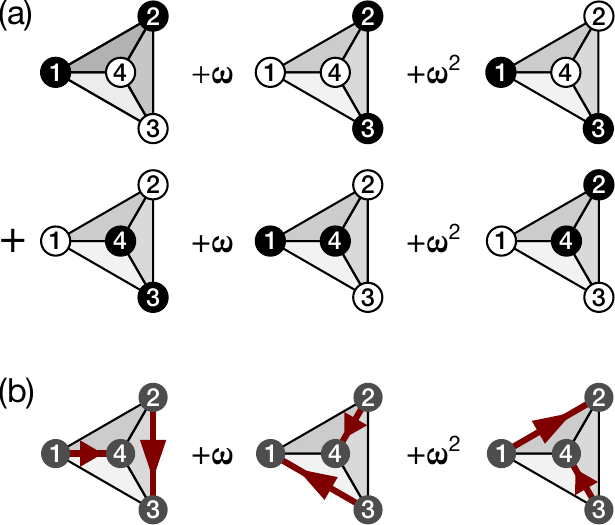}
\caption{
    The chiral tetrahedral singlet $\ket{+}$, Eq.~(\ref{eq:chistate+}), shown on a tetrahedron projected as a triangle with one interior vertex.
(a)~Superposition of the six $S^z$-basis states with two up and two
down spins, with relative phases $1$, $\omega$, and $\omega^2$. Filled and open
circles denote the two spin states.
(b)~The same state represented as a superposition of the three
oriented valence-bond coverings with relative phases
$1$, $\omega$, and $\omega^2$
[Eq.~(\ref{eq:plus_dimers_oriented})], up to an overall phase.
An arrow from $i$ to $j$ denotes the oriented singlet valence bond 
$[ij]$.
}
    \label{fig:chiral_singlet}
\end{figure}

The spin-$1/2$ case makes the construction particularly transparent.
The local Hilbert space $\mathscr H^{1/2}$ has dimension $2^4=16$,
whereas the construction above yields
\begin{equation}
  \dim\mathscr V^{1/2}=15 .
\end{equation}
Its orthogonal complement is therefore one dimensional.  To identify
the missing state, we consider the quantum tetrahedral-chirality
operator $\hat\chi_{\mathrm{tet}}$ corresponding to
Eq.~(\ref{eq:tetrahedronchirality}).
For four spin-$1/2$ degrees of freedom, $\hat\chi_{\mathrm{tet}}$ has
eigenvalues $0$ and $\pm\sqrt{3}$.  The two nonzero eigenvalues belong
to a time-reversal pair of total-spin singlets \cite{Tsunetsugu2001},
\begin{equation}
  \hat\chi_{\mathrm{tet}}\ket{\pm}
  =
  \pm\sqrt{3}\,\ket{\pm},
  \label{eq:chipm}
\end{equation}
where
\begin{subequations}
\label{eq:chistates}
\begin{align}
\ket{+}
&=
\frac{1}{\sqrt{6}}
\left[
\ket{\downarrow\downarrow\uparrow\uparrow}
+
\ket{\uparrow\uparrow\downarrow\downarrow}
+
\omega
\left(
\ket{\downarrow\uparrow\uparrow\downarrow}
+
\ket{\uparrow\downarrow\downarrow\uparrow}
\right)
\right.
\nonumber\\
&\hphantom{=\frac{1}{\sqrt{6}}\Bigl[}
\left.
+
\omega^2
\left(
\ket{\downarrow\uparrow\downarrow\uparrow}
+
\ket{\uparrow\downarrow\uparrow\downarrow}
\right)
\right],
\label{eq:chistate+}\\
\ket{-}
&=
\ket{+}^{*},
\end{align}
\end{subequations}
with $\omega=e^{2\pi i/3}$.  The arrows label the spin states on
sites $1$--$4$ from left to right.

From the viewpoint of site permutations, the two-dimensional singlet
sector carries the $[2,2]$ irreducible representation of $S_4$,
denoted $\mathsf E$ in Table~\ref{tab:T_character_table}.  Upon
restriction to the proper tetrahedral subgroup $A_4$, this
representation decomposes as
\begin{equation}
  \mathsf E\!\downarrow_{A_4}
  =
  \mathsf E_+\oplus\mathsf E_-,
\end{equation}
with $\ket{+}$ and $\ket{-}$ spanning the two complex-conjugate
one-dimensional components.

The chiral singlets take a particularly transparent form in the
valence-bond representation.  We define a singlet on bond $(ij)$ by
\begin{equation}
  [ij]=-[ji]
  =
  \frac{1}{\sqrt2}
  \left(
    \ket{\uparrow_i\downarrow_j}
    -
    \ket{\downarrow_i\uparrow_j}
  \right).
\end{equation}
The three valence-bond coverings satisfy
\begin{equation}
  [14][23]+[24][31]+[34][12]=0 .
  \label{eq:dimer_relation}
\end{equation}
Let $\hat P_{(123)}$ denote the cyclic site permutation that transports
the spins as $1\to2\to3\to1$.  It acts on the three coverings as
\begin{equation}
  [14][23]
  \mapsto
  [24][31]
  \mapsto
  [34][12]
  \mapsto
  [14][23].
  \label{eq:P_on_matchings}
\end{equation}
Consequently,
\begin{subequations}
\label{eq:chistatesVB}
\begin{align}
  \ket{+}
  &=
  \frac{\sqrt2}{3}e^{i\pi/6}
  \left(
  [14][23]+\omega[24][31]+\omega^2[34][12]
  \right),
  \label{eq:plus_dimers_oriented}
  \\
  \ket{-}
  &=
  \frac{\sqrt2}{3}e^{-i\pi/6}
  \left(
  [14][23]+\omega^2[24][31]+\omega[34][12]
  \right),
  \label{eq:minus_dimers_oriented}
\end{align}
\end{subequations}
which are eigenstates of $\hat P_{(123)}$ with eigenvalues
$\omega^2$ and $\omega$, respectively.  Thus the two chiralities are
the two nontrivial cyclic modes of the three valence-bond coverings, with their handedness encoded in the
relative phases $1,\omega,\omega^2$, as illustrated in
Fig.~\ref{fig:chiral_singlet}. They provide a minimal four-site
analog of the chiral spin states of
Refs.~\cite{Kalmeyer1987,WWZ1989}.

For any local CCI state in the same chiral sector as the reference
state of Eq.~(\ref{eq:Psi_CCI}), including all even site permutations
and all global SU(2) rotations, one finds
\begin{subequations}
\begin{align}
  \langle + \ket{\Psi_{\rm CCI}}
  &=0,
  \\
  \bigl|\langle - \ket{\Psi_{\rm CCI}}\bigr|
  &=\frac{\sqrt{2}}{3}.
\end{align}
\end{subequations}
The first relation shows that the entire set of rotated
fixed-chirality CCI states
is orthogonal to the $\chi_{\mathrm{tet}}=+\sqrt3$ singlet $\ket{+}$.
Since $\dim\mathscr V^{1/2}=15$, this set spans the full
orthogonal complement of $\ket{+}$ in the 16-dimensional local
Hilbert space.
The local parent term is therefore the rank-one projector
\begin{equation}
  \mathcal H_{\mathrm{CCI}}^{(1/2)}
  =
  \ket{+}\bra{+}.
  \label{eq:proj++}
\end{equation}
The projector selects the chosen CCI handedness while preserving full
spin-rotation invariance, since $\ket{+}$ is a total-spin singlet.

Since $\hat\chi_{\mathrm{tet}}$ has eigenvalues $0$ and
$\pm\sqrt{3}$, and $\ket{+}$ is the unique eigenstate with eigenvalue
$+\sqrt{3}$, the projector can equivalently be written as
\begin{equation}
  \mathcal H_{\mathrm{CCI}}^{(1/2)}
  =
  \frac{1}{6}
  \left(
  \hat\chi_{\mathrm{tet}}
  +
  \sqrt{3}
  \right)
  \hat\chi_{\mathrm{tet}}.
  \label{eq:spin_1_2_hamiltonian_with_chirality}
\end{equation}
Expanding this polynomial in $\hat\chi_{\mathrm{tet}}$ in terms of spin
operators gives
\begin{multline}
  \mathcal H_{\mathrm{CCI}}^{(1/2)}
  =
  \frac{1}{16}
  +
  \frac{1}{2\sqrt{3}}
  \sum_{\langle ijk\rangle}
  \hat{\mathbf S}_{i}\cdot
  \left(
  \hat{\mathbf S}_j\times\hat{\mathbf S}_k
  \right)
  \\
  -\frac{1}{12}
  \sum_{\langle ij\rangle}
  \hat{\mathbf S}_i\cdot\hat{\mathbf S}_j
  +
  \frac{1}{3}
  \sum_{\langle ijkl\rangle}
  \left(
  \hat{\mathbf S}_i\cdot\hat{\mathbf S}_j
  \right)
  \left(
  \hat{\mathbf S}_k\cdot\hat{\mathbf S}_l
  \right).
  \label{eq:spin_1_2_hamiltonian}
\end{multline}
Here $\langle ijk\rangle$ runs over the four oriented faces of the
tetrahedron with the handedness convention of
Eq.~(\ref{eq:tetrahedronchirality}), $\langle ij\rangle$ over its six
bonds, and $\langle ijkl\rangle$ over the three pairings of opposite
bonds, $(12,34)$, $(13,24)$, and $(14,23)$.  The parent term thus
involves only scalar chiralities, Heisenberg exchanges, and products
of Heisenberg exchanges.  Useful identities relating these operators
for $S=1/2$ are collected in Appendix~\ref{app:chirality}.

\subsection{Arbitrary spin}
\label{sec:spinS}

We now turn to $S>1/2$.  For $S=1$, the local Hilbert space has
dimension $3^4=81$, whereas the globally rotated CCI product states
span a 65-dimensional subspace.  Repeating the construction for
$S=3/2$, $2$, and $5/2$ gives the dimensions and total-spin
decompositions collected in Table~\ref{tab:Stot_vs_S}.  
For all spins studied, the dimensions follow the simple pattern
\begin{equation}
  \dim \mathscr V^S
  =
  (2S+1)^4-(2S)^4 .
  \label{eq:dimV}
\end{equation}
We show the origin of this relation for arbitrary $S$ in
Sec.~\ref{sec:schwinger}.

Guided by the spin-$1/2$ parent term
(\ref{eq:spin_1_2_hamiltonian}), we search within the same operator
basis---bilinear exchanges, scalar chiralities, and products of
bilinear exchanges---for a positive-semidefinite operator that
annihilates $\mathscr V^S$.  A particularly simple solution, valid for
arbitrary $S$, is
\begin{multline}
  \mathcal H_{\mathrm{CCI}}^{(S)}
  =
  \frac{S}{\sqrt{3}}
  \sum_{\langle ijk\rangle}
  \hat{\mathbf S}_{i}\cdot
  \left(
  \hat{\mathbf S}_{j}\times\hat{\mathbf S}_{k}
  \right)
  \\
  +\frac{1}{3}
  \sum_{\langle ijkl\rangle}
  \bigl(
  S^2-\hat{\mathbf S}_i\cdot\hat{\mathbf S}_j
  \bigr)
  \bigl(
  S^2-\hat{\mathbf S}_k\cdot\hat{\mathbf S}_l
  \bigr).
  \label{eq:spin_S_hamiltonian}
\end{multline}
We derive the general result
analytically in Appendix~\ref{app:HamS}.
Its classical large-$S$ limit and complete coherent zero-mode
variety are discussed in Appendix~\ref{app:classical}.

\subsection{Exact quantum handedness of the local kernel}
\label{sec:exact_quantum_handedness}

For $S=1/2$, Eq.~(\ref{eq:proj++}) shows directly that the local
parent Hamiltonian projects out the chiral singlet of the opposite
handedness. The form of Eq.~(\ref{eq:spin_S_hamiltonian}) allows this
chirality-sign selection to be extended to arbitrary spin $S$ and to
the complete local quantum kernel.

For the reference chirality, write the local parent term as
\begin{equation}
  \mathcal H_{\mathrm{CCI}}^{(S)}
  =
  \frac{S}{\sqrt3}\hat\chi_{\mathrm{tet}}
  +
  \frac13\mathcal R ,
  \label{eq:H_R_chi}
\end{equation}
where
\begin{equation}
  \mathcal R
  =
  \sum_{\langle ijkl\rangle}
  \left(
    S^2-\hat{\mathbf S}_i\cdot\hat{\mathbf S}_j
  \right)
  \left(
    S^2-\hat{\mathbf S}_k\cdot\hat{\mathbf S}_l
  \right),
  \label{eq:R_def}
\end{equation}
and the sum runs over the three pairings of opposite bonds,
$(12,34)$, $(13,24)$, and $(14,23)$.

Each factor appearing in Eq.~(\ref{eq:R_def}) is positive
semidefinite. Indeed, for two spin-$S$ sites in a sector of pair
total spin $J=0,\ldots,2S$,
\begin{align}
  S^2-\hat{\mathbf S}_i\cdot\hat{\mathbf S}_j
  & =
  \frac12
  \left[
    2S(2S+1)-J(J+1)
  \right] \nonumber\\
  & =
  \frac12(2S-J)(2S+J+1)
  \geq0 .
  \label{eq:bond_factor_positive}
\end{align}
The two factors in each term of Eq.~(\ref{eq:R_def}) act on disjoint
pairs of sites and therefore commute. Their product is consequently
positive semidefinite, and hence
\begin{equation}
  \mathcal R\succeq0 .
  \label{eq:R_positive}
\end{equation}

Let $\ket{\Psi}$ be any normalized state in the local zero-energy
kernel, $  \mathcal H_{\mathrm{CCI}}^{(S)}\ket{\Psi}=0 $.
Taking the expectation value of Eq.~(\ref{eq:H_R_chi}) gives
\begin{equation}
  \frac{S}{\sqrt3}
  \langle\Psi|\hat\chi_{\mathrm{tet}}|\Psi\rangle
  =
  -\frac13
  \langle\Psi|\mathcal R|\Psi\rangle .
\end{equation}
Since $\mathcal R\succeq0$,
\begin{equation}
  \langle\Psi|\hat\chi_{\mathrm{tet}}|\Psi\rangle
  \leq0
    \label{eq:local_chirality_sign}
\end{equation}
Thus, for four spins of arbitrary length $S$, the complete quantum
kernel of the local CCI parent Hamiltonian obeys the same
chirality-sign constraint as its coherent-product states. Zero
chirality is allowed, and states in the kernel need not themselves
be eigenstates of $\hat\chi_{\mathrm{tet}}$.

For the time-reversed parent term,
$\mathcal T\mathcal H_{\mathrm{CCI}}^{(S)}\mathcal T^{-1}$, the operator
$\mathcal R$ is unchanged while
$\hat\chi_{\mathrm{tet}}\to-\hat\chi_{\mathrm{tet}}$, and the inequality in
Eq.~(\ref{eq:local_chirality_sign}) is therefore reversed.

\section{Singlet-annihilation formulation}
\label{sec:schwinger}

In this section, we show that the parent Hamiltonian takes a particularly simple form in the Schwinger-boson representation: the local term factorizes as
\(\mathcal H_{\mathrm{CCI}}^{(S)}=\mathcal B^\dagger\mathcal B\),
making its positive-semidefinite character explicit. Here \(\mathcal B\) is a four-site singlet-annihilation operator, so that the zero-energy condition is simply \(\mathcal B\ket{\Psi}=0\). This formulation also provides a direct route to the kernel's dimension and total-spin content, as well as the exact local excitation gap.

\subsection{The operator $\mathcal{B}$}

The homogeneous scaling of all the terms in
Eq.~(\ref{eq:spin_S_hamiltonian}) with the spin length makes the
Schwinger-boson representation particularly natural, since $S$
enters through the local boson-number constraint. Introducing two
bosons $a_i$ and $b_i$ at each site, the spin operators are
represented as
\begin{subequations}
\label{eq:Schwinger_bosons}
\begin{align}
  \hat S^x_i
  &=
  \frac{1}{2}
  \left(
  a^\dagger_i b^{\vphantom{\dagger}}_i
  +
  b^\dagger_i a^{\vphantom{\dagger}}_i
  \right),
  \\
  \hat S^y_i
  &=
  \frac{1}{2i}
  \left(
  a^\dagger_i b^{\vphantom{\dagger}}_i
  -
  b^\dagger_i a^{\vphantom{\dagger}}_i
  \right),
  \\
  \hat S^z_i
  &=
  \frac{1}{2}
  \left(
  a^\dagger_i a^{\vphantom{\dagger}}_i
  -
  b^\dagger_i b^{\vphantom{\dagger}}_i
  \right),
\end{align}
with the local constraint
\begin{equation}
  a^\dagger_i a^{\vphantom{\dagger}}_i
  +
  b^\dagger_i b^{\vphantom{\dagger}}_i
  =
  2S .
\end{equation}
\end{subequations}

We introduce the SU(2)-invariant singlet pair-annihilation operators
\begin{equation}
  F_{ij}=a_i b_j-b_i a_j,
  \label{eq:Fij}
\end{equation}
which remove one boson from each of sites $i$ and $j$ in the singlet channel \cite{Arovas_Auerbach_PRB.38.316_1988,*Auerbach_Arovas_PhysRevLett.61.617_1988} and satisfy $F_{ij}=-F_{ji}$.
Next, we define the three four-site singlet-annihilation operators corresponding to the three perfect matchings of the tetrahedron,
\begin{equation}
  X=F_{14}F_{23},
  \qquad
  Y=F_{24}F_{31},
  \qquad
  Z=F_{34}F_{12}.
  \label{eq:XYZ}
\end{equation}
These correspond to the singlet coverings appearing in
Eqs.~(\ref{eq:plus_dimers_oriented}) and
(\ref{eq:minus_dimers_oriented}) and satisfy
\begin{equation}
  X+Y+Z=0,
  \label{eq:XYZlindep}
\end{equation}
in analogy with Eq.~(\ref{eq:dimer_relation}).

Motivated by the chiral phase patterns of the $S=1/2$ singlets, we
form the two complex-conjugate combinations
\begin{subequations}
\label{eq:B_in_F}
\begin{align}
  \mathcal B
  &=
  \frac{e^{-i\pi/6}}{\sqrt{18}}
  \left(
    X+\omega^2Y+\omega Z
  \right),
  \\
  \mathcal B^{*}
  &=
  \frac{e^{i\pi/6}}{\sqrt{18}}
  \left(
    X+\omega Y+\omega^2 Z
  \right).
\end{align}
\end{subequations}
Here the star denotes complex conjugation of the coefficients. Their
Hermitian conjugates create the two $S=1/2$ chiral singlets,
\begin{equation}
  \mathcal{B}^{\dagger}\ket{\mathrm{vac}}=\ket{+},
  \qquad
  \mathcal{B}^{*\dagger}\ket{\mathrm{vac}}=\ket{-},
  \label{eq:Bdag_creates_chiral}
\end{equation}
shown in Eqs.~(\ref{eq:chistatesVB}) and Fig.~\ref{fig:chiral_singlet}(a). Correspondingly,
\begin{equation}
  \mathcal{B}\ket{-}=0,
  \qquad
  \mathcal{B}^{*}\ket{+}=0.
  \label{eq:B_kills_minus}
\end{equation}
Under the cyclic permutation $\hat P_{(123)}$, for which
$X\to Y\to Z\to X$ [cf.~Eq.~(\ref{eq:P_on_matchings})], the
annihilation operator transforms as
\begin{equation}
  \hat P_{(123)}\,
  \mathcal B\,
  \hat P_{(123)}^{-1}
  =
  \omega\,\mathcal B .
\end{equation}
Consequently,
$\mathcal B^\dagger\ket{\mathrm{vac}}=\ket{+}$ transforms with
eigenvalue $\omega^2$, while
$\mathcal B^{*\dagger}\ket{\mathrm{vac}}=\ket{-}$ transforms with
eigenvalue $\omega$, in agreement with
Eq.~(\ref{eq:chistatesVB}). Thus, with the convention of
Table~\ref{tab:T_character_table}, the singlet belonging to the CCI
kernel transforms as $\mathsf E_+$ under this three-cycle.

For $S=1/2$, the projector in Eq.~(\ref{eq:proj++}) suggests the
generalization
\begin{equation}
  \mathcal H_{\mathrm{CCI}}^{(S)}
  =
  \mathcal{B}^{\dagger}\mathcal{B}
  \label{eq:BdB}
\end{equation}
to arbitrary $S$. Remarkably, this factorization is exact. Expanding
Eq.~(\ref{eq:B_in_F}) gives
\begin{multline}
  \mathcal{B}
  =
  \frac{1}{\sqrt{6}}
  \bigl[
  a_4a_3b_2b_1+b_4b_3a_2a_1
  +
  \omega
  \left(
  a_4b_3a_2b_1+b_4a_3b_2a_1
  \right)
  \\
  +
  \omega^2
  \left(
  a_4b_3b_2a_1+b_4a_3a_2b_1
  \right)
  \bigr].
  \label{eq:Bop}
\end{multline}
Substituting this expression into
$\mathcal{B}^{\dagger}\mathcal{B}$ and using the local constraint in
Eq.~(\ref{eq:Schwinger_bosons}) reproduces
Eq.~(\ref{eq:spin_S_hamiltonian}). The positive-semidefinite
character of the local Hamiltonian is therefore explicit.

Since each of $X$, $Y$, and $Z$ is an SU(2) scalar and removes
exactly one boson from each site, the same is true of
$\mathcal{B}$ and $\mathcal{B}^{*}$. In particular,
\begin{equation}
  \bigl[\hat S_{\mathrm{tet}}^\alpha,\mathcal{B}\bigr]=0,
  \qquad
  \hat S_{\mathrm{tet}}^\alpha
  =
  \sum_{i=1}^{4}\hat S_i^\alpha,
  \qquad
  \alpha=x,y,z.
  \label{eq:B_singlet}
\end{equation}
Thus, $\mathcal{B}$ transforms as a scalar under global spin
rotations and maps the Hilbert space with spin $S$ on every site to
that with spin $S-\tfrac{1}{2}$,
\begin{equation}
  \mathcal{B}:
  \mathscr{H}^{S}
  \longrightarrow
  \mathscr{H}^{S-1/2},
\end{equation}
while preserving the total spin. Its Hermitian conjugate
$\mathcal{B}^{\dagger}$ acts in the reverse direction.
Furthermore, since $\mathcal{B}$ and $\mathcal{B}^{*}$ contain only
boson annihilation operators,
\begin{equation}
  [\mathcal{B},\mathcal{B}^{*}]=0.
\end{equation}

For arbitrary $S$, the factorization in Eq.~(\ref{eq:BdB}) implies
that a state $\ket{\Psi}$ has zero local energy if and only if
\begin{equation}
  \mathcal{B}\ket{\Psi}=0.
  \label{eq:Bpsi0}
\end{equation}
This \emph{singlet-annihilation constraint} provides the algebraic
characterization of the local zero-energy states.

The transformation of $\mathcal{B}$ and $\mathcal{B}^{*}$ under
permutations of the four sites makes the chirality selection
explicit. An even permutation maps
$\mathcal{B}\to\omega^{k}\mathcal{B}$, whereas an odd permutation
maps $\mathcal{B}\to\omega^{k}\mathcal{B}^{*}$, with
$k\in\{0,1,2\}$ depending on the permutation. Consequently,
$\mathcal{H}^{(S)}_{\mathrm{CCI}}$ in Eq.~(\ref{eq:BdB}) is
invariant under the alternating group $A_4$ of even permutations,
whereas an odd permutation maps it to the parent Hamiltonian of the
opposite chirality,
$\mathcal{B}^{*\dagger}\mathcal{B}^{*}$. The construction therefore
selects one of the two time-reversal-related chiral sectors while
preserving full spin-rotation symmetry.

\subsection{Parent Hamiltonian for the full color-ice subspace}
\label{sec:CIparent}

Table~\ref{tab:Stot_vs_S} also gives the decomposition of the full
color-ice subspace $\mathscr W^S$, generated by all $24$ site
permutations and global spin rotations. Since its two chiral sectors
lie in the kernels of $\mathcal B$ and $\mathcal B^{*}$,
respectively, and $[\mathcal B,\mathcal B^{*}]=0$, the product
\begin{equation}
  \mathcal C
  =
  \mathcal B^{*}\mathcal B
  =
  \mathcal B\mathcal B^{*}
  \label{eq:CBB}
\end{equation}
annihilates the entire subspace $\mathscr W^S$.
$\mathcal C$ removes two Schwinger bosons from each site
and is an SU(2) scalar, since both $\mathcal B$ and
$\mathcal B^{*}$ are scalars.  A positive-semidefinite parent
Hamiltonian for the full color-ice subspace is consequently
\begin{equation}
  \mathcal H_\mathrm{CI}^{(S)}
  =
  \mathcal C^\dagger\mathcal C
  =
  \bigl(\mathcal B^{*}\mathcal B\bigr)^\dagger
  \bigl(\mathcal B^{*}\mathcal B\bigr).
  \label{eq:HCI}
\end{equation}

The absence of chirality selection becomes particularly transparent
in the valence-bond representation.  Using
Eqs.~(\ref{eq:XYZlindep}) and (\ref{eq:B_in_F}), one obtains
\begin{equation}
  \mathcal C
  =
  \frac{1}{12}
  \left(
    X^2+Y^2+Z^2
  \right).
  \label{eq:C_symmetric}
\end{equation}
Thus, the complex phase pattern distinguishing $\mathcal B$ from
$\mathcal B^{*}$ drops out. A permutation of the four sites merely
permutes $X$, $Y$, and $Z$, up to signs from
$F_{ij}=-F_{ji}$ that disappear upon squaring. Consequently,
$\mathcal C$, and hence $\mathcal H_\mathrm{CI}^{(S)}$, is invariant
under the full permutation group $S_4$.

The lowest-spin cases illustrate the construction directly. For
$S=1/2$, the two chiral sectors together span the full tetrahedron
Hilbert space,
$\mathscr W^{1/2}=\mathscr H^{1/2}$. Hence $\mathcal C$ vanishes
identically in the physical Hilbert space, and no nontrivial parent
Hamiltonian exists for the full color-ice subspace.

For $S=1$, by contrast, Table~\ref{tab:Stot_vs_S} gives
$\dim\mathscr W^1=80$ in the $3^4=81$-dimensional Hilbert space.
The one-dimensional complement is the achiral singlet
$\mathcal C^\dagger\ket{\mathrm{vac}}$. Since the three-dimensional
total-spin-singlet sector consists of this state, with
$\hat\chi_{\mathrm{tet}}=0$, and two chiral singlets with
$\hat\chi_{\mathrm{tet}}=\pm4\sqrt3$ belonging to $\mathscr W^1$, we can
write
\begin{equation}
  \mathcal{H}_\mathrm{CI}^{(1)}
  \propto
  \mathcal{P}_{0}
  \left(
    48-\hat{\chi}_\mathrm{tet}^{\,2}
  \right)
  \mathcal{P}_{0},
  \label{eq:HCI1}
\end{equation}
where $\mathcal P_{0}$ projects onto the total-spin-singlet subspace.
Thus $\mathcal H_\mathrm{CI}^{(1)}$ annihilates the two chiral
singlets and assigns positive energy only to the achiral one.

\subsection{Zero-energy states and total-spin multiplets}
\label{sec:kernel_dims}

The factorization
$\mathcal H_{\mathrm{CCI}}^{(S)}=\mathcal B^\dagger\mathcal B$
also gives a simple way to count the local zero-energy states.
The operator $\mathcal B$ removes one Schwinger boson from each site,
and therefore maps four spin-$S$ sites onto four
spin-$(S-\frac12)$ sites. Here and throughout, $S_{\mathrm{tet}}$
denotes the total spin of the four spins of a single tetrahedron,
while $J$ is reserved for the total spin of a finite lattice
cluster. Since $\mathcal B$ commutes with the total
spin, it acts separately in each $S_{\mathrm{tet}}$ sector,
\begin{equation}
  \mathcal B:
  \mathscr H^{S}_{S_{\mathrm{tet}}}
  \longrightarrow
  \mathscr H^{S-\frac12}_{S_{\mathrm{tet}}},
  \label{eq:Bmap}
\end{equation}
where $\mathscr H^{S}_{S_{\mathrm{tet}}}$ is the full subspace with
total spin $S_{\mathrm{tet}}$.

As shown in Appendix~\ref{app:surjectivity}, this map is surjective in every total-spin sector, meaning that every state in the target space is obtained from some state in the original space.
 The number of states
annihilated by $\mathcal B$ is therefore simply the difference between
the dimensions of the two spaces,
\begin{equation}
  \dim\ker\!\left(
  \mathcal B\big|_{S_{\mathrm{tet}}}
  \right)
  =
  \dim\mathscr H^{S}_{S_{\mathrm{tet}}}
  -
  \dim\mathscr H^{S-\frac12}_{S_{\mathrm{tet}}}.
  \label{eq:dim_recursion}
\end{equation}
In particular, the largest total spin in the target space is
$4S-2$. Hence the complete $S_{\mathrm{tet}}=4S$ and $4S-1$
sectors are automatically annihilated by $\mathcal B$.

Since a spin-$S_{\mathrm{tet}}$ multiplet contains
$2S_{\mathrm{tet}}+1$ states, dividing
Eq.~(\ref{eq:dim_recursion}) by this factor gives the number of such
multiplets. Summing over all total spins gives
\begin{equation}
  \dim\ker\mathcal B
  =
  (2S+1)^4-(2S)^4 .
  \label{eq:dim_kerB}
\end{equation}

We now identify the kernel of the local parent term. Since every
globally rotated CCI state has zero energy and
\begin{equation}
  \langle\Psi|
  \mathcal H_{\mathrm{CCI}}^{(S)}
  |\Psi\rangle
  =
  \|\mathcal B|\Psi\rangle\|^2,
\end{equation}
all states in $\mathscr V^S$ are annihilated by $\mathcal B$,
\begin{equation}
  \mathscr V^S\subseteq\ker\mathcal B .
  \label{eq:V_in_kerB}
\end{equation}
Appendix~\ref{app:kernel_span} shows that there are no additional
zero modes: every state orthogonal to $\mathscr V^S$ can be generated
by acting with $\mathcal B^\dagger$. Hence
\begin{equation}
  \mathscr V^S=\ker\mathcal B
  \label{eq:V_equals_kerB}
\end{equation}
for arbitrary $S$.

Equation~(\ref{eq:dim_kerB}) therefore proves Eq.~(\ref{eq:dimV}),
while the recursion relation (\ref{eq:dim_recursion}) gives the
total-spin multiplicities of $\mathscr V^S$ listed in
Table~\ref{tab:Stot_vs_S}. For $S=1/2$, this reduces to the
rank-one projector discussed above.

The same argument applies to the full color-ice space
$\mathscr W^S$. The operator
$\mathcal C=\mathcal B^{*}\mathcal B$ removes two bosons from each
site and preserves the total spin,
\begin{equation}
  \mathcal C:
  \mathscr H^{S}_{S_{\mathrm{tet}}}
  \longrightarrow
  \mathscr H^{S-1}_{S_{\mathrm{tet}}}.
\end{equation}
For $S\geq1$, both steps in this map reach their complete target
spaces, as shown in Appendix~\ref{app:surjectivity}. Hence
\begin{equation}
  \dim\ker\!\left(
  \mathcal C\big|_{S_{\mathrm{tet}}}
  \right)
  =
  \dim\mathscr H^{S}_{S_{\mathrm{tet}}}
  -
  \dim\mathscr H^{S-1}_{S_{\mathrm{tet}}}.
  \label{eq:dim_recursion_C}
\end{equation}
Moreover, $\mathcal C$ annihilates all rotated colorings of both
handednesses, so that
$\mathscr W^S\subseteq\ker\mathcal C$.
The Appendix again shows that there are no additional zero-energy
states. Therefore
\begin{equation}
  \mathscr W^S=\ker\mathcal C,
  \qquad S\geq1 .
\end{equation}
It follows that
\begin{equation}
  \dim\mathscr W^S
  =
  (2S+1)^4-(2S-1)^4 ,
  \qquad S\geq1 ,
\end{equation}
and Eq.~(\ref{eq:dim_recursion_C}) gives the multiplets listed for
$\mathscr W^S$ in Table~\ref{tab:Stot_vs_S}. For $S=1/2$,
$\mathcal C$ vanishes identically and
$\mathscr W^{1/2}=\mathscr H^{1/2}$, as discussed above.

\subsection{Spectral pairing and local gap}
\label{sec:spectral_pairing}

The factorization
$\mathcal H_{\mathrm{CCI}}^{(S)}=\mathcal B^\dagger\mathcal B$
also determines the nonzero local spectrum. Since $\mathcal B$
removes one boson from each site,
\begin{equation}
  \mathcal B:
  \mathscr H^S
  \longrightarrow
  \mathscr H^{S-\frac12},
\end{equation}
the spectral partner of $\mathcal B^\dagger\mathcal B$ on
$\mathscr H^S$ is $\mathcal B\mathcal B^\dagger$ on
$\mathscr H^{S-\frac12}$.

Indeed, if
$\mathcal B^\dagger\mathcal B\ket{\Psi}=E\ket{\Psi}$ with $E>0$,
then $\mathcal B\ket{\Psi}\neq0$ and
\begin{equation}
  \mathcal B\mathcal B^\dagger
  \bigl(\mathcal B\ket{\Psi}\bigr)
  =
  E\,\mathcal B\ket{\Psi}.
\end{equation}
Conversely, $\mathcal B^\dagger$ maps every $E>0$ eigenstate of
$\mathcal B\mathcal B^\dagger$ back to an eigenstate of
$\mathcal B^\dagger\mathcal B$ with the same eigenvalue. Thus the two
operators have identical nonzero spectra, including multiplicities.

As shown in Appendix~\ref{app:surjectivity},
\begin{equation}
  \mathcal B^\dagger:
  \mathscr H^{S-\frac12}
  \longrightarrow
  \mathscr H^S
\end{equation}
is injective. Consequently,
\begin{equation}
  \langle\Phi|
  \mathcal B\mathcal B^\dagger
  |\Phi\rangle
  =
  \|\mathcal B^\dagger\ket{\Phi}\|^2
  >0
\end{equation}
for every nonzero
$\ket{\Phi}\in\mathscr H^{S-\frac12}$.
The entire space $\mathscr H^{S-\frac12}$ is therefore paired with
the nonzero-energy sector of $\mathcal H_{\mathrm{CCI}}^{(S)}$.
The remaining states in $\mathscr H^S$ form
$\ker\mathcal B=\mathscr V^S$, giving
\begin{equation}
  \dim\mathscr V^S
  =
  \dim\mathscr H^S
  -
  \dim\mathscr H^{S-\frac12},
\end{equation}
in agreement with Eq.~(\ref{eq:dimV}).

The same pairing determines the local excitation gap,
\begin{equation}
  \Delta_{\rm loc}(S)
  =
  \lambda_{\min}
  \left(
    \mathcal B\mathcal B^\dagger
    \big|_{\mathscr H^{S-\frac12}}
  \right).
\end{equation}
Appendix~\ref{app:local_gap} shows that
$\mathcal B\mathcal B^\dagger\succeq4S^2\mathbb I$ on
$\mathscr H^{S-\frac12}$ and that the bound is saturated by its
maximal-total-spin multiplet, with $S_{\mathrm{tet}}=4S-2$. Hence
\begin{equation}
  \Delta_{\rm loc}(S)=4S^2 .
  \label{eq:local_gap}
\end{equation}
This is the energy required to leave the local zero-energy space.
It sets a local energy scale, but does not by itself imply a finite
many-body gap when the tetrahedral terms are coupled on a lattice.

The multiplet at the gap is obtained directly by spectral pairing.
The unique $S_{\mathrm{tet}}=4S-2$ multiplet in
$\mathscr H^{S-\frac12}$ is mapped
by the SU(2) scalar $\mathcal B^\dagger$ onto an
$S_{\mathrm{tet}}=4S-2$ multiplet
in $\mathscr H^S$ with energy $4S^2$. In particular, if $\ket{F}$ is
its fully polarized highest-weight state, then
$\mathcal B^\dagger\ket{F}$ is the highest-weight state of the
multiplet at the gap. In the spin basis, this state is a superposition
of configurations obtained from the fully polarized spin-$S$ state by
lowering two of the four spins by one unit, with relative phases fixed
by $\mathcal B^\dagger$.

For $S\geq1$, $\mathscr H^S$ contains six $S_{\mathrm{tet}}=4S-2$
multiplets.
Equation~(\ref{eq:dim_recursion}) shows that five belong to the
zero-energy space $\mathscr V^S$, while the remaining one is the
multiplet at energy $4S^2$ identified above.

\section{Coherent-state zero modes and the M\"obius completion rule}\label{sec:mobius}

As shown above, the globally rotated CCI product states span the exact
quantum kernel of the local term $\mathcal H_{\mathrm{CCI}}^{(S)}$. Here we
characterize
the zero modes that are themselves products of spin coherent states.
Their four spin directions obey an $S$-independent completion rule:
once three are specified, the fourth is fixed by a M\"obius
transformation. We now derive this rule from the singlet-annihilation
constraint Eq.~(\ref{eq:Bpsi0}).

We use the stereographic parametrization of the spin-coherent state
$\ket{\mathbf n}$ defined in Eq.~(\ref{eq:coherent_def}). In terms of
Schwinger bosons,
\begin{equation}
  \ket{z;S}
  =
  \frac{(a^\dagger+z b^\dagger)^{2S}}
  {(1+|z|^2)^S\sqrt{(2S)!}}\ket{0},
  \label{eq:coherent_z}
\end{equation}
with the corresponding spin direction
\begin{equation}
  \mathbf n(z)
  =
  \frac{(2\Re z,\,2\Im z,\,1-|z|^2)}
  {1+|z|^2}.
\end{equation}
The stereographic coordinate
$z\in\hat{\mathbb C}=\mathbb C\cup\{\infty\}$ parametrizes the
Bloch sphere, with $\hat{\mathbb C}$ the Riemann sphere. The action of
a single boson annihilation operator is then
\begin{subequations}
\label{eq:coherent_annihilation}
\begin{align}
  a\ket{z;S}
  &=
  \frac{\sqrt{2S}}{\sqrt{1+|z|^2}}
  \ket{z;S-\tfrac12},
  \\
  b\ket{z;S}
  &=
  \frac{\sqrt{2S}\,z}{\sqrt{1+|z|^2}}
  \ket{z;S-\tfrac12}.
\end{align}
\end{subequations}

It is convenient to introduce the two complex-conjugate quadratic
polynomials
\begin{subequations}
\label{eq:p_pm}
\begin{align}
  p_+
  &=
  \frac{1}{\sqrt6}
  \left[
    z_1z_2+z_3z_4
    +\omega^2(z_1z_3+z_2z_4)
    +\omega(z_1z_4+z_2z_3)
  \right],
  \label{eq:p_p}
  \\
  p_-
  &=
  \frac{1}{\sqrt6}
  \left[
    z_1z_2+z_3z_4
    +\omega(z_1z_3+z_2z_4)
    +\omega^2(z_1z_4+z_2z_3)
  \right],
  \label{eq:p_m}
\end{align}
\end{subequations}
related by $\omega\leftrightarrow\omega^2$.

Since every term in $\mathcal B$ contains one boson annihilation
operator on each site, its action on a coherent product state preserves
the four spin directions while reducing the spin length by $1/2$:
\begin{equation}
  \mathcal B
  \bigotimes_{i=1}^4\ket{z_i;S}
  =
  B(z_1,z_2,z_3,z_4)
  \bigotimes_{i=1}^4\ket{z_i;S-\tfrac12},
  \label{eq:B_on_coherent}
\end{equation}
with
\begin{equation}
  B(z_1,z_2,z_3,z_4)
  =
  \frac{(2S)^2}
  {\sqrt{\prod_{i=1}^4(1+|z_i|^2)}}\,
  p_-(z_1,z_2,z_3,z_4).
  \label{eq:Bzzzz}
\end{equation}
The energy expectation value is therefore
\begin{equation}
  \bra{z_1z_2z_3z_4}
  \mathcal H_{\mathrm{CCI}}^{(S)}
  \ket{z_1z_2z_3z_4}
  =
  |B(z_1,z_2,z_3,z_4)|^2.
  \label{eq:EexpB}
\end{equation}
The zero-mode condition is therefore independent of $S$: for every
$S\geq 1/2$, including $S=1/2$, the same equation
\begin{equation}
  p_-(z_1,z_2,z_3,z_4)=0
  \label{eq:Bsceq0}
\end{equation}
selects the allowed configurations of the four coherent-state
directions.

As a check, the four tetrahedral directions
Eq.~(\ref{eq:color_directions}) have stereographic coordinates
\begin{align}
 z^{A}&=w_-e^{i\pi/4},\quad
 z^{B}=w_-e^{5i\pi/4},\nonumber\\
 z^{C}&=w_+e^{3i\pi/4},\quad
 z^{D}=w_+e^{7i\pi/4},\nonumber\\
 w_\pm&=\sqrt{2\pm\sqrt3},
 \label{eq:z_color_directions}
\end{align}
and satisfy Eq.~(\ref{eq:Bsceq0}), as do their even color
permutations and global spin rotations.  Odd permutations exchange
$\omega$ and $\omega^2$ and therefore satisfy the coefficient-conjugated
constraint associated with $\mathcal B^*$.  In addition, $p_-$
vanishes whenever three coordinates coincide, with the fourth
arbitrary.  These polarized solutions are the only locus on which the
fourth spin is not uniquely fixed by the other three; the degeneracies
are classified in Appendix~\ref{app:specialloci}.

A common framework for classical spin liquids describes the ground-state
manifold in terms of local constrainers that are linear in the spin
variables
\cite{Balla2019,Yan2024typology,Yan2024formalism,Davier2023,Davier2026}.
Equation~(\ref{eq:Bsceq0}) has a different structure. Although it is
quadratic in the four stereographic coordinates, it is
\emph{multi-affine}: it is linear in any one $z_i$ when the other three
are held fixed. The local constraint can therefore be solved as an exact
completion rule. Solving for the fourth coordinate gives
\begin{equation}
  z_4
  =
  -\frac{z_1z_2+\omega z_1z_3+\omega^2z_2z_3}
  {z_3+\omega z_2+\omega^2z_1}.
  \label{eq:Mobius}
\end{equation}
For fixed $z_1\neq z_2$, this defines the invertible M\"obius
transformation \(   z_4=T_{z_1,z_2}(z_3)\), with
\begin{equation}
  T_{z_1,z_2}(z)
  =
  -\frac{z(\omega z_1+\omega^2z_2)+z_1z_2}
  {z+\omega z_2+\omega^2z_1}.
  \label{eq:Tmap}
\end{equation}
A vanishing denominator simply corresponds to $z_4=\infty$ and is
therefore regular on the Riemann sphere. The exceptional configurations
for which the fourth direction is not uniquely determined by the other
three occur when spin directions coincide. In fact, any zero mode
containing a coincident pair necessarily contains at least three
coincident spins, and
\begin{equation}
  (z_1,z_2,z_3,z_4)=(z,z,z,w)
  \label{eq:zzzw}
\end{equation}
together with its site permutations satisfies
Eq.~(\ref{eq:Bsceq0}) for arbitrary $w$. These special loci are
classified in Appendix~\ref{app:specialloci}.

M\"obius transformations are familiar in spin physics because the
Bloch sphere is naturally identified with $\mathbb{CP}^{1}$: in
stereographic coordinates an SU(2) rotation acts fractionally
linearly on the coherent-state coordinate
\cite{Radcliffe1971,Perelomov1977}.  More explicitly, M\"obius
transformations have been used to describe the time evolution of
classical spins in stereographic coordinates
\cite{GaldaVinokur2017,Jacimovic2018}, and they also appear in the
Majorana representation of symmetric multiqubit states, where
SL(2,$\mathbb C$) transformations act as M\"obius maps on the
Majorana constellation \cite{RibeiroMosseri2011}.

Their role here is different.  The transformation $T_{z_1,z_2}$ is
neither an imposed spin rotation nor a time-evolution operator.
Instead, it is generated by the local zero-energy constraint itself:
for fixed $z_1$ and $z_2$, solving $p_-(z_1,z_2,z_3,z_4)=0$
determines the spin at one site from that at another,
$z_4=T_{z_1,z_2}(z_3)$ [Eqs.~(\ref{eq:Mobius}) and (\ref{eq:Tmap})].
The completion maps are generically M\"obius transformations in
PSL(2,$\mathbb C$) and need not belong to the SU(2) subgroup
corresponding to rigid rotations of the Bloch sphere.  They
therefore act as spatial propagation rules for the local
ground-state constraint.  Their compositions along sequences of
tetrahedra, and in particular around closed loops, define the
corresponding constraint holonomy.  In this way the projective
geometry of the local zero-energy condition directly controls the
existence and dimensionality of coherent zero-mode families on the
lattice.

\begin{figure}[tbp]
    \centering
    \includegraphics[width=0.75\linewidth]{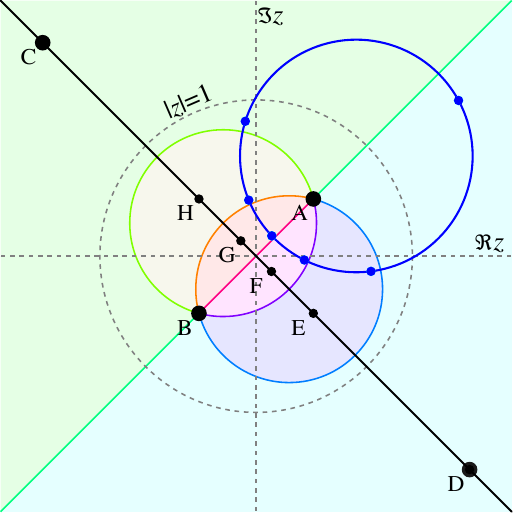}
    \caption{
    Local M\"obius propagation for the reference tetrahedral
    coordinates. The points $A,B,C,D$ are the stereographic images of
    the four color directions given in Eq.~(\ref{eq:z_color_directions}).
    Keeping the pair $(A,B)$ fixed, the completion map
    $T_{z_A,z_B}$ leaves $A$ and $B$ invariant and generates the
    six-cycle
    $C\to D\to E\to F\to G\to H\to C$, which passes through the two
    remaining color points and the four additional points
    $E$--$H$.
    Every consecutive pair $X$, $T_{z_A,z_B}(X)$ of the
    orbit---whether or not $X$ is a color point---yields an exact
    coherent zero mode with the four spins at $A$, $B$, $X$, and
    $T_{z_A,z_B}(X)$.
    The blue points show a second six-cycle generated from a generic
    initial point. Its orbit lies on the corresponding Apollonius
    circle, on which the ratio
    $|z-z_A|/|z-z_B|$ is constant.
    }
    \label{fig:Riemann_sphere}
\end{figure}

The action of a single completion map is particularly transparent from
the identity
\begin{equation}
  \frac{T_{z_1,z_2}(z)-z_1}
       {T_{z_1,z_2}(z)-z_2}
  =
  -\omega\,
  \frac{z-z_1}{z-z_2}.
  \label{eq:crossratio_id}
\end{equation}
Thus $z_1$ and $z_2$ are its two fixed points, while in the projective
coordinate $(z-z_1)/(z-z_2)$ the map acts as multiplication by
$-\omega=e^{-i\pi/3}$. Hence
\begin{equation}
  T_{z_1,z_2}^{\,6}(z)=z,
\end{equation}
and every point other than the two fixed points belongs to a six-cycle.
Figure~\ref{fig:Riemann_sphere} illustrates this local propagation for
the reference tetrahedral coloring. We emphasize that this order-six
property applies to repeated application of the \emph{same} local map
with the same fixed pair. 

The coherent zero modes obey a sharp handedness constraint. For the
sector defined by $\mathcal B$, the local zero-mode condition implies
\begin{equation}
  \chi_{\mathrm{tet}}\leq0 ,
  \label{eq:chi_sign_main}
\end{equation}
as shown in Appendix~\ref{app:equilateral}. This is the coherent-state
counterpart of the exact quantum handedness constraint established for
the complete local kernel in
Sec.~\ref{sec:exact_quantum_handedness}.
Independently of the zero-mode constraint, any four unit spin
directions satisfy the universal bound
\begin{equation}
  |\chi_{\mathrm{tet}}|
  \leq
  \frac{16}{3\sqrt3},
  \label{eq:chi_bound_main}
\end{equation}
with equality if and only if the four directions form a regular
tetrahedral frame; see Appendix~\ref{app:chirality_bounds}. Together,
these results give the exact chirality window of the coherent
zero-mode variety,
\begin{equation}
  -\frac{16}{3\sqrt3}
  \leq
  \chi_{\mathrm{tet}}
  \leq
  0 .
  \label{eq:chi_window}
\end{equation}
The lower bound is saturated precisely by the negatively oriented
regular tetrahedral colorings, up to global spin rotations and even
site permutations, while the upper bound is reached on the
three-coincident branch, which includes the fully polarized state.
Thus the local constraint fixes the \emph{handedness}, but not the
magnitude, of the tetrahedral chirality: the coherent zero-mode
variety contains continuous paths from the maximally chiral regular
tetrahedron to achiral configurations, including polarized ones,
without entering the opposite-handedness sector. For the
coefficient-conjugated constraint associated with $\mathcal B^*$,
all chirality inequalities are reversed.

\section{From local constraints to lattice zero modes}
\label{sec:lattice_construction}

We now use the single-tetrahedron results derived above as local
building blocks for extended lattices. Assign to each tetrahedron
$t$ a chirality label $\varepsilon_t=\pm1$ and define the lattice
Hamiltonian
\begin{equation}
  \mathcal H
  \equiv
  \mathcal H[\{\varepsilon_t\}]
  =
  \sum_t
  \mathcal H_{\mathrm{CCI}}^{(S),\varepsilon_t}(t),
  \label{eq:lattice_H}
\end{equation}
where
$\mathcal H_{\mathrm{CCI}}^{(S),+}
=\mathcal H_{\mathrm{CCI}}^{(S)}$
denotes the reference local parent term and
$\mathcal H_{\mathrm{CCI}}^{(S),-}
=\mathcal T\mathcal H_{\mathrm{CCI}}^{(S)}\mathcal T^{-1}$
its time-reversed partner.

For lattices in which the tetrahedra fall into two orientation
classes $\mathcal O_1$ and
$\mathcal O_2$ (Sec.~\ref{sec:cci_classical}), we take
\begin{equation}
  \varepsilon_t
  =
  \begin{cases}
    +1, & t\in\mathcal O_1,\\
    \varepsilon, & t\in\mathcal O_2,
  \end{cases}
  \qquad
  \varepsilon=\pm1 .
  \label{eq:chirality_assignment}
\end{equation}
Thus $\varepsilon=+1$ gives a uniform assignment of the
fixed-site-ordering chirality label, whereas $\varepsilon=-1$ gives
an alternating one. This convention will be used below for the
checkerboard, pyrochlore, fcc, square, and honeycomb constructions.

For any fixed assignment, time reversal reverses all local chirality
labels,
\begin{equation}
  \mathcal T
  \mathcal H[\{\varepsilon_t\}]
  \mathcal T^{-1}
  =
  \mathcal H[\{-\varepsilon_t\}] .
  \label{eq:lattice_time_reversal}
\end{equation}
Thus every chirality assignment and its sign-reversed partner define
time-reversal-conjugate Hamiltonians with identical spectra. The
handedness of the zero-energy states is therefore selected explicitly
by the CCI parent Hamiltonian rather than spontaneously. Spontaneous
selection would instead require a parent Hamiltonian that treats the
two local chiral sectors symmetrically; the full color-ice parent
Hamiltonian of Sec.~\ref{sec:CIparent}, which is time-reversal
invariant, provides such an example.

Since every local term in Eq.~(\ref{eq:lattice_H}) is positive
semidefinite, the zero-energy space of the lattice Hamiltonian is
precisely the common kernel
\begin{equation}
  \mathscr K
  \equiv
  \bigcap_t
  \ker
  \mathcal H_{\mathrm{CCI}}^{(S),\varepsilon_t}(t)
  =
  \ker\mathcal H .
  \label{eq:common_kernel}
\end{equation}
Whenever the prescribed chirality assignment admits a compatible
global CCI coloring, the corresponding product state and all of its
global spin rotations belong to $\mathscr K$. The Hamiltonian is
then frustration free, with $\mathscr K$ as its exact zero-energy
ground-state space.

The exact quantum handedness established on a single tetrahedron in
Sec.~\ref{sec:exact_quantum_handedness} immediately carries over to
the many-body common kernel. Indeed, every
$\ket{\Psi}\in\mathscr K$ belongs to the kernel of every local term
separately. Hence, for any normalized $\ket{\Psi}\in\mathscr K$,
\begin{equation}
  \varepsilon_t
  \langle\Psi|
  \hat\chi_{\mathrm{tet}}(t)
  |\Psi\rangle
  \leq0
  \label{eq:common_kernel_chirality_sign}
\end{equation}
for every tetrahedron $t$, where $\hat\chi_{\mathrm{tet}}(t)$ is defined
in the fixed site ordering of the corresponding local parent term.
This is an exact statement about the complete quantum common kernel,
including entangled zero-energy states, and is independent of spin
length, system size, and lattice connectivity. In the geometric
orientation convention of Sec.~\ref{sec:cci_classical}, the
alternating assignment gives the same chirality sign on the two
tetrahedron orientation classes, whereas for the uniform assignment
the geometric chirality changes sign between them.

Equation~(\ref{eq:common_kernel_chirality_sign}) constrains the sign
of the tetrahedral chirality throughout the full quantum common
kernel, but not its magnitude. In particular, the coherent-state
window Eq.~(\ref{eq:chi_window}) does not bound the spectrum of the
chirality operator within $\mathscr K$; projected chirality
eigenvalues can exceed the coherent-state values in magnitude, as
illustrated in Sec.~\ref{sec:quantum_chirality}.

It is useful at this point to distinguish three nested classes of
zero-energy states. The smallest consists of the compatible CCI
colorings, promoted to coherent product states, together with their
global spin rotations. The second is the \emph{coherent zero-mode
variety}: the set of coherent product states
$\bigotimes_i\ket{z_i;S}$ whose stereographic coordinates satisfy
the corresponding local coherent-state constraint on every
tetrahedron. For the reference sector this is $p_-=0$,
Eq.~(\ref{eq:Bsceq0}), while the time-reversed sector obeys its
coefficient-conjugated counterpart $p_+=0$. The largest set is the
full quantum common kernel $\mathscr K$, regarded here as its set of
state vectors. Thus
\begin{multline}
  \{\text{rotated CCI colorings}\}
  \\
  \subset
  \{\text{coherent zero-mode states}\}
  \subset
  \mathscr K .
  \label{eq:hierarchy}
\end{multline}
The first two sets contain only product states, whereas $\mathscr K$
is a linear subspace of the many-body Hilbert space. The first
inclusion is already proper on a single tetrahedron because the
coherent constraint also admits non-coloring solutions, such as the
three-coincident configurations of Eq.~(\ref{eq:zzzw}). The second
inclusion can likewise be proper on an extended lattice because
$\mathscr K$ may contain genuinely entangled states.

Here ``variety'' is used in its mathematical sense as the joint zero
set of the local polynomial constraints. It need not be a smooth
manifold; in particular, the three-coincident configurations of
Eq.~(\ref{eq:zzzw}) are singular loci at which the local completion
rule ceases to determine the fourth spin uniquely. Throughout the
following discussion, completion maps and holonomies characterize
the coherent zero-mode variety, whereas exact-diagonalization
spectra and quantum degeneracy counts refer to the full common kernel
$\mathscr K$.

The coherent subset admits a particularly direct local-to-global
construction. Suppose a sequence of tetrahedra is traversed such
that, at each step, the already specified spin directions determine
a local completion map $T_i$ for the remaining spin. Its
stereographic coordinate is then propagated by successive M\"obius
transformations. The order-six property derived above concerns
repeated application of a single completion map with the same fixed
pair. On an extended lattice the successive maps $T_i$ generally
differ, and their composition need not have finite order.

Around a closed loop of length $L$, the net propagation is described
by the M\"obius holonomy
\begin{equation}
  \mathcal M_L
  =
  T_L\circ T_{L-1}\circ\cdots\circ T_1 .
  \label{eq:loop_holonomy}
\end{equation}
A coherent loop configuration closes precisely when its initial
stereographic coordinate is a fixed point of $\mathcal M_L$. A
generic nonidentity M\"obius transformation has two distinct fixed
points, while a parabolic transformation has a single double fixed
point. Such loops therefore admit only isolated coherent solutions.
In contrast, when
\begin{equation}
  \mathcal M_L=\mathrm{id},
  \label{eq:identity_holonomy_general}
\end{equation}
every initial coordinate closes consistently, producing a continuous
family of exact coherent zero modes.

The holonomy therefore measures the coherent flexibility generated
by the way in which the local tetrahedral constraints are connected.
It does not, by itself, determine the dimension or structure of the
full quantum common kernel $\mathscr K$. The distinction between
these two levels will be important below: different lattice
connectivities can produce extensive, subextensive, or only global
coherent deformations, while the corresponding quantum kernels may
contain additional entangled zero modes.

More generally, the large quantum common kernel can be viewed as the
price of allowing families of exact noncollinear coherent product
states. For the Heisenberg ferromagnet, an arbitrary global spin
direction gives the coherent family
$\bigotimes_i\ket{z;S}_i$, whose linear span is the familiar
maximal-spin multiplet. Here the local constraint is more flexible:
the coherent-state coordinates of a tetrahedron need not coincide
but are related through the M\"obius completion rule, with three
directions determining the fourth. This permits exact product states
with noncollinear spin directions. Since every member of such a
coherent family is annihilated by the same positive-semidefinite
Hamiltonian, the linear span of the entire family necessarily belongs
to $\mathscr K$. The coherent-state freedom therefore provides a
direct mechanism for generating quantum ground-state degeneracy,
although it need not exhaust it: the full kernel can be substantially
larger and may also contain genuinely entangled zero modes.

We now turn to the corner-sharing checkerboard and pyrochlore
lattices, where the overlap of the local constraints produces an
extensive zero-energy degeneracy.

\section{Corner-sharing lattices: checkerboard and pyrochlore}
\label{sec:degeneracy}

\begin{table*}[tb]
\caption{
\label{tab:planar16}
Zero-energy multiplet counts for spin-$1/2$ models on periodic
16-site clusters.  For each total spin $J$, ``full'' is
the number of multiplets in the complete 16-spin Hilbert space.
For the checkerboard, square, and honeycomb lattices, ``alt.'' and
``unif.'' denote the alternating and uniform assignments of the
fixed-ordering chirality label $\varepsilon_t$, respectively.  The
triangular face-sharing construction has a single chirality pattern.
The checkerboard columns also apply, for the corresponding chirality
assignments, to the periodic 16-site cubic pyrochlore cluster, since
the two clusters are isomorphic as tetrahedron-incidence structures
with dual graph $K_{4,4}$.  The final row gives the total number of
zero-energy states after weighting each multiplet by
$2J+1$.
}
\begin{ruledtabular}
\begin{tabular}{c c cc cc cc c}
& &
\multicolumn{2}{c}{Checkerboard} &
\multicolumn{2}{c}{Square} &
\multicolumn{2}{c}{Honeycomb} &
Triangular\\
$J$ & full
& alt. & unif.
& alt. & unif.
& alt. & unif.
& \\ \hline
0 & 1430 & 662  & 528  & 101 & 56  & 114 & 89  & 1  \\
1 & 3432 & 1632 & 1368 & 244 & 118 & 263 & 205 & 3  \\
2 & 3640 & 1888 & 1672 & 382 & 312 & 364 & 332 & 5  \\
3 & 2548 & 1484 & 1400 & 486 & 476 & 478 & 476 & 7  \\
4 & 1260 & 852  & 840  & 453 & 452 & 452 & 452 & 21 \\
5 & 440  & 352  & 352  & 264 & 264 & 264 & 264 & 45 \\
6 & 104  & 96   & 96   & 88  & 88  & 88  & 88  & 56 \\
7 & 15   & 15   & 15   & 15  & 15  & 15  & 15  & 15 \\
8 & 1    & 1    & 1    & 1   & 1   & 1   & 1   & 1  \\ \hline
Total states
& $65536$
& $38416$ & $35714$
& $14512$ & $13660$
& $14427$ & $14054$
& $1738$
\end{tabular}
\end{ruledtabular}
\end{table*}

We now place the local term (\ref{eq:BdB}) on the checkerboard and
pyrochlore lattices.  In both cases the tetrahedra are
corner-sharing: every spin belongs to two local terms, while
tetrahedra of the same family are mutually disjoint.  This geometry
leaves a large common kernel $\mathscr K$ and leads to an extensive
ground-state degeneracy.  Each crossed plaquette of the checkerboard lattice, or
each tetrahedron of the pyrochlore lattice, carries one local parent
term with chirality label $\varepsilon_t$.  We consider both the
uniform assignment and the alternating assignment, in which
$\varepsilon_t$ changes sign between the two tetrahedron families.

\subsection{Exact diagonalization on the checkerboard lattice}
\label{sec:checkerboard}

We first examine the many-body kernel by exact diagonalization of the
spin-$1/2$ model on the periodic 16-site checkerboard cluster for both
chirality assignments. The zero-energy space is already very large:
it contains $38416$ states for the alternating assignment and $35714$
for the uniform one, in both cases more than half of the full
$2^{16}$-dimensional Hilbert space. The multiplet-resolved
degeneracies, together with those of the other 16-site geometries, are
listed in Table~\ref{tab:planar16}. For comparison at the level of
classical configurations, exhaustive enumeration gives $576$
four-colorings on the same cluster, of which $36$ satisfy the
prescribed alternating chirality and $36$ the uniform chirality.

Several features of the checkerboard spectrum are worth emphasizing.
For $S=1/2$, each local term projects onto one of the two chiral
four-spin singlets. In any fixed quantization axis, these singlets
contain two up and two down spins and are therefore orthogonal to every
local basis state with zero or one spin flipped relative to full
polarization. Consequently, the fully polarized $J=8$
multiplet and all $J=7$ multiplets belong to the common
kernel independently of the chirality assignment. Numerically, the
alternating and uniform spectra remain identical also in the
$J=6$ and $5$ sectors and first differ at
$J=4$, with the alternating pattern producing
additional zero modes at lower total spin. Thus the local handedness
pattern affects the strongly overlapping part of the many-body kernel
but leaves the polarized high-spin sector unchanged. The coexistence
of the ferromagnetic multiplet with the chirality-carrying zero-energy
states is
a special feature of the frustration-free parent-Hamiltonian point.
Since the ferromagnetic sector already lies in the local kernel,
perturbations favoring polarization can select it without overcoming a
local energy cost, making a ferromagnetic phase a natural competitor
near the solvable point. A related one-dimensional example is the
multicritical point of the zigzag chain with anisotropic $\Gamma$
exchange, where the Hamiltonian also reduces to a sum of local simplex
terms with a common kernel containing the ferromagnet and with a
ground-state degeneracy growing polynomially with system size
\cite{Saito2024}. The analogy is structural: in the present
SU(2)-symmetric models, corner sharing produces a much larger,
exponentially degenerate kernel.

The periodic $4\times4$ checkerboard cluster and the 16-site cubic
pyrochlore cluster define the same finite constraint problem up to a
relabeling of sites. Each contains eight tetrahedra arranged into two
families of four. Tetrahedra within a given family are site-disjoint,
while every tetrahedron in one family shares exactly one spin with
every tetrahedron in the other. The dual graph, whose vertices
represent tetrahedra and whose edges represent shared spins, is
therefore $K_{4,4}$ in both cases. The two tetrahedral hypergraphs are
consequently isomorphic, and the corresponding Hamiltonians have
identical spectra and zero-energy multiplet counts for matching
uniform or alternating chirality assignments. This equivalence is a
special property of these finite clusters.

The remainder of this section explains the origin of the large kernel
generated by corner sharing. We first construct exact loop zero modes
in polarized backgrounds and then map the polarized product-state
sector to matchings on the dual graph. The resulting combinatorial
bounds establish an exponential lower bound on the ground-state
degeneracy in the thermodynamic limit.

\subsection{Loop zero modes}
\label{sec:loops}

The local completion rule becomes especially transparent on an
embedded one-dimensional loop.  Fixing all spins outside the loop
turns the constraint on each tetrahedron into a relation between two
successive loop spins.  The resulting loop problem depends strongly
on the background.  We distinguish three cases: a uniform polarized
background, a generic coherent background, and a four-coloring
background.

\begin{figure}[tbp]
    \centering
    \includegraphics[width=0.95\linewidth]{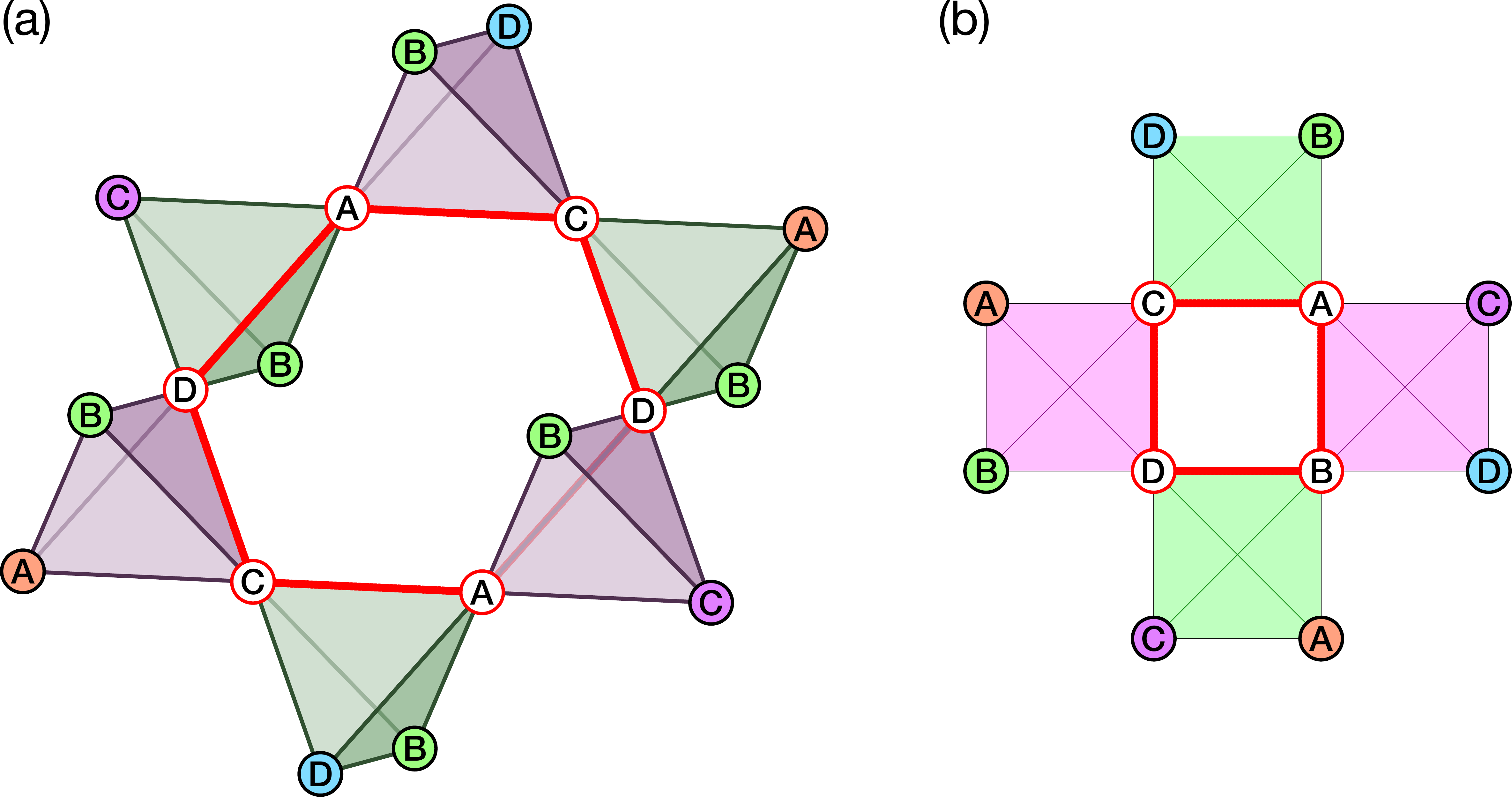}
    \caption{Embedded loops supporting local zero modes.
    (a)~A hexagonal loop of the pyrochlore lattice (bold red), sharing
    one site with each of six surrounding tetrahedra.
    (b)~The analogous square loop of the checkerboard lattice.}
    \label{fig:local_mode}
\end{figure}

\subsubsection{Uniform-color background}
\label{sec:loops_uniform}

Consider first a square loop of the checkerboard lattice or a
hexagonal loop of the pyrochlore lattice embedded in a fully
polarized environment [Fig.~\ref{fig:local_mode}].  Each tetrahedron
along the loop contains two consecutive loop sites and two
environment sites polarized along the same quantization axis.
As shown in Appendix~\ref{app:loops}, all but one term in
$\mathcal B$ then vanish, and the exact zero-mode condition reduces
to
\begin{equation}
  b_i b_j\ket{\psi}=0
    \label{eq:bibj}
\end{equation}
for every loop bond $(ij)$.
Thus two neighboring loop sites cannot both deviate from full
polarization.

The transfer-matrix counting is elementary and is left to
Appendix~\ref{app:loops}.  The resulting loop dimensions are
\begin{subequations}
\label{eq:N4N6}
\begin{align}
  N_4(S)&=1+8S+8S^2,\\
  N_6(S)&=1+12S+36S^2+16S^3 .
\end{align}
\end{subequations}
Accordingly, the square loop has $N_4=7$, $17$, and $31$ zero modes
for $S=\tfrac12$, $1$, and $\tfrac32$, while the hexagon has
$N_6=18$, $65$, and $154$.  Direct diagonalization of the embedded
loop reproduces these numbers.  For $S=\tfrac12$ the same counting
is equivalent to the monomer--dimer partition function of the cycle
graph, anticipating the matching construction of
Sec.~\ref{sec:matching}.

\subsubsection{Generic coherent background}
\label{sec:generic_loops}

Away from the polarized background, the reduction
(\ref{eq:bibj}) no longer applies and the full M\"obius propagation
rule of Sec.~\ref{sec:mobius} must be used. For a loop of length $L$,
the local completion maps $T_i$, including the appropriate local site
orderings and fixed environment spins, combine into the holonomy
$\mathcal M_L$ of Eq.~(\ref{eq:loop_holonomy}). The loop closes when
the initial stereographic coordinate is a fixed point of
$\mathcal M_L$.

For a generic background, $\mathcal M_L$ is a nonidentity M\"obius
transformation with two distinct fixed points on
$\hat{\mathbb C}$. The coherent-product problem therefore has two
isolated loop solutions rather than a continuous zero mode. A
parabolic holonomy is exceptional and has a single double fixed point,
whereas identity holonomy gives the continuous family discussed below.

Exact diagonalization of the embedded square and hexagonal problems
gives a two-dimensional zero-energy kernel for generic backgrounds at
$S=\tfrac12$, $1$, and $\tfrac32$, consistent with the two
fixed-point solutions. This finite-size agreement provides a check of
the M\"obius propagation picture. The fixed-point argument itself,
however, characterizes only the coherent-product zero modes and does
not establish the complete kernel of the loop problem for arbitrary
$S$.

\subsubsection{Four-coloring background}
\label{sec:coloring_loops}

A four-coloring background is special because every tetrahedron
contains all four colors $A,B,C,D$ exactly once. Since the background
itself satisfies the local constraint, the colors
$c_1,c_2,\ldots,c_L$ encountered along a closed loop form an orbit of
the corresponding completion maps,
\begin{equation}
  T_i(c_i)=c_{i+1},
  \qquad
  c_{L+1}\equiv c_1 .
  \label{eq:color_orbit}
\end{equation}
Thus the background color $c_1$ is always a fixed point of the loop
holonomy $\mathcal M_L$. Whether this isolated solution extends to a
continuous zero mode is determined by the full holonomy.

For an elementary hexagon of the four-sublattice pyrochlore coloring,
the spins follow the three-color sequence
\begin{equation}
  XYZXYZ ,
  \label{eq:XYZXYZ}
\end{equation}
as illustrated in Fig.~\ref{fig:local_mode}(a). For this sequence, the
six completion maps have identity holonomy,
\begin{equation}
  \mathcal M_6=\mathrm{id}.
  \label{eq:pyrochlore_hexagon_identity}
\end{equation}
An arbitrary deformation of one spin can therefore be propagated
around the hexagon and closes after one circuit, producing a
continuous chiral analog of a weathervane mode. The derivation of
Eq.~(\ref{eq:pyrochlore_hexagon_identity}) is given in
Appendix~\ref{app:identity_holonomy}.

Identity holonomy also has an exact quantum consequence. Consider an
embedded loop of length $L$ with its coherent environment fixed, and
suppose that every local completion map is well defined and
invertible. If
\begin{equation}
  \mathcal M_L=\mathrm{id},
\end{equation}
then the complete quantum kernel $\mathscr K_{\rm loop}$ of the
reduced loop constraints is spanned by the continuously propagated
coherent states,
\begin{equation}
  \mathscr K_{\rm loop}
  =
  \operatorname{span}
  \left\{
    \ket{\widetilde{\psi}(z)}
    \,\middle|\,
    z\in\widehat{\mathbb C}
  \right\},
\end{equation}
and has dimension
\begin{equation}
  \dim \mathscr K_{\rm loop}=2SL+1 .
  \label{eq:loop_kernel_exact}
\end{equation}
A derivation is given in Appendix~\ref{app:identity_holonomy_kernel}. There, identity holonomy
allows the loop constraints to be transformed into nearest-neighbor
singlet-annihilation constraints, which force the connected loop into
its maximal-spin $LS$ multiplet.

Identity holonomy is a property of the specific color sequence around
the loop, rather than of four-coloring backgrounds as such. For
example, the four-color sequence $ABCD$ around an empty square of the
checkerboard lattice, Fig.~\ref{fig:local_mode}(b), has nonidentity
holonomy and therefore admits only isolated coherent fixed-point
solutions, whereas the $ABCABC$ sequence around a pyrochlore hexagon
has identity holonomy and supports a continuous family.

The linearized coherent-state analysis of
Appendix~\ref{app:rigidity} reveals a further distinction between
first-order and finite deformations. Around the four-coloring
background, the linearized constraint matrix $\mathsf R(\bm k)$ has
nullity two at every wavevector on both the checkerboard and
pyrochlore lattices, giving $N/2+1$ complex infinitesimal zero modes.
On the checkerboard lattice these include modes localized on
individual empty squares. Their holonomy is parabolic: the linearized
multiplier around the square is unity even though the full M\"obius
transformation is not the identity. The square modes therefore
satisfy the constraints to first order but are obstructed at higher
order. By contrast, the identity holonomy of the pyrochlore hexagon
allows the corresponding infinitesimal modes to extend to exact
continuous weathervane families.

Equation~(\ref{eq:loop_kernel_exact}) applies to the embedded-loop
problem with the coherent environment fixed. It does not imply that
the full checkerboard or pyrochlore many-body kernel is exhausted by
coherent states; additional entangled zero modes are present.

\subsection{Matching representation of the polarized sector}
\label{sec:matching}

For $S=\tfrac12$, a large polarized subset of the common kernel
$\mathscr K$
has a direct formulation as a matching problem. We choose a fixed
quantization axis and introduce the dual graph $G$, whose vertices
represent tetrahedra and whose edges represent the physical spin sites
shared by neighboring tetrahedra. The dual graphs of the checkerboard
and pyrochlore lattices are, respectively, the square and diamond
lattices. In both cases, $G$ is a $4$-regular bipartite graph; for a
system of $N$ spins, it has $N$ edges and $N/2$ vertices.

We associate a down spin with an occupied edge of $G$. For
$S=\tfrac12$, the operator $\mathcal B$ annihilates every four-site
basis state containing at most one down spin. Hence any spin
configuration in which no two down spins belong to the same
tetrahedron is an exact zero-energy state. On the dual graph, this
means that no two occupied edges share a vertex. The occupied edges
therefore form a \emph{matching} of $G$.

The number of such polarized product zero modes is consequently the
matching partition function at unit activity,
\begin{equation}
  \mathcal D_{\rm match}(G)
  =
  Z_G(1)
  =
  \sum_k m_k(G),
  \label{eq:matching_partition}
\end{equation}
where $m_k(G)$ denotes the number of matchings containing $k$ edges.

Rigorous results for matchings of $4$-regular bipartite graphs then give
exponential bounds on this polarized zero-energy sector,
\begin{equation}
  1.3919^N
  \lesssim
  \mathcal D_{\rm match}
  \le
  209^{N/16}
  \simeq
  1.3964^N .
  \label{eq:match_bounds}
\end{equation}
In particular, the lower bound alone establishes an exponentially large
CCI ground-state degeneracy. The graph-theoretic derivation of these
bounds, including the origin of the constant $209$, is given in
Appendix~\ref{app:matching}.

The matching construction describes only a polarized subset of the
full common kernel. The kernel $\mathscr K$ also contains coherent
zero
modes away from the polarized sector as well as genuinely entangled
states. Equation~(\ref{eq:match_bounds}) therefore provides a rigorous
lower bound on the ground-state degeneracy rather than a calculation
of the full residual entropy.

\subsection{Comparison of degeneracies}
\label{sec:degeneracy_comparison}

The loop and matching constructions identify explicitly countable
subsets of the common kernel $\mathscr K$ and therefore provide lower
bounds on its
ground-state degeneracy. This allows a direct comparison with familiar
classical ice manifolds. Here $N$ denotes the number of physical spin
sites, and we compare the exponential degeneracy per spin,
$N^{-1}\ln\dim\mathscr K$, with the corresponding classical quantity
$N^{-1}\ln\mathcal N_{\rm ice}$.

\subsubsection{Checkerboard lattice}
\label{sec:degeneracy_checkerboard}

On compatible periodic checkerboard clusters, one can choose $N/8$
square loops such that no two share a tetrahedron. Their zero modes
can therefore be chosen independently, giving
\begin{equation}
  \dim\mathscr K_{\rm cb}
  \ge
  [N_4(S)]^{N/8}.
  \label{eq:loopbound_checkerboard}
\end{equation}
For $S=\tfrac12$, where $N_4=7$, this yields
\begin{equation}
  \dim\mathscr K_{\rm cb}
  \ge
  7^{N/8}
  \simeq
  1.2754^N .
  \label{eq:loopbound_checkerboard_half}
\end{equation}
This already exceeds the exact asymptotic degeneracy of square ice
\cite{Lieb1967},
\begin{equation}
  W_{\rm sq.\,ice}
  =
  \left(\frac43\right)^{3N/4}
  \simeq
  1.2408^N.
  \label{eq:square_ice_entropy}
\end{equation}
The matching construction of Sec.~\ref{sec:matching} gives a much
stronger bound,
\begin{equation}
  \dim\mathscr K_{\rm cb}
  \ge
  \mathcal D_{\rm match}
  \gtrsim
  1.3919^N .
  \label{eq:checkerboard_matching_bound}
\end{equation}
Thus even explicitly constructed subsets of $\mathscr K$ already
have a degeneracy density substantially larger than that of square
ice.

\subsubsection{Pyrochlore lattice}
\label{sec:degeneracy_pyrochlore}

On compatible periodic pyrochlore clusters, one can choose $N/12$
elementary hexagons such that their surrounding tetrahedra are
disjoint. Their zero modes can therefore be chosen independently,
giving
\begin{equation}
  \dim\mathscr K_{\rm pyr}
  \ge
  [N_6(S)]^{N/12}.
  \label{eq:loopbound_pyrochlore}
\end{equation}
For $S=1/2$, where $N_6=18$, this yields
\begin{equation}
  \dim\mathscr K_{\rm pyr}
  \ge
  18^{N/12}
  \simeq
  1.2723^N .
  \label{eq:loopbound_pyrochlore_half}
\end{equation}
This already exceeds the conventional Pauling estimate for
pyrochlore spin ice \cite{Pauling-1935},
\begin{equation}
  W_{\rm Pauling}
  =
  \left(\frac32\right)^{N/2}
  \simeq
  1.2247^N .
  \label{eq:pauling_entropy}
\end{equation}

As for the checkerboard lattice, the matching construction of
Sec.~\ref{sec:matching} gives a substantially stronger bound,
\begin{equation}
  \dim\mathscr K_{\rm pyr}
  \ge
  \mathcal D_{\rm match}
  \gtrsim
  1.3919^N .
  \label{eq:pyrochlore_matching_bound}
\end{equation}
Thus the explicitly constructed CCI subsets already exhibit a
degeneracy density substantially larger than the conventional spin-ice
benchmark. We emphasize that, unlike the square-ice result above, the
Pauling value is an estimate rather than the exact pyrochlore-ice
degeneracy.

\section{Edge-sharing tetrahedra: fcc, square, and honeycomb lattices}
\label{sec:otherlattices}

We next place the same local parent Hamiltonian,
Eq.~(\ref{eq:spin_S_hamiltonian}), on lattices of edge-sharing
tetrahedra. The increased overlap between neighboring constraints
qualitatively changes the structure of the zero-energy kernel. On
the corner-sharing checkerboard and pyrochlore lattices, neighboring
tetrahedra share only one spin, allowing the local completion rule to
generate an extensively degenerate family of coherent zero modes. For edge-sharing
tetrahedra, two spins are shared and the corresponding constraints are
therefore much more strongly coupled, substantially restricting the
available zero-mode deformations. We first discuss the three-dimensional
fcc lattice and then its two-dimensional square- and honeycomb-lattice
descendants.

\subsection{fcc lattice}
\label{sec:fcc}

\begin{figure}[tbp]
    \centering
    \includegraphics[width=0.9\linewidth]{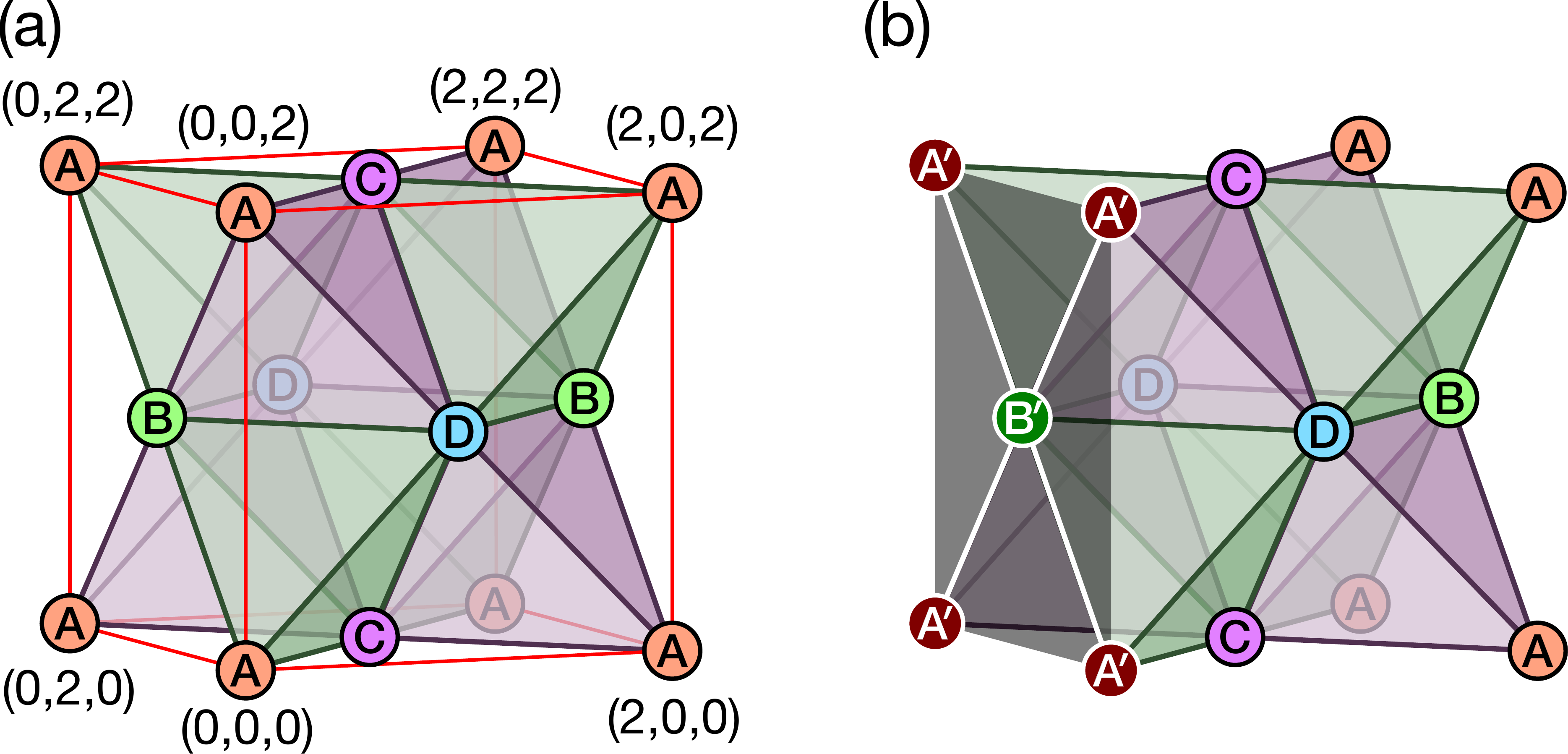}
\caption{
Tetrahedral order and a planar zero mode on the fcc lattice.
(a) The elementary tetrahedra form two orientation classes exchanged by
inversion (green and purple). Assigning opposite local chirality labels
to the two classes is compatible with the four-sublattice tetrahedral
coloring shown. In the coordinate convention of the figure, the cubic
magnetic unit cell (red) is generated by $(2,0,0)$, $(0,2,0)$, and
$(0,0,2)$. For the alternating chirality assignment, inversion $\mathcal I$ and
time reversal $\mathcal T$ separately exchange the two local chiral
sectors, whereas their product $\mathcal{IT}$ is a symmetry. The
four-sublattice tetrahedral ordered state shown preserves this
combined symmetry.
(b) A planar zero-energy deformation of the tetrahedral state. The spin
directions $A'$ and $B'$ within a single $(100)$ plane can be varied
continuously, with the local M\"obius completion rule fixing one in
terms of the other, while all spins outside the plane remain at their
tetrahedral directions.}
\label{fig:fcc_lattice}
\end{figure}

\begin{figure}[tbp]
    \centering
    \includegraphics[width=0.7\linewidth]{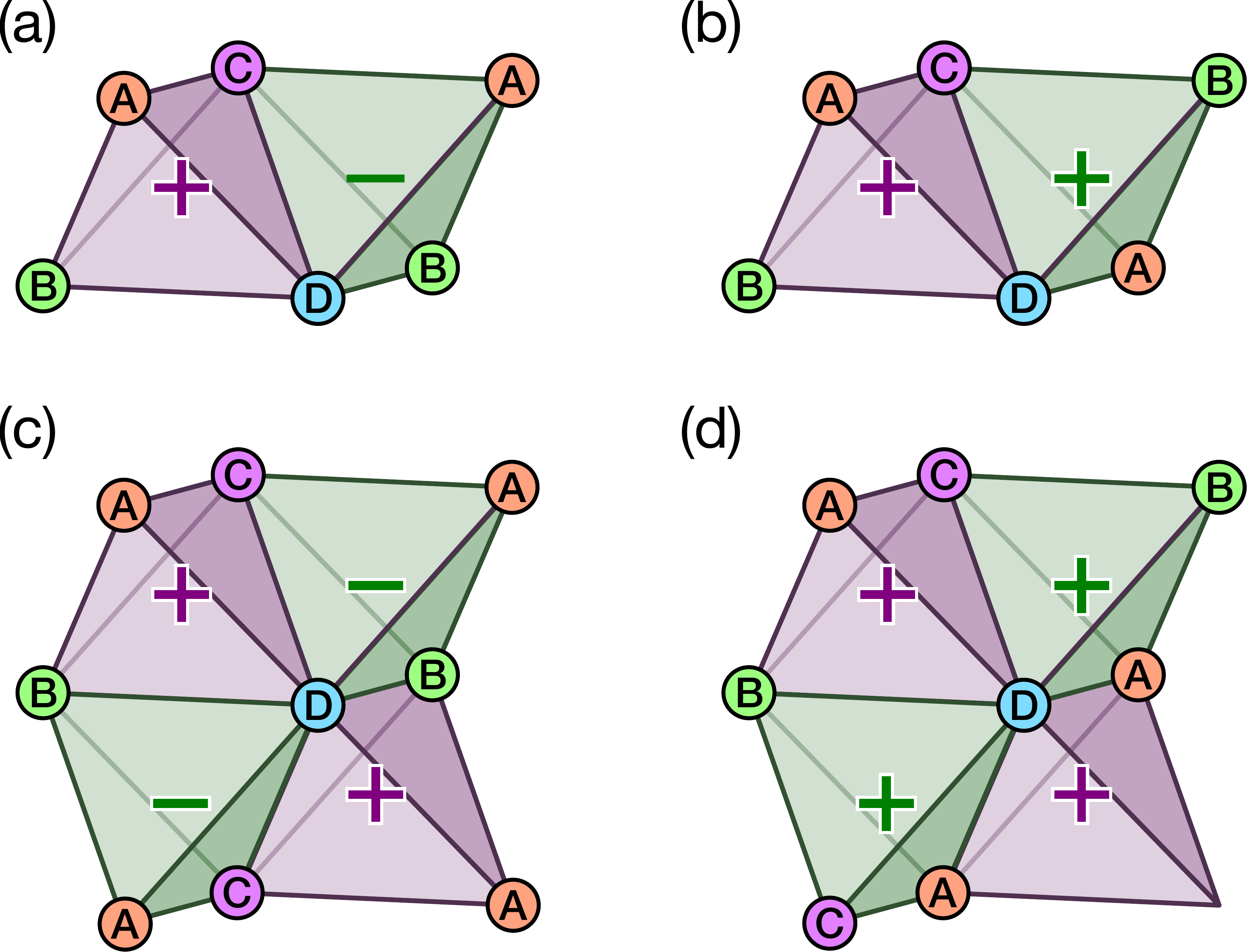}
\caption{
Propagation of the four-coloring constraint across shared edges for
(a) the alternating and (b) the uniform chirality assignments; the
corresponding $\varepsilon_t$ are indicated in the figure. Once the
colors on one tetrahedron are fixed, an edge-sharing neighbor already
contains two of them. Its remaining two sites must therefore carry the
other two colors, with their ordering uniquely determined by the
prescribed chirality parity.
(c) For the alternating assignment, repeated propagation around the
closed sequence is consistent and reproduces the four-sublattice
coloring of Fig.~\ref{fig:fcc_lattice}(a).
(d) For the uniform assignment, the same procedure returns an
incompatible color assignment, showing that a global four-coloring is
frustrated.}
\label{fig:fcc_frustration}
\end{figure}

The elementary nearest-neighbor tetrahedra of the fcc lattice fall
into two orientation classes exchanged by inversion
[Fig.~\ref{fig:fcc_lattice}(a)]. Tetrahedra overlap along edges, and
every fcc site belongs to eight elementary tetrahedra. We place one
local CCI term on every tetrahedron and consider two choices of the
chirality label $\varepsilon_t$: a uniform assignment, with the same
$\varepsilon_t$ on the two orientation classes, and an alternating
assignment, with opposite $\varepsilon_t$.

Here and below, ``uniform'' and ``alternating'' refer to the
fixed-ordering chirality labels $\varepsilon_t$, rather than directly
to the sign of the geometric tetrahedral chirality of
Sec.~\ref{sec:cci_classical}; in the geometric convention, the
alternating assignment is precisely the one that enforces the same
handedness on every tetrahedron. For the
alternating assignment, inversion $\mathcal I$ exchanges the two
tetrahedron classes while time reversal $\mathcal T$ exchanges the two
local chiral sectors. Neither is a symmetry separately, whereas their
product $\mathcal{IT}$ leaves the Hamiltonian invariant. The
four-sublattice tetrahedral state shown in
Fig.~\ref{fig:fcc_lattice}(a) also preserves this combined symmetry.

Edge sharing makes the discrete four-coloring constraint particularly
restrictive. Once the colors on one tetrahedron are fixed, a
neighboring tetrahedron sharing an edge already contains two of the
four colors. Its remaining sites must carry the other two, and the
prescribed $\varepsilon_t$ fixes their ordering. The coloring therefore
propagates uniquely from one tetrahedron to the next, as illustrated
in Figs.~\ref{fig:fcc_frustration}(a) and
\ref{fig:fcc_frustration}(b).

For the alternating assignment, this propagation is globally
consistent. As shown in Fig.~\ref{fig:fcc_frustration}(c), it closes
around the elementary sequence and generates the four-sublattice
tetrahedral coloring of Fig.~\ref{fig:fcc_lattice}(a). Since the
network of edge-sharing tetrahedra is connected, specifying the
coloring on one tetrahedron fixes it throughout the lattice. The
discrete CCI four-coloring is therefore unique up to even permutations
of the four colors and lattice symmetries. Its global SO(3) rotations
generate the corresponding orbit of tetrahedrally ordered states.

For the uniform assignment, by contrast, the same propagation is
inconsistent: following the constraint around the closed sequence in
Fig.~\ref{fig:fcc_frustration}(d) returns a different color to an
already fixed site. This provides a local obstruction to a global
four-coloring, rather than a finite-size effect. We therefore focus
below on the alternating, order-supporting branch.

The uniqueness of the discrete four-coloring does not imply that the
coherent zero-mode variety is discrete. Starting from the
four-sublattice state, consider a single $(100)$ plane as in
Fig.~\ref{fig:fcc_lattice}(b), keeping all spins outside the plane
fixed. The sites in the plane belong to two sublattices. Choosing a
common spin direction $A'$ on the $A$ sublattice---here and below, a
primed color label denotes the common direction of that sublattice
after a continuous deformation away from its reference value---the
local M\"obius
completion rule determines a corresponding direction $B'$ on the $B$
sublattice. The same pair $(A', B')$ satisfies the constraint on every
tetrahedron intersecting the plane. Varying $A'$ therefore generates
a continuous family of exact coherent zero-energy states localized on
that plane.

The construction can be repeated on sufficiently separated parallel
$(100)$ planes, so that no tetrahedron intersects two deformed planes.
The corresponding deformations are then independent. Since a number
proportional to the linear system size $L$ of such planes can be
chosen, the fcc model has at least $O(L)$ independent continuous
zero-mode parameters around the four-sublattice state. These planar
modes thus provide a subextensive family of exact zero-energy
deformations. 
The construction gives a lower bound on the dimension of the
coherent zero-mode variety, while the linearized analysis of
Appendix~\ref{app:rigidity} shows that it is exhaustive at first
order. The kernel of $\mathsf R(\mathbf k)$ is nontrivial only on
the three $\langle100\rangle$ axes of the Brillouin zone, giving
$6L-3$ complex infinitesimal modes on an $L^3$ torus of magnetic
cells. These modes are exhausted by the planar deformations together
with the three global M\"obius modes. Thus there are no additional
infinitesimal coherent deformations around the four-sublattice state.

Restricting the fcc construction to appropriate bilayers turns these
planar zero modes into the line defects of the square and honeycomb
models discussed next.

\subsection{Square and honeycomb lattices}
\label{sec:square_honeycomb}

\begin{figure}[tp]
    \centering
    \includegraphics[width=0.8\linewidth]
    {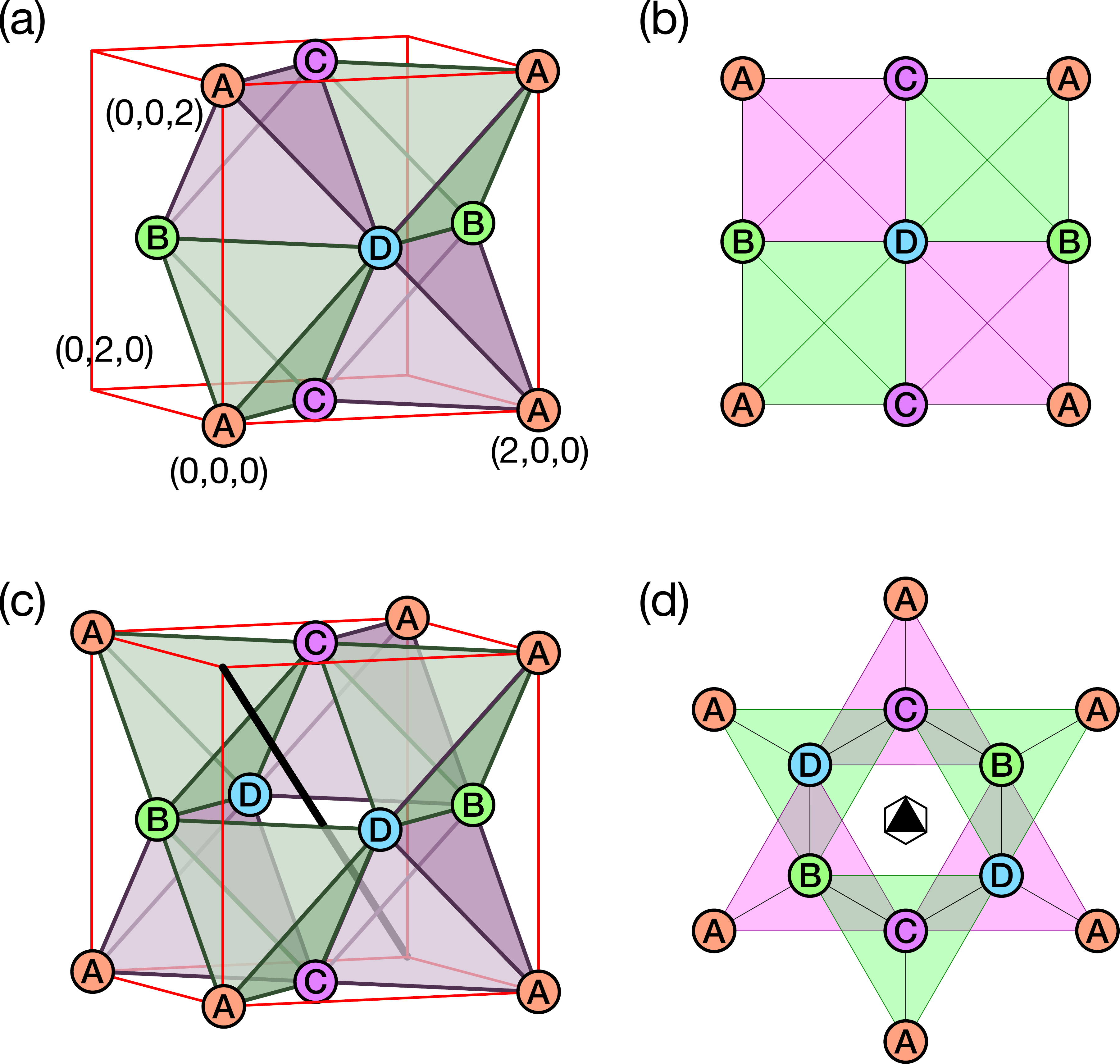}
    \caption{
    Square- and honeycomb-lattice models obtained from fcc bilayers.
    (a,b) Two adjacent $(010)$ planes project onto a square lattice,
    where the edge-sharing tetrahedra become crossed plaquettes.
    (c,d) Two adjacent $(\bar1\bar11)$ triangular planes project onto
    a honeycomb lattice, where each tetrahedron becomes a site together
    with its three nearest neighbors. The two tetrahedron orientation
    classes are shown in green and magenta. The red frame indicates
    the projection of the cubic fcc unit cell, and the black line in
    (c,d) marks the $[\bar1\bar11]$ axis of the $S_6$ rotoreflection.}
    \label{fig:fcc_lattice_square_honey}
\end{figure}

\begin{figure}[bp]
    \centering
    \includegraphics[width=0.7\linewidth]{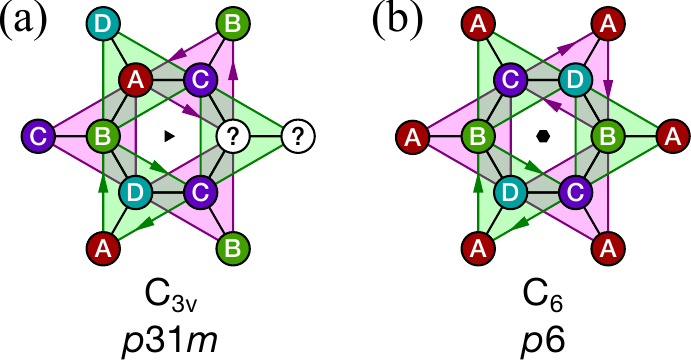}
    \caption{
    Wallpaper symmetries of the chirality-decorated honeycomb models.
    Arrows indicate the circulation of the projected chirality term
    with respect to a fixed normal to the plane.
    (a) A uniform assignment of the fixed-ordering chirality labels
    $\varepsilon_t$ produces alternating planar circulation, with point
    group $C_{3v}$ and wallpaper group $p31m$.
    (b) An alternating assignment produces uniform planar circulation,
    with point group $C_6$ and wallpaper group $p6$.}
    \label{fig:wallpaper_honeycomb}
\end{figure}

\begin{figure}[bp]
    \centering
    \includegraphics[width=0.9\linewidth]{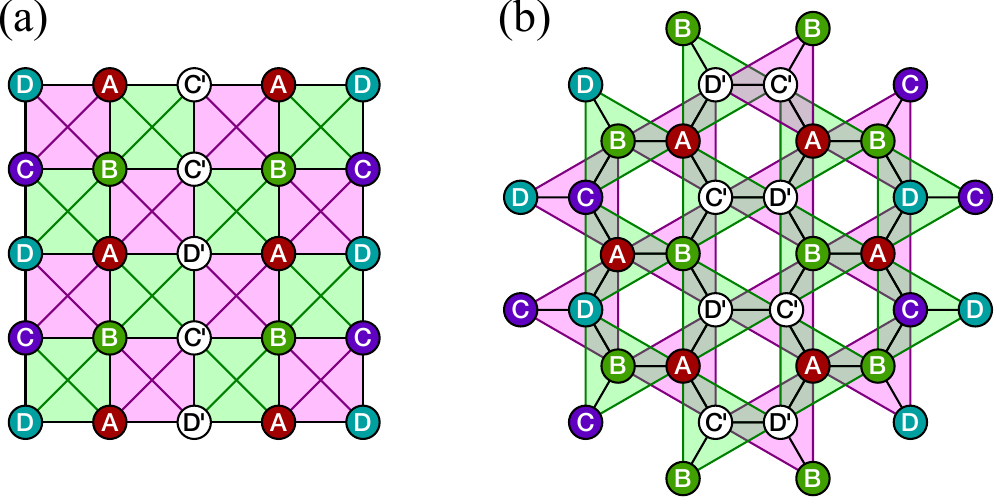}
    \caption{
    Exact zero-energy line deformations on (a) the square and
    (b) the honeycomb lattice for the alternating, order-supporting
    chirality assignment. The green and magenta tetrahedra belong
    to the two orientation classes and carry opposite
    $\varepsilon_t$. Along the indicated line, the spin directions
    $C'$ and $D'$ can be varied continuously according to the
    M\"obius completion rule, while all spins outside the line remain
    fixed.}
    \label{fig:linear-defect}
\end{figure}

The square and honeycomb models are two-dimensional descendants of
the fcc construction obtained by collapsing the bilayers shown in
Fig.~\ref{fig:fcc_lattice_square_honey}. In both cases, the two
tetrahedron orientation classes are inherited from the fcc parent.
Using the fixed-site-ordering convention introduced above, both
Hamiltonians take the form of Eq.~(\ref{eq:lattice_H}), with
$\mathcal O_1$ and $\mathcal O_2$ corresponding to the green and
magenta tetrahedra. Thus $\varepsilon=+1$ gives the uniform assignment
of chirality labels between the two classes, while $\varepsilon=-1$
gives the alternating assignment.

The projection also makes the distinction between the fixed-ordering
chirality label and the apparent planar circulation particularly
transparent. Since the two tetrahedron classes are viewed from
opposite sides of the collapsed bilayer, the projected circulation
acquires an additional orientation-dependent sign. A uniform
$\varepsilon_t$ assignment therefore appears as alternating planar
circulation, whereas an alternating assignment appears as uniform
circulation. The corresponding honeycomb symmetry patterns are shown
in Fig.~\ref{fig:wallpaper_honeycomb}.

In both the square and honeycomb geometries, only the alternating
branch supports the four-sublattice tetrahedral coloring inherited
from the fcc parent. For the uniform branch, propagation of the
coloring constraint through the edge-sharing tetrahedra leads to an
inconsistency, as in Fig.~\ref{fig:fcc_frustration}. On the honeycomb
lattice, the same incompatibility is also reflected directly in the
projected chirality pattern of Fig.~\ref{fig:wallpaper_honeycomb}.

Classical four-coloring enumeration provides an independent check.
Before imposing the chirality constraint, the $16$-site square and
honeycomb clusters contain, respectively, $168$ and $96$ color-ice
configurations. In each case, $12$ satisfy the alternating CCI
assignment, forming the orbit of a single tetrahedral coloring under
the proper tetrahedral color permutations, whereas none satisfy the
uniform assignment. For the square lattice, the absence of a
uniform-branch CCI coloring was also verified on periodic tori up to
$8\times6$ sites.

The alternating branch also inherits the subdimensional zero modes of
the fcc construction. The planar fcc deformation descends to the line
modes shown in Fig.~\ref{fig:linear-defect}. Mutually separated
parallel lines can be deformed independently, giving at least $O(L)$
continuous zero-mode parameters on an $L\times L$ sample. These line
modes therefore form a subextensive family of exact coherent
zero-energy deformations. As for the planar fcc modes, this construction provides a lower bound. The linearized analysis of Appendix~\ref{app:rigidity} shows
that it is exhaustive at first order: the kernel of
$\mathsf R(\mathbf k)$ is confined to two lines in the Brillouin
zone for the square lattice and three symmetry-related lines for the
honeycomb lattice. This gives $4L-1$ and $6L-3$ complex
infinitesimal modes, respectively, including the three global
M\"obius modes, in one-to-one correspondence with the line
deformations of Fig.~\ref{fig:linear-defect}.

Exact diagonalization on the $16$-site clusters further illustrates
the distinction between the polarized and chirality-sensitive parts
of the zero-energy kernel (Table~\ref{tab:planar16}). The uniform
and alternating branches have identical zero-mode counts in the
high-total-spin sectors but separate at lower total spin.

\begin{figure*}[bt]
    \centering
    \includegraphics[width=0.8\linewidth]
    {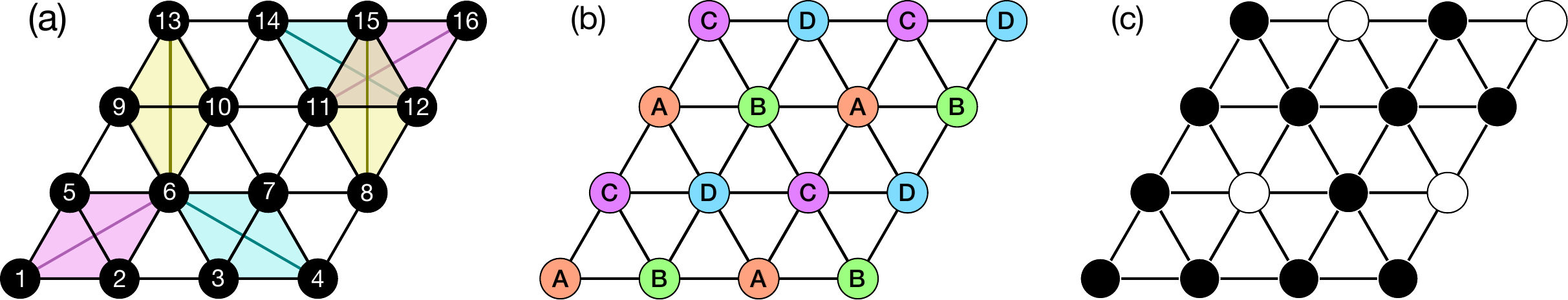}
    \caption{
    Triangular-lattice construction and representative zero-energy
    states.
    (a) A $16$-site cluster. The colored rhombi, interpreted as
    flattened tetrahedra, are the four-site supports of the local
    Hamiltonian in Eq.~(\ref{eq:spin_1_2_hamiltonian}). They occur in
    three orientations (cyan, magenta, and yellow), and the full
    Hamiltonian sums the local term over all translations of all three
    orientations. Consequently, every elementary triangle belongs to
    three tetrahedra; for example, the triangle $(11,12,15)$ is shared
    by $(14,11,12,15)$, $(8,12,15,11)$, and $(11,12,16,15)$.
    (b) A four-sublattice coherent zero-energy state. The regular
    tetrahedral configuration is one member of this family; more
    generally, the four sublattice spin directions may be continuously
    deformed subject to the local M\"obius constraint $p_-=0$,
    Eq.~(\ref{eq:Bsceq0}).
    (c) A maximal-density polarized zero mode built from local
    $\uparrow\uparrow\uparrow\downarrow$ motifs (filled and open circles
    denote $\uparrow$ and $\downarrow$). At lower densities, the same
    local constraint admits disordered polarized zero-energy
    configurations.}
    \label{fig:triangular_lattice}
\end{figure*}

The simplest chirality-blind states are the polarized $4{:}0$ and
$3{:}1$ configurations, as they are annihilated by either local chiral
parent term. In the coherent-state description, they belong to the
three-coincident locus in Eq.~(\ref{eq:zzzw}), which is common to the
conditions $p_-=0$ and $p_+=0$ and therefore does not distinguish the
two local chiralities.
On the corner-sharing lattices, this same local $4{:}0$/$3{:}1$ condition
underlies the matching construction of Sec.~\ref{sec:matching}.

The high-spin kernel is, however, larger than this polarized
product-state subset. Starting in the two-down-spin sector,
zero-energy states can contain amplitudes for configurations with two
down spins on the same tetrahedron, provided that the many-body wave
function remains orthogonal to the forbidden chiral singlet on every
tetrahedron. As long as the corresponding local constraint vectors
are linearly independent, their total rank is fixed simply by their
number and is therefore independent of the chirality assignment.
This explains why the uniform and alternating spectra remain
identical near full polarization even beyond the polarized
product-state sector.

Farther from saturation, linear dependencies develop among the local
constraints and can depend on the relative chiralities of overlapping
tetrahedra. The two branches then begin to separate. For the
$16$-site square cluster the first difference occurs at
$J=4$, whereas for the honeycomb cluster it occurs at
$J=3$. The detailed fixed-$J^z$ counting and its relation
to the multiplet-resolved ED spectrum are given in
Appendix~\ref{app:highspin_count}.

\section{Face-sharing tetrahedra: The triangular lattice and chiral magnetic order}
\label{sec:triangular_lattice}

The triangular-lattice construction provides the strongest overlap of
local tetrahedra considered here: neighboring four-site terms can share
three spins, corresponding to an entire face of the tetrahedron. This
strongly constrains the propagation of local zero modes and suppresses
the extensive degeneracies found on the corner-sharing lattices. The zero-energy kernel nevertheless preserves a four-sublattice chiral ordered component, together with additional polarized zero modes.

\begin{table*}[tbp]
\caption{
Zero-energy multiplet content of the $N=12$ and $N=16$
triangular-lattice clusters, resolved by total spin $J$ and
space-group irreducible representation. For point group $C_6$, the
little co-groups at $\Gamma$, $K$, and $M$ are $C_6$, $C_3$, and
$C_2$. We choose the counterclockwise sixfold rotation
$\bm a_1\!\to\!\bm a_2$,
$\bm a_2\!\to\!\bm a_2-\bm a_1$; reversing its sense, or the
chirality, complex-conjugates the labels.
The irreps $\Gamma_m$, $K_{0,\pm}$, and $M_\pm$ are labeled by
eigenvalues $e^{im\pi/3}$, $1,e^{\pm2\pi i/3}$, and $\pm1$,
respectively. Their momentum stars contain $1$, $2$, and $3$ points;
the generic $Q$ points belong to six-point stars with trivial little
co-group. Blank entries denote zero multiplicity, and a dash denotes
a momentum absent from the finite-cluster Brillouin zone.
``Total'' gives the number of zero-energy spin multiplets after
accounting for the space-group irrep dimension. The final column gives
the number $\mathscr V^{\widetilde S}_J$ of spin-$J$ multiplets in
the tetrahedral Anderson tower, with $\widetilde S=3/2$ for $N=12$
and $\widetilde S=2$ for $N=16$; see
Table~\ref{tab:tetrahedral_tower}.
}
\label{tab:degeneracy_triangular_clusters}
\centering
\setlength{\tabcolsep}{2.5pt}
\renewcommand{\arraystretch}{1.08}
\begin{ruledtabular}
\begin{tabular}{cc cccccc ccc cc ccc cc}
& & \multicolumn{6}{c}{$C_6\quad(d=1)$} & \multicolumn{3}{c}{$C_3\quad(d=2)$} & \multicolumn{2}{c}{$C_2\quad(d=3)$} & 
\multicolumn{3}{c}{$C_1\quad(d=6)$} & &
\\
$N$
&
$J$
&
$\Gamma_0$
&
$\Gamma_{+1}$
&
$\Gamma_{-1}$
&
$\Gamma_{+2}$
&
$\Gamma_{-2}$
&
$\Gamma_3$
& $K_0$ & $K_+$ & $K_-$ & $M_+$ & $M_-$ & $Q_{12}$ & $Q_{16,1}$ & $Q_{16,2}$ & Total & $\mathscr{V}^{\widetilde S}_{J}$
\\
\hline
\multirow{7}{*}{$12$}
& 0
& 1 &  &  &  &  & 
&  &  & 
&  & 
&  & -- & --
& 1   & 1
\\
& 1
&  &  &  &  &  & 
&  &  & 
& 1 & 
&  & -- & --
& 3   & 3
\\
& 2
&  &  &  & 1 & 1 & 
&  &  & 
& 1 & 
&  & -- & --
& 5   & 5
\\
& 3
& 1 &  &  &  &  & 
&  & 1 & 
& 2 & 
& 1 & -- & --
& 15  & 7
\\
& 4
& 1 &  &  & 1 & 1 & 
& 1 & 1 & 1
& 1 & 
& 2 & -- & --
& 24  & 5
\\
& 5
&  &  &  &  &  & 
& 1 &  & 
& 1 & 
& 1 & -- & --
& 11  & 3
\\
& 6
& 1 &  &  &  &  & 
&  &  & 
&  & 
&  & -- & --
& 1   & 1
\\
\hline
\multirow{9}{*}{$16$}
& 0
&  &  &  &  & 1 & 
& -- & -- & --
&  & 
& -- &  & 
& 1   & 1
\\
& 1
&  &  &  &  &  & 
& -- & -- & --
& 1 & 
& -- &  & 
& 3   & 3
\\
& 2
& 1 &  &  & 1 &  & 
& -- & -- & --
& 1 & 
& -- &  & 
& 5   & 5
\\
& 3
&  &  &  &  & 1 & 
& -- & -- & --
& 2 & 
& -- &  & 
& 7   & 7
\\
& 4
& 1 &  &  & 1 & 1 & 
& -- & -- & --
& 2 & 
& -- & 1 & 1
& 21  & 9
\\
& 5
& 2 &  & 1 & 1 &  & 2
& -- & -- & --
& 4 & 1
& -- & 2 & 2
& 45  & 7
\\
& 6
& 1 &  &  & 2 & 2 & 
& -- & -- & --
& 3 & 2
& -- & 3 & 3
& 56  & 5
\\
& 7
&  &  &  &  &  & 
& -- & -- & --
& 1 & 
& -- & 1 & 1
& 15  & 3
\\
& 8
& 1 &  &  &  &  & 
& -- & -- & --
&  & 
& -- &  & 
& 1   & 1
\\
\end{tabular}
\end{ruledtabular}
\end{table*}

\subsection{Ground-state structure and Anderson tower}

The local constraints remain compatible with a four-sublattice
family of coherent zero modes containing the regular tetrahedral state
shown in Fig.~\ref{fig:triangular_lattice}(b). With the site ordering
of Fig.~\ref{fig:triangular_lattice}(a), the sublattice labels on every
rhombus differ from those of a reference rhombus only by an even
permutation. Consequently, any four-sublattice configuration satisfying
the local CCI constraint on one reference rhombus satisfies it on every
rhombus.

The face-sharing geometry strongly restricts deformations of this
ordered family. In contrast to the fcc, square, and honeycomb
constructions, we find no continuous local, line, or plane zero modes.
The linearized analysis of Appendix~\ref{app:rigidity} makes this
statement exhaustive at first order: $\mathsf R(\bm k)$ has no kernel
at nonzero momentum, while its kernel at $\Gamma$ consists only of the
three global M\"obius deformations. Thus the four-sublattice state has
no nonuniform infinitesimal coherent deformation. Additional exact
product zero modes nevertheless survive in the polarized sector:
since a local term annihilates every configuration with at most one
down spin on a rhombus, any global configuration satisfying this
condition is an exact zero mode. A maximal-density example is shown in
Fig.~\ref{fig:triangular_lattice}(c).

To identify the ordered component of the full quantum kernel, we
performed exact diagonalization on periodic $N=12$ and $N=16$
clusters preserving the full $C_6$ point symmetry. The zero modes were
resolved by total spin and by the irreducible representations of the
chiral wallpaper group $p6$; the results are summarized in
Table~\ref{tab:degeneracy_triangular_clusters}.

A characteristic finite-size signature of four-sublattice magnetic
order is its Anderson tower of states
\cite{Bernu_PRL.69.2590_1992,Bernu_PRB.50.10048_1994,
Lecheminant1995,Wietek2017,Khatua2026}.
Grouping the $N/4$ spin-$1/2$ sites of each sublattice into a
collective spin of maximal length
\begin{equation}
  \widetilde S=\frac{N}{8},
\end{equation}
the ordered-state subspace is the finite-spin realization
$\mathscr V^{\widetilde S}$ of four collective spins introduced for a
single tetrahedron.

For a rigid tetrahedral rotor, the symmetry content at fixed total
spin $J$ is obtained by restricting the spin-$J$ representation of
SO(3) to the proper tetrahedral group,
\begin{equation}
  \mathcal M_J
  =
  D^J\!\downarrow_{A_4}.
  \label{eq:tower_rotor_rep}
\end{equation}
Its decomposition into
$\mathsf A$, $\mathsf E_\pm$, and $\mathsf T$ irreducible
representations is derived in Appendix~\ref{app:tower}. Primitive
translations map the one-dimensional tetrahedral irreps to the
$\Gamma$ point, whereas the triplet $\mathsf T$ carries the three
momenta in the $M$-point star. With the rotation convention of
Table~\ref{tab:degeneracy_triangular_clusters},
\begin{equation}
  \mathsf A\leftrightarrow\Gamma_0,\qquad
  \mathsf E_+\leftrightarrow\Gamma_{-2},\qquad
  \mathsf E_-\leftrightarrow\Gamma_{+2},\qquad
  \mathsf T\leftrightarrow M_+ .
  \label{eq:tower_spacegroup_mapping}
\end{equation}
Reversing the sense of $C_6$, or equivalently the chiral sector,
interchanges $\Gamma_{+2}$ and $\Gamma_{-2}$.

At finite $\widetilde S$, the rotor can additionally acquire a
one-dimensional stabilizer character under the discrete tetrahedral
identifications of the order-parameter space. As shown in
Appendix~\ref{app:tower},
\begin{equation}
  \eta_{\widetilde S}
  =
  \mathsf E_+^{\otimes N/4},
  \qquad
  \mathscr V^{\widetilde S}_{J}
  \simeq
  \mathcal M_J\otimes\eta_{\widetilde S},
  \qquad
  0\leq J\leq2\widetilde S .
  \label{eq:tower_twist}
\end{equation}
The twist leaves the rotor multiplicities unchanged but cyclically
permutes the one-dimensional $\Gamma$-point irreps. Thus the $N=12$
cluster is untwisted,
$\eta_{3/2}=\mathsf A$, whereas the $N=16$ cluster carries
$\eta_2=\mathsf E_+$. In particular, the tower singlet changes from
$\Gamma_0$ for $N=12$ to $\Gamma_{-2}$ for $N=16$, while the
$J=1$ tower remains in the $M_+$ sector.

The exact spectra reproduce both the rotor multiplicities and these
cluster-dependent symmetry labels. For $N=12$, the tower exhausts the
zero-energy sectors for $J=0,1,2$, with multiplicities $1,3,5$; at
$J=3$ the expected tower states remain present together with additional
zero modes. For $N=16$, the tower similarly exhausts the
$J=0,1,2,3$ sectors, with multiplicities $1,3,5,7$, while additional
states first appear at $J=4$. The agreement of both multiplicities and
space-group quantum numbers provides a finite-size symmetry fingerprint
of the tetrahedral ordered component of the zero-energy kernel.

For $J>2\widetilde S$, the finite-spin space
$\mathscr V^{\widetilde S}$ no longer follows the rigid-rotor
multiplicities, and additional zero modes become increasingly
important. In particular, the complete $J=N/2$ and $J=N/2-1$ sectors
are at zero energy because their highest-weight states contain,
respectively, zero and one down spin and are annihilated by every local
term. The full kernel is therefore strictly larger than the
tetrahedral rotor sector.

For a generic Hamiltonian in a tetrahedrally ordered phase, the
Anderson-tower levels acquire finite-size splittings that collapse in
the thermodynamic limit. At the present frustration-free point, the
entire finite-size realization of this tower is instead pinned exactly
at zero energy. This is an exact representation-theoretic embedding of
the chiral four-sublattice ordered component in the zero-energy kernel;
by itself it does not establish thermodynamic long-range order, since
the kernel also contains additional polarized and entangled states.

\subsection{Chirality spectra in the quantum common kernel}
\label{sec:quantum_chirality}

\begin{figure}[tbp]
    \centering
    \includegraphics[width=0.95\linewidth]{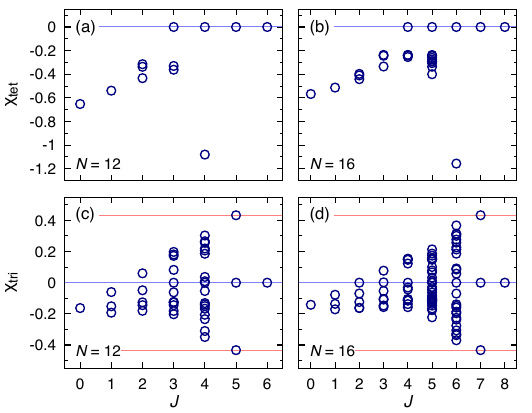}
\caption{\label{fig:tri_chirality}
Chirality spectra within the zero-energy kernels of the
(a,c) $12$-site and (b,d) $16$-site triangular clusters, resolved by
total spin.
(a,b) Eigenvalues of the projected tetrahedral chirality
$\mathcal P_0\hat\chi_{\mathrm{tet}}(t)\mathcal P_0$ on a single rhombus.
All eigenvalues are nonpositive.
(c,d) Eigenvalues of
$\mathcal P_0\hat\chi_{ijk}\mathcal P_0$ on a single elementary
triangle. In contrast to the tetrahedral chirality, the triangle
chirality is not sign definite and reaches the extrema
$\pm\sqrt3/4$ available to three spin-$1/2$ degrees of freedom,
indicated by the red horizontal lines.}
\end{figure}

The local result of Sec.~\ref{sec:exact_quantum_handedness} implies
that every state in the common kernel obeys the chirality-sign
constraint in Eq.~(\ref{eq:common_kernel_chirality_sign}). We now
examine the spectrum of the local chirality operator within this
many-body subspace.

Let $\mathcal P_0$ denote the projector onto $\mathscr K$. The
expectation-value inequality
Eq.~(\ref{eq:common_kernel_chirality_sign}), valid for every state in
$\mathscr K$, is equivalently the operator statement
\begin{equation}
  \varepsilon_t
  \mathcal P_0
  \hat\chi_{\mathrm{tet}}(t)
  \mathcal P_0
  \preceq0 .
  \label{eq:global_chirality_semidefinite}
\end{equation}
For the chirality sector used in the triangular-lattice calculations
below, $\varepsilon_t=+1$, so the projected tetrahedral chirality is
negative semidefinite on every rhombus.

We consider two local observables. The first is the
tetrahedral chirality $\hat\chi_{\mathrm{tet}}(t)$ on a rhombus, the direct
quantum analogue of the quantity constrained by the parent
Hamiltonian. The second is the chirality $\hat\chi_{ijk}$ of a single
triangular face. Comparing the two determines whether the constraint
selects handedness only at the tetrahedral level or, more strongly,
also fixes the chirality sign of each individual face.

The finite-cluster spectra in
Figs.~\ref{fig:tri_chirality}(a,b) therefore illustrate an exact
property of the complete common kernel rather than providing
finite-size evidence for it; beyond the sign, they reveal how the
eigenvalues are distributed within the sign-constrained sector.

The total-spin-resolved spectra of the projected local tetrahedral
chirality are shown in Figs.~\ref{fig:tri_chirality}(a,b). On the
$12$-site cluster, the spectrum progressively contracts toward zero
with increasing total spin and vanishes identically for $J=5$ and
$6$. These sectors lie entirely in the near-polarized part of the
zero-energy kernel and represent the quantum counterpart of the
$\chi_{\mathrm{tet}}=0$ end of the classical zero-mode family.

At the opposite end, the unique $J=0$ zero-energy state carries a
large, spatially uniform tetrahedral chirality. For $N=12$,
$\langle\hat\chi_{\mathrm{tet}}(t)\rangle_{J=0}=-0.6518$ on every rhombus,
compared with
$\chi_{\mathrm{tet}}^{\rm coh}=-2/(3\sqrt3)\simeq-0.3849$
for the maximally chiral tetrahedral spin-$1/2$ coherent-product
state. The singlet magnitude is therefore larger by a factor of
approximately $1.69$. For the $16$-site singlet we similarly find
$\langle\hat\chi_{\mathrm{tet}}(t)\rangle=-0.5672$ on every rhombus, about
$1.47$ times the coherent-product value. Quantum fluctuations within
$\mathscr K$ therefore preserve the selected handedness
while allowing the local chirality to exceed the maximal magnitude
accessible to a spin-$1/2$ coherent product state.

The chirality of an individual elementary triangle behaves quite
differently. The projected local operator
$\mathcal P_0\hat\chi_{ijk}\mathcal P_0$ has eigenvalues of both signs
and reaches $\pm\sqrt3/4$
[Figs.~\ref{fig:tri_chirality}(c,d)]. This mirrors the coherent
zero-mode variety, where the chiralities of individual tetrahedral
faces are not sign definite and can change sign along a M\"obius
orbit. The sign-definite handedness
imposed by the CCI constraint is therefore a property of the
tetrahedrally oriented combination
in Eq.~(\ref{eq:tetrahedronchirality}), rather than of its individual
triangular faces.

\subsection{Alternative edge-sharing construction on the triangular lattice}
\label{sec:triangular_edge_sharing}

The triangular lattice also admits an edge-sharing realization of the
CCI constraint that is obtained naturally as a dimensional reduction
of the fcc construction.  Recall that the fcc lattice viewed along a
$[111]$ direction consists of triangular layers in the usual
$abc\,abc\ldots$ stacking (lowercase letters denote stacking
registries, to avoid confusion with the color labels).  We retain one
complete $abc$ period and
impose periodic boundary conditions in the $[111]$ direction, so that
the layer following $c$ is identified with the original $a$ layer.
Equivalently, we quotient the fcc lattice by the shortest lattice
translation parallel to $[111]$ that returns a triangular layer to
the same registry.  Collapsing the $[111]$ coordinate then superposes
the three $a$, $b$, and $c$ registries.  Their union forms a
triangular
lattice whose lattice constant is smaller by a factor $\sqrt{3}$ than
that of an individual $(111)$ layer.

\begin{figure}[tb]
    \centering
    \includegraphics[width=0.85\linewidth]
    {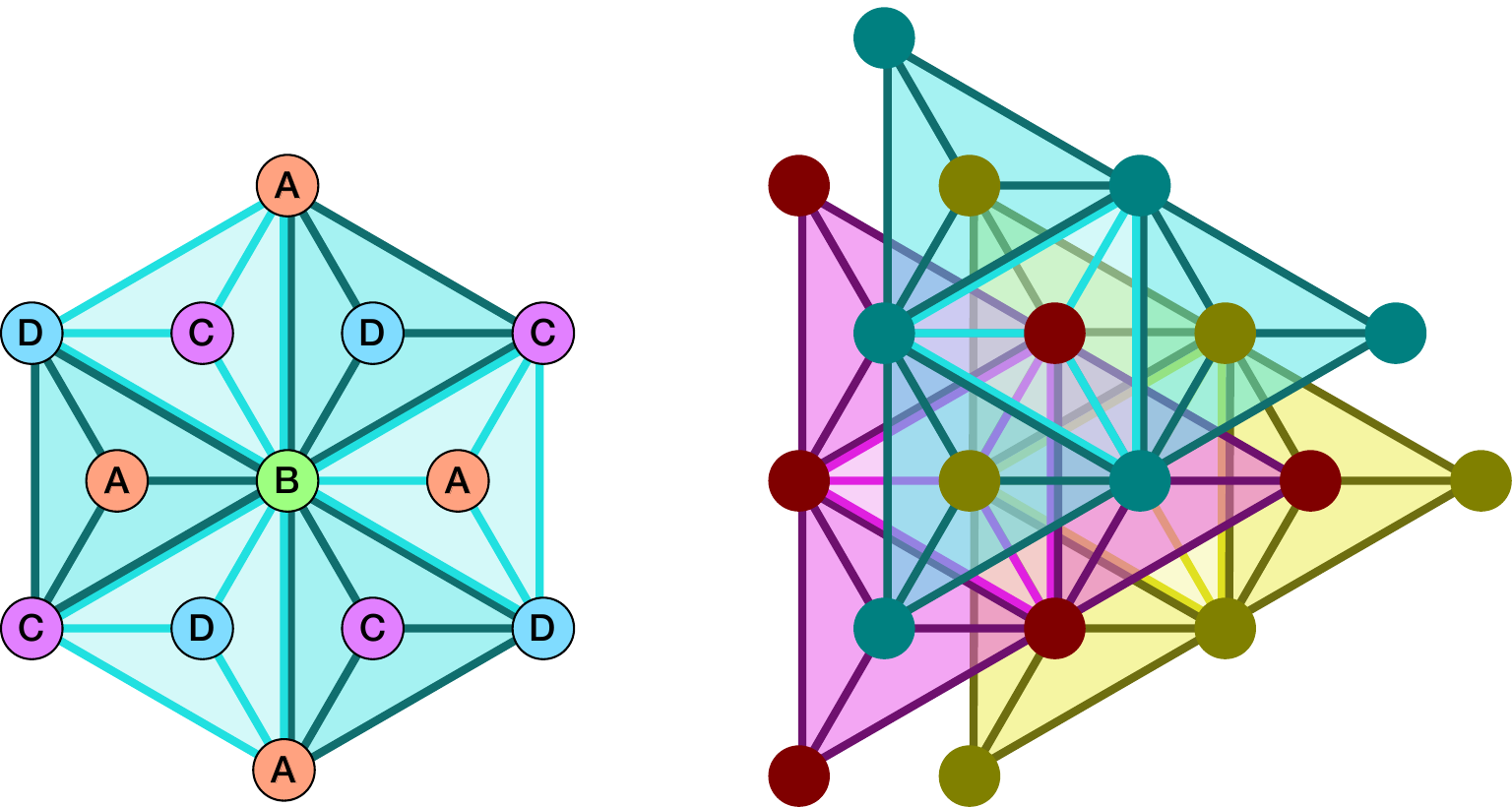}
    \caption{
    Alternative edge-sharing construction on the triangular lattice.
    Each four-site motif consists of a central site and three
    alternating nearest neighbors. Of the six edges of the associated
    tetrahedron, three are nearest-neighbor bonds of the triangular
    lattice and the remaining three connect second neighbors.
    (a)~Representative motifs of the two orientations, which are
    exchanged by a $C_6$ rotation.
    (b)~The motifs can be organized into three symmetry-related
    rhombille families; together they generate the full $C_6$-symmetric
    set of $2N$ four-site terms.
    }
    \label{fig:triangular_lattice_alt}
\end{figure}

An elementary fcc tetrahedron has a $1+3$ structure with respect to
two neighboring $(111)$ layers: one vertex lies in one layer and the
other three form a triangle in the adjacent layer.  After the
three-layer quotient and projection, such a tetrahedron therefore
becomes a four-site motif consisting of one site and three alternating
nearest neighbors of the resulting triangular lattice, as shown in
Fig.~\ref{fig:triangular_lattice_alt}.  The three bonds connecting the
central site to the outer sites become nearest-neighbor bonds, whereas
the three edges of the outer triangle connect second neighbors.
The two orientations of the fcc tetrahedron project to the two motifs
related by a $C_6$ rotation.  Since the periodic three-layer fcc
quotient contains two elementary tetrahedra per projected site, the
complete construction contains $2N$ local CCI terms, precisely the
full $C_6$-symmetric set used below.

This fcc origin also clarifies the fate of the planar zero modes of
the three-dimensional model.  On the fcc lattice, a representative
zero mode is obtained by deforming the two spin directions carried by
a single $(100)$ plane, while the remaining two sublattice directions
are held fixed; the M\"obius constraint fixes one of the two deformed
directions in terms of the other
[Fig.~\ref{fig:fcc_lattice}(b)].  In the square- and honeycomb-lattice
bilayers, the intersection of such a plane with the retained slab is
one dimensional, and the planar mode consequently descends to the
line defects discussed above.

The $[111]$ reduction leading to the triangular lattice behaves
differently because the transverse direction is periodic.  A single
$(100)$ plane is not invariant under the three-layer identification;
the periodic zero mode must therefore contain all of its images under
the $[111]$ compactification.  After projection, this periodically
repeated family occupies two of the four tetrahedral sublattices
throughout the two-dimensional system.  Thus the fcc planar mode does
not become a localized line defect on the triangular lattice.  It
instead reduces to a spatially periodic four-sublattice deformation,
\begin{equation}
  (A,B,C,D)
  \longrightarrow
  (A',B',C,D),
  \qquad
  B'=T_{z_C,z_D}(A'),
  \label{eq:triangular_projected_planar_mode}
\end{equation}
up to permutations of the four sublattice labels.  The $(010)$ and
$(001)$ planar families give the corresponding deformations of the
other pairings of tetrahedral sublattices.  These projected modes are
therefore contained in the continuously deformable four-sublattice
family of coherent zero modes, rather than producing additional
subdimensional defects. At linear order this projected family is precisely a global
M\"obius deformation---its tangent
$\delta z_i\propto(z_i-z_C)(z_i-z_D)$ vanishes on the two fixed
sublattices---consistent with the linearized analysis of
Appendix~\ref{app:rigidity}, which finds no nonuniform infinitesimal
coherent modes for either triangular covering.

The important difference from the preceding face-sharing construction is the incidence structure of the local terms. Two distinct motifs (tetrahedra) of Fig.~\ref{fig:triangular_lattice_alt} share at most two sites in the infinite lattice, and hence are edge-sharing, whereas two rhombi of Fig.~\ref{fig:triangular_lattice} can share three sites, corresponding to a common tetrahedral face. The triangular lattice therefore provides both edge- and face-sharing realizations of the same local chiral constraint.

The four-sublattice ordered state remains an exact zero-energy state of the alternative construction. With the site ordering indicated in Fig.~\ref{fig:triangular_lattice_alt}, the four sublattice labels on each motif are related to those of the reference tetrahedron by an even permutation, so every local term selects the same CCI handedness. The complete finite-size ordered subspace $\mathscr V^{\widetilde S}$, including the twisted Anderson-tower content discussed above, is therefore again contained in the common kernel. What changes is the number of additional zero modes allowed around this ordered component.

We tested this directly by exact diagonalization. The $N=16$ cluster provides a clean realization of the edge-sharing incidence structure: no two distinct motifs share more than two sites. The corresponding zero-energy multiplet content is given in Table~\ref{tab:degeneracy_triangular_claw}. The $N=12$ torus is exceptional: periodic wrapping causes twelve pairs of motifs to share three sites, so that this cluster acquires an artificial face-sharing component. We therefore use the $N=16$ results for the comparison below. The decomposition into the three rhombille families is a property of the infinite lattice; on the $N=16$ torus the Hamiltonian is defined directly from the full $C_6$-symmetric set of $2N$ motifs.

\begin{table*}[tp]
\caption{
Zero-energy multiplet content of the alternative edge-sharing
triangular-lattice construction, resolved using the same space-group
conventions as Table~\ref{tab:degeneracy_triangular_clusters}.
Blank entries denote zero multiplicity and a dash denotes a momentum
not contained in the finite-cluster Brillouin zone. The ``Total''
column gives the number of zero-energy spin multiplets after weighting
by the dimension of the corresponding space-group irrep. The
$N=12$ data are included for completeness, but periodic wrapping
produces twelve pairs of motifs with a three-site overlap; the
$N=16$ cluster realizes the intended edge-sharing geometry without
such overlaps.
}
\label{tab:degeneracy_triangular_claw}
\centering
\setlength{\tabcolsep}{2.5pt}
\renewcommand{\arraystretch}{1.08}
\begin{ruledtabular}
\begin{tabular}{cc cccccc ccc cc ccc c}
& & \multicolumn{6}{c}{$C_6\quad(d=1)$}
& \multicolumn{3}{c}{$C_3\quad(d=2)$}
& \multicolumn{2}{c}{$C_2\quad(d=3)$}
& \multicolumn{3}{c}{$C_1\quad(d=6)$}
& \\
$N$ & $J$
& $\Gamma_0$ & $\Gamma_{+1}$ & $\Gamma_{-1}$
& $\Gamma_{+2}$ & $\Gamma_{-2}$ & $\Gamma_3$
& $K_0$ & $K_+$ & $K_-$
& $M_+$ & $M_-$
& $Q_{12}$ & $Q_{16,1}$ & $Q_{16,2}$
& Total
\\
\hline
\multirow{7}{*}{$12$}
& 0
& 3 &  &  &  & 1 &
&  & 1 &
&  &
&  & -- & --
& 6
\\

& 1
&  & 1 &  &  &  & 1
& 1 & 1 &
& 1 &
&  & -- & --
& 9
\\

& 2
& 1 &  &  & 2 & 2 &
& 2 & 1 & 1
& 1 &
&  & -- & --
& 16
\\

& 3
& 2 & 1 &  &  &  & 1
& 1 & 2 &
& 2 &
& 1 & -- & --
& 22
\\

& 4
& 2 &  &  & 1 & 2 &
& 2 & 2 &
& 1 & 1
& 2 & -- & --
& 31
\\

& 5
&  &  &  &  &  &
& 1 &  &
& 1 &
& 1 & -- & --
& 11
\\

& 6
& 1 &  &  &  &  &
&  &  &
&  &
&  & -- & --
& 1
\\
\hline
\multirow{9}{*}{$16$}
& 0
& 2 &  &  & 2 & 3 &
& -- & -- & --
&  & 2
& -- &  &
& 13
\\

& 1
&  &  &  &  &  &
& -- & -- & --
& 1 &
& -- & 1 & 1
& 15
\\

& 2
& 4 &  &  & 3 & 2 &
& -- & -- & --
& 2 & 2
& -- & 1 & 1
& 33
\\

& 3
& 1 &  &  &  & 2 & 1
& -- & -- & --
& 4 &
& -- & 3 & 3
& 52
\\

& 4
& 4 &  &  & 4 & 4 &
& -- & -- & --
& 5 & 2
& -- & 4 & 4
& 81
\\

& 5
& 3 & 1 & 1 & 1 &  & 2
& -- & -- & --
& 5 &
& -- & 7 & 7
& 107
\\

& 6
& 2 &  &  & 2 & 3 &
& -- & -- & --
& 3 & 3
& -- & 4 & 4
& 73
\\

& 7
&  &  &  &  &  &
& -- & -- & --
& 1 &
& -- & 1 & 1
& 15
\\

& 8
& 1 &  &  &  &  &
& -- & -- & --
&  &
& -- &  &
& 1
\\
\end{tabular}
\end{ruledtabular}
\end{table*}

The difference from the face-sharing model is already pronounced in the low-spin spectrum. For $N=16$, the edge-sharing construction has $13$, $15$, $33$, and $52$ zero-energy multiplets in the $J=0,1,2,3$ sectors, respectively. In the face-sharing construction, the corresponding sectors contain only $1$, $3$, $5$, and $7$ multiplets and are exhausted by the tetrahedral rotor. Nevertheless, the symmetry sectors required by the twisted ordered component remain present in Table~\ref{tab:degeneracy_triangular_claw}: the $\Gamma_{-2}$ tower singlet occurs at $J=0$, the $M_+$ triplet at $J=1$, and the higher-$J$ tower irreps are contained in the corresponding zero-energy sectors. Thus the four-sublattice ordered component survives, but the weaker overlap leaves a much larger kernel around it.

The difference is also substantial in the total dimension of the kernel. Weighting the multiplets by $2J+1$, the $N=16$ edge-sharing model has $3684$ zero-energy states, compared with $1738$ for the face-sharing rhombus construction. By contrast, the high-spin polarized sectors are unchanged: the complete $J=8$ and $J=7$ sectors remain at zero energy, giving respectively one and $15$ multiplets, as required by the zero- and one-down-spin argument.

The linearized analysis sharpens this comparison
(Appendix~\ref{app:rigidity}): around the four-sublattice state,
neither triangular construction has a nonuniform infinitesimal
coherent zero mode. In both cases $\mathsf R(\mathbf k)$ has trivial
kernel for every $\mathbf k\neq0$, even though the edge-sharing claw
construction has a much larger quantum zero-energy kernel. The
number of infinitesimal coherent modes and the dimension of the full
quantum kernel can therefore vary independently, and the appearance
of line modes is not an automatic consequence of edge sharing.

This comparison separates the effect of constraint connectivity from that of the underlying lattice. The same triangular lattice, the same local CCI projector, and the same four-sublattice chiral order give very different zero-energy kernels depending on how strongly the four-site terms overlap. Face sharing suppresses the additional low-spin zero modes and leaves a rigid ordered component, whereas edge sharing permits a much larger residual kernel. The hierarchy between corner-, edge-, and face-sharing constructions is therefore controlled by the overlap of the local constraints rather than by the lattice geometry alone.
At the level of coherent deformations, this hierarchy is particularly
transparent: the zero set of $\mathsf R(\mathbf k)$ extends over the
full Brillouin zone for the corner-sharing checkerboard and
pyrochlore lattices, is restricted to lines for the fcc, square, and
honeycomb constructions, and reduces to $\Gamma$ alone for both
triangular coverings, as shown in Appendix~\ref{app:rigidity}.

\section{Summary of results}
\label{sec:summary_results}

For reference, we collect here the principal results established in the
preceding sections.  The purpose of this section is to provide a compact
map of the construction and its consequences; their physical
interpretation, relation to other constrained spin systems, and possible
extensions are discussed separately in Sec.~\ref{sec:discussion}.
The results naturally separate into three levels: the local quantum
kernel of a single tetrahedron, the corresponding coherent-state
zero-mode variety, and the common kernel obtained by imposing the local
constraints on extended lattices [cf.~Eq.~(\ref{eq:hierarchy})].
Tables~\ref{tab:summary_local}--\ref{tab:summary_lattices} summarize
these statements, one table per level, and give
the sections in which they are derived.

\begin{table*}[t]
\caption{
Summary of the principal results for the single-tetrahedron quantum
problem.  The second column states the result, and the last column
gives the corresponding derivation or numerical characterization in
the main text; additional proofs and technical details are given in
the appendices cited in the corresponding sections.
}
\label{tab:summary_local}
\centering
\small
\setlength{\tabcolsep}{5pt}
\renewcommand{\arraystretch}{1.15}
\begin{ruledtabular}
\begin{tabular}{
p{0.15\textwidth}
p{0.61\textwidth}
p{0.18\textwidth}
}

\textbf{Scope}
&
\textbf{Result}
&
\textbf{Location}
\\
\hline

Spin-$1/2$ parent term
&
The rotated CCI product states of one handedness span $15$ of the
$16$ dimensions of the four-spin Hilbert space; the local parent
Hamiltonian is the rank-one projector onto the complementary chiral
tetrahedral singlet,
$\mathcal H_{\mathrm{CCI}}^{(1/2)}=\ket{+}\bra{+}$, expressible
entirely in terms of Heisenberg exchange, scalar spin chirality, and
products of exchanges.
&
Sec.~\ref{sec:spinhalf};
Eqs.~(\ref{eq:proj++}) and
(\ref{eq:spin_1_2_hamiltonian})
\\[1mm]

Arbitrary spin
&
For four spins of arbitrary length $S$, an SU(2)-invariant,
positive-semidefinite parent Hamiltonian can be constructed in the same
local operator basis.  It annihilates all global rotations of the CCI
product states of the selected handedness.
&
Sec.~\ref{sec:spinS};
Eq.~(\ref{eq:spin_S_hamiltonian})
\\[1mm]

Singlet-annihilation form
&
The arbitrary-$S$ parent Hamiltonian factorizes exactly as
\(
\mathcal H_{\mathrm{CCI}}^{(S)}
=
\mathcal B^\dagger\mathcal B
\),
where $\mathcal B$ is a four-site SU(2)-singlet annihilation operator
that removes one Schwinger boson from every site.  The local zero-energy
condition is therefore exactly
$\mathcal B\ket{\Psi}=0$.
&
Sec.~\ref{sec:schwinger};
Eqs.~(\ref{eq:BdB}) and
(\ref{eq:Bpsi0})
\\[1mm]

Complete local kernel
&
The kernel of $\mathcal B$ is exactly the subspace $\mathscr V^S$
spanned by the globally rotated CCI product states,
\(
\mathscr V^S=\ker\mathcal B
\).
Its dimension is
\(
(2S+1)^4-(2S)^4
\),
and its total-spin content follows from a sector-by-sector
recursion between four spin-$S$ and four spin-$(S-\tfrac12)$
degrees of freedom.
&
Sec.~\ref{sec:kernel_dims};
Eqs.~(\ref{eq:dim_kerB}) and
(\ref{eq:V_equals_kerB});
Table~\ref{tab:Stot_vs_S}
\\[1mm]

Quantum handedness
&
Every normalized state in the complete local quantum kernel, including
entangled states, obeys a definite sign constraint on the tetrahedral
chirality.  In the reference convention,
\(
\langle\hat\chi_{\mathrm{tet}}\rangle\leq0
\);
the time-reversed parent Hamiltonian reverses this inequality.
&
Sec.~\ref{sec:exact_quantum_handedness};
Eq.~(\ref{eq:local_chirality_sign})
\\[1mm]

Both chiral sectors
&
Combining the two complex-conjugate singlet-annihilation operators gives
a positive-semidefinite parent Hamiltonian for the full color-ice
subspace containing both handedness sectors.  The resulting operator is
invariant under the full site-permutation group $S_4$.  For $S\geq1$,
its complete kernel is the span of the globally rotated color-ice states
of both chiralities.
&
Secs.~\ref{sec:CIparent} and
\ref{sec:kernel_dims};
Eq.~(\ref{eq:HCI})
\\[1mm]

Local spectrum
&
The nonzero spectra of $\mathcal B^\dagger\mathcal B$ (spin $S$)
and $\mathcal B\mathcal B^\dagger$ [spin $(S-\tfrac12)$] are
paired, giving the exact local gap
\(
\Delta_{\mathrm{loc}}(S)=4S^2
\)---a local excitation scale that does not by itself imply a
many-body spectral gap.
&
Sec.~\ref{sec:spectral_pairing};
Eq.~(\ref{eq:local_gap})
\\
\end{tabular}
\end{ruledtabular}
\end{table*}

\begin{table*}[t]
\caption{
Summary of the principal results for the coherent-state geometry.
Columns as in Table~\ref{tab:summary_local}.
}
\label{tab:summary_coherent}
\centering
\small
\setlength{\tabcolsep}{5pt}
\renewcommand{\arraystretch}{1.15}
\begin{ruledtabular}
\begin{tabular}{
p{0.15\textwidth}
p{0.61\textwidth}
p{0.18\textwidth}
}

\textbf{Scope}
&
\textbf{Result}
&
\textbf{Location}
\\
\hline

Coherent zero modes
&
For products of spin-coherent states, the zero-energy condition is
independent of $S$ and reduces to the multi-affine polynomial equation
\(
p_-(z_1,z_2,z_3,z_4)=0
\)
on the Riemann sphere.  Away from the exceptional locus, three spin
directions determine the fourth uniquely.
&
Sec.~\ref{sec:mobius};
Eq.~(\ref{eq:Bsceq0})
\\[1mm]

M\"obius completion
&
Solving the coherent zero-mode constraint for one spin gives an
invertible M\"obius transformation,
\(
z_4=T_{z_1,z_2}(z_3)
\).
For a fixed noncoincident pair $(z_1,z_2)$, these two points are the
fixed points of the map and the remaining points form six-cycles.
Unique completion fails only on the three-coincident branch.
&
Sec.~\ref{sec:mobius};
Eqs.~(\ref{eq:Mobius}) and
(\ref{eq:zzzw});
Appendix~\ref{app:specialloci}
\\[1mm]

Coherent chirality
&
The coherent zero-mode variety has a definite handedness but a
continuously varying chirality magnitude,
\[
-\frac{16}{3\sqrt3}
\leq
\chi_{\mathrm{tet}}
\leq
0
\]
for the reference sector.  The lower endpoint is attained by the
regular tetrahedral frame, whereas the achiral endpoint occurs on the
three-coincident branch.  Independently of the zero-mode constraint, $16/(3\sqrt3)$ bounds
the tetrahedral chirality of any four unit spin directions.
&
Sec.~\ref{sec:mobius};
Eq.~(\ref{eq:chi_window});
Appendices~\ref{app:equilateral} and
\ref{app:chirality_bounds}
\\[1mm]

Loop holonomy
&
Successive local completion maps around a closed loop define a M\"obius
holonomy $\mathcal M_L$.  Generic nonidentity holonomy gives isolated
coherent loop solutions, whereas identity holonomy produces a
continuous family.  For identity holonomy this statement extends to the
complete quantum loop problem: the kernel is spanned by the propagated
coherent states and has dimension $2SL+1$.
&
Secs.~\ref{sec:lattice_construction},
\ref{sec:loops}, and
\ref{sec:coloring_loops};
Eq.~(\ref{eq:loop_kernel_exact})
\\
\end{tabular}
\end{ruledtabular}
\end{table*}

\begin{table*}[t]
\caption{
Summary of the principal results for the extended lattices.
Columns as in Table~\ref{tab:summary_local}.
}
\label{tab:summary_lattices}
\centering
\small
\setlength{\tabcolsep}{5pt}
\renewcommand{\arraystretch}{1.15}
\begin{ruledtabular}
\begin{tabular}{
p{0.15\textwidth}
p{0.61\textwidth}
p{0.18\textwidth}
}

\textbf{Scope}
&
\textbf{Result}
&
\textbf{Location}
\\
\hline

Many-body common kernel
&
Summing the local positive-semidefinite terms gives a lattice
Hamiltonian whose ground-state space is their common kernel;
whenever a compatible CCI coloring exists, the model is frustration
free.  The handedness condition extends to the complete many-body
kernel:
\(
\varepsilon_t
\langle\hat\chi_{\mathrm{tet}}(t)\rangle
\leq0
\)
on every tetrahedron and for every ground state.
&
Sec.~\ref{sec:lattice_construction};
Eqs.~(\ref{eq:common_kernel}) and
(\ref{eq:common_kernel_chirality_sign})
\\[1mm]

Corner sharing
&
The checkerboard and pyrochlore constructions have extensively
degenerate quantum ground-state spaces.  Embedded loops generate
explicit coherent and quantum zero modes, while for $S=1/2$ a polarized
subset of the common kernel maps onto matchings of the dual square or
diamond lattice.
&
Secs.~\ref{sec:degeneracy},
\ref{sec:loops}, and
\ref{sec:matching};
Table~\ref{tab:planar16}
\\[1mm]

Extensive degeneracy
&
For $S=1/2$, rigorous bounds on the matching partition function give
\(
1.3919^N
\lesssim
\mathcal D_{\mathrm{match}}
\lesssim
1.3964^N
\)
for an explicitly constructed polarized subset of the
corner-sharing ground-state space; the lower bound alone establishes
exponential growth of the full quantum degeneracy, and independent
loop constructions give complementary bounds for arbitrary $S$.
&
Sec.~\ref{sec:degeneracy_comparison};
Eqs.~(\ref{eq:match_bounds}),
(\ref{eq:loopbound_checkerboard}), and
(\ref{eq:loopbound_pyrochlore})
\\[1mm]

fcc lattice
&
For the alternating assignment, constraint propagation fixes the
four-sublattice tetrahedral coloring uniquely once one tetrahedron
is specified; the uniform assignment is frustrated.  Exact coherent
zero-energy deformations localized on $(100)$ planes yield a
subextensive family of zero modes.
&
Sec.~\ref{sec:fcc}
\\[1mm]

Square and honeycomb
&
The alternating branch supports the four-sublattice state, the
uniform branch admits no compatible coloring, and the fcc planar
modes descend to exact line deformations, with $4L-1$ (square) and
$6L-3$ (honeycomb) infinitesimal coherent zero modes.
&
Sec.~\ref{sec:square_honeycomb};
Eqs.~(\ref{eq:Nlin_square}) and
(\ref{eq:Nlin_honeycomb})
\\[1mm]

Triangular lattice:
face sharing
&
The face-sharing rhombus construction retains the four-sublattice
tetrahedral zero-energy family but has no nonuniform infinitesimal
coherent deformation about it: only the three global M\"obius
directions survive.  Exact diagonalization identifies the corresponding
finite-size tetrahedral Anderson tower inside the common kernel,
together with additional zero modes at higher total spin.
&
Sec.~\ref{sec:triangular_lattice};
Eq.~(\ref{eq:Nlin_triangular_rhombi});
Table~\ref{tab:degeneracy_triangular_clusters}
\\[1mm]

Tetrahedral Anderson tower
&
The finite-size ordered component can be identified with the
four-collective-spin space $\mathscr V^{\widetilde S}$.  Its spin-$J$
multiplicities follow from restricting the SO(3) representation to
the proper tetrahedral group $A_4$.  A finite-spin stabilizer character
can shift the discrete lattice quantum numbers without changing the
multiplicities.  For the triangular clusters, the predicted
multiplicities and space-group sectors occur exactly at zero energy.
&
Sec.~\ref{sec:triangular_lattice};
Eqs.~(\ref{eq:tower_spacegroup_mapping}) and
(\ref{eq:tower_twist});
Appendix~\ref{app:tower}
\\[1mm]

Quantum chirality spectra
&
Within the complete triangular-lattice common kernel, the projected
tetrahedral chirality operator is negative semidefinite in the selected
sector, whereas the chirality of an individual triangular face is not
sign definite.  Low-spin quantum zero modes can also carry a
tetrahedral-chirality magnitude exceeding the maximal value available
to a spin-$1/2$ coherent product state.
&
Sec.~\ref{sec:quantum_chirality};
Eq.~(\ref{eq:global_chirality_semidefinite})
\\[1mm]

Triangular lattice:
edge sharing
&
An alternative edge-sharing realization on the same triangular lattice
contains the same four-sublattice ordered component and has the same
harmonic zero-mode content---only the global M\"obius modes remain---
but a substantially larger full quantum zero-energy kernel.  On the
$N=16$ cluster, the edge-sharing and face-sharing constructions contain
$3684$ and $1738$ zero-energy states, respectively.
&
Sec.~\ref{sec:triangular_edge_sharing};
Eq.~(\ref{eq:Nlin_triangular_claw});
Table~\ref{tab:degeneracy_triangular_claw}
\\[1mm]

Constraint-overlap
hierarchy
&
The linearized coherent zero modes occupy the full Brillouin zone for
the corner-sharing checkerboard and pyrochlore lattices,
one-dimensional momentum-space sets for the fcc, square, and honeycomb
constructions, and only $\Gamma$ for the two triangular constructions.
The corresponding numbers of infinitesimal coherent deformations are
therefore extensive, subextensive, and $O(1)$, respectively.
&
Appendix~\ref{app:rigidity};
Table~\ref{tab:rigidity}
\\[1mm]

Harmonic versus
quantum zero modes
&
The two triangular-lattice constructions have identical
infinitesimal coherent zero-mode spaces around the four-sublattice
state but substantially different full quantum kernels; coherent
flexibility and quantum kernel dimension are therefore independent
characteristics of the constraint problem.
&
Sec.~\ref{sec:triangular_edge_sharing};
Appendix~\ref{app:rigidity}
\\[1mm]

Polarized/high-spin
sector
&
Fully polarized states and the one-spin-flip sector are annihilated by
every local CCI term independently of the chirality assignment.
Consequently, the complete $J=N/2$ and $J=N/2-1$ multiplet sectors are
at zero energy.  Differences between chirality assignments appear only
at lower total spin, where linear dependencies can develop among the
overlapping local constraints.
&
Secs.~\ref{sec:checkerboard},
\ref{sec:square_honeycomb}, and
\ref{sec:triangular_lattice};
Appendix~\ref{app:highspin_count}
\end{tabular}
\end{ruledtabular}
\end{table*}

Several distinctions summarized in
Tables~\ref{tab:summary_local}--\ref{tab:summary_lattices} are useful in organizing the results.
First, the rotated CCI colorings form only a subset of the coherent
zero-mode variety, which in turn is contained in the complete quantum
common kernel [Eq.~(\ref{eq:hierarchy})].  The M\"obius completion rule
and its holonomies characterize the coherent product-state sector,
whereas exact-diagonalization spectra, kernel dimensions, and matching
bounds refer to the full quantum zero-energy space.

Second, the local CCI constraint fixes the handedness of the
tetrahedrally oriented combination of face chiralities, but does not fix
its magnitude and does not impose a definite chirality sign on each
individual triangular face.  This distinction persists in the complete
many-body quantum kernel, as demonstrated explicitly by the projected
chirality spectra in Sec.~\ref{sec:quantum_chirality}.

Finally, the lattice constructions separate coherent flexibility from
quantum ground-state degeneracy.  Around the same four-sublattice
tetrahedral state, the support of the linearized zero modes changes from
the full Brillouin zone on the corner-sharing lattices, to
one-dimensional momentum-space sets in the fcc, square, and honeycomb
models, and finally to $\Gamma$ alone in the triangular constructions.
At the same time, the comparison of the two triangular coverings shows
that identical harmonic zero-mode content does not imply identical
quantum kernel dimensions.  The overlap structure of the local tetrahedral
constraints therefore controls several distinct properties of the
zero-energy space that need not vary together.

\section{Outlook}
\label{sec:discussion}

We set out to answer a concrete question: can a local quantum spin
Hamiltonian enforce an ice-type rule that knows about handedness?
The models constructed here answer it in the affirmative, and in a
stronger form than we had anticipated.  A single four-site term,
built only from scalar chiralities and products of Heisenberg
exchanges, annihilates every chiral four-coloring state and all of
its global rotations, for any spin length.  Three technical facts
carry most of the weight.  For $S=1/2$ the term is the projector
onto one member of the time-reversal pair of tetrahedral singlets,
so selecting a handedness costs exactly one state in sixteen.  For
general $S$ the same statement takes the form
$\mathcal B^\dagger\mathcal B$ with a singlet-annihilation condition
whose classical solutions obey a M\"obius completion rule: three
spins fix the fourth, loops fix their own consistency, and the
ground-state counting reduces to fixed points and matchings, with
residual-entropy bounds that exceed the Pauling and Lieb values for
ice.  Finally, the chirality window
$-16/(3\sqrt3)\le\chi_{\mathrm{tet}}\le0$ pins down what the constraint
actually enforces: the sign of the local handedness, exactly, and
nothing else.  In this precise sense the coherent zero-mode variety
is the deformable
envelope of the rigid classical chiral states of
Ref.~\cite{Lozano2024}.

The large degeneracy is not a defect of the construction; it is
forced.  A local, spin-rotation-invariant, positive-semidefinite
term whose kernel contains one coherent product state must annihilate
the entire rotated family, and the singlet-annihilation structure
then admits the polarized states and their matching descendants as
well.  Every parent Hamiltonian of this kind therefore sits at a
degenerate multiphase point adjacent to a ferromagnetic phase, as the
kagome three-coloring point does in its own setting
\cite{Changlani2018}.  The productive attitude toward such a point is
to perturb away from it and map the surrounding phase diagram;
quantum melting of noncoplanar tetrahedral order is indeed known to
produce chiral spin liquids \cite{Hickey2017}, and the field
theories of such chiral-spin-liquid--to--spin-crystal transitions
are now being formulated \cite{Bose_2025}.  Here
we should be precise about what is and is not in hand.  On the
corner-sharing lattices we know the product-state sector of the
kernel completely---the M\"obius coordinates parametrize it---and we
know the polarized matching states, but we do not have a closed-form
basis for the full kernel of the pyrochlore model; controlled
degenerate perturbation theory over the entire kernel is therefore
still out of reach in general.  What is well posed already is the
exploration of the neighborhood: weak antiferromagnetic exchange,
biquadratic terms, or a chirality field
$\lambda\hat\chi_{\mathrm{tet}}$ each define a definite path away from
the solvable point, and one can ask which zero-energy states
they favor.  It is natural to expect a selection effect---by energy,
by fluctuations \cite{Villain1980,Chalker1992}, or by the
environment in the pointer-state sense discussed
below---and the chiral color-ice states themselves are the obvious
candidates to be selected.  Whether that happens, or the ferromagnet
wins, or an intermediate regime intervenes, is open on both the
classical and the quantum side.

What do correlations look like within the constrained ensemble
itself?
Ice-type manifolds form Coulomb phases: an emergent divergence-free
field produces pinch points in the structure factor
\cite{Henley2010}, and the classification of
Refs.~\cite{Yan2024typology,Yan2024formalism} rests on the linear
constraints that generate them.  The condition $p_-=0$ is not
linear, so chiral color ice falls outside this classification, and
whether its correlations retain any pinch-point-like structure is a
sharp open question.  The M\"obius completion rule offers a concrete
route to it: because the fourth spin of every tetrahedron is
determined by the other three, ensembles of configurations that are
\emph{exactly at zero energy} can be generated by growing the
lattice tetrahedron by tetrahedron and completing each new
tetrahedron by Eq.~(\ref{eq:Mobius}). Turning such ensembles into a
statement about the correlations of the coherent zero-mode variety,
however, requires care that we have not undertaken here: a declared
probability measure (the variety carries no canonical one), a
treatment of loop closure and M\"obius holonomy on periodic
clusters, and controlled system-size and error analyses. We
therefore leave the correlation pattern of chiral color ice as an
open problem.  We caution, moreover, against expecting the
correlation pattern of a stable phase.  The solvable point sits on
the boundary of the ferromagnetic phase, and its closest relative,
the multicritical point of the zigzag chain \cite{Saito2024}, is a
critical point rather than a phase; the appropriate question is not
the field theory of a chiral-color-ice \emph{liquid}, but the nature
of this multiphase point itself---whether it is critical, what
gapless structure and entanglement it carries, and what controls the
transitions out of it.  The completion-rule sampling and the exact
zero-energy kernels of the quantum clusters give
this question an unusually concrete starting point.

The ordered side of the phase diagram comes with its own exact
statement, and with a caveat worth repeating.  The four-sublattice
coloring survives only when the chirality assignment
\emph{alternates} between
the two inversion-related families of tetrahedra: on the fcc, square,
and honeycomb lattices with this assignment (and on the triangular
lattice, whose face-sharing covering admits the coloring directly),
the
kernel contains the chiral four-sublattice states and the complete
set of multiplets of their Anderson tower, all at zero energy. A
weak perturbation that favors this ordered component would split
these multiplets into the tower from which the
symmetry-broken state is assembled in the thermodynamic limit;
which part of the kernel a given perturbation selects is a question
we leave open.  The
kernel is not exhausted by the ordered component---a subextensive
family of planar (fcc) or line (square, honeycomb) deformations
survives, along with the discrete polarized motifs, and only on the
triangular lattice do the continuous deformations disappear
entirely---but this residual degeneracy is parametrically smaller
than the extensive one of the corner-sharing lattices, and it leaves
the tower content intact.  For the uniform assignment, the coloring
rule is frustrated
and this entire structure is absent---we verified by exhaustive
enumeration that not a single four-coloring survives on the fcc,
square, or honeycomb lattices.  This puts the tower-of-states analysis of
Ref.~\cite{Sotnikov2023} into an unusually clean setting.  There, the
decomposition of classical order into tower eigenstates, and the
selection of low-entanglement pointer states by an environment, had
to be extracted numerically from split spectra; here the kernel is
exactly degenerate, its product-state members are classified
completely, and each of them carries a definite handedness.  The
emergence of a classical chirality from a quantum-degenerate
kernel can therefore be studied here without any of the usual
finite-size ambiguity.  On the technical side, the matrix-product
construction of Ref.~\cite{Saito2024} for degenerate
frustration-free points applies to our models on strips and
cylinders. The zero-energy kernel, as the common kernel of all
local terms, remains pinned at zero energy under arbitrary
reweighting of those terms---the structural prerequisite for
protected embeddings inside thermalizing Hamiltonians
\cite{Chertkov2021,Moudgalya2022}; whether such embeddings produce
genuine scar phenomenology, such as long-lived chirality
oscillations \cite{Serbyn2021}, is left open here.

We should be honest about where such Hamiltonians might exist.  We
know of no material whose microscopic couplings realize
Eq.~(\ref{eq:spin_S_hamiltonian}) as written.  What exists in nature
are the ingredients: scalar-chirality interactions generated by ring
exchange and orbital fields, applied fields, or circular driving in
Mott insulators
\cite{WWZ1989,Momoi1997,Sen1995,Motrunich2006,Kitamura2017},
the tetrahedral triple-$\mathbf q$ order that itinerant frustrated
magnets select, with its spontaneous topological Hall response
\cite{Martin2008,Batista2016}, and tetrahedral spin frames in the
cyclic phase of spin-2 condensates \cite{Barnett2006}.  The most
direct route would be engineered, and we state its status carefully.
As an always-on analog Hamiltonian,
Eq.~(\ref{eq:spin_S_hamiltonian}) requires a specific
SU(2)-invariant combination of three- and four-spin couplings that
no current platform provides natively---although programmable three-
and four-body interactions have been demonstrated for small
trapped-ion registers \cite{Katz2023}---and the inverse
quantum-simulation framework of Ref.~\cite{Kokail2026}, which
reconstructs a parent Hamiltonian for a variationally prepared
target state, is at present a theoretical proposal.  For $S=1/2$, a
digital route is more immediate: each local term is the rank-one
projector of Eq.~(\ref{eq:proj++}), so exponentiating or measuring
a single tetrahedron term reduces to a four-qubit controlled-phase
primitive conjugated by a fixed local basis change, a gate class
now demonstrated with high fidelity in neutral-atom arrays
\cite{Evered2023}.  When such capabilities mature, exactly solvable points
with fully computable ground-state spaces are the natural
benchmarks against which they will be validated.  Meanwhile, the
solvable point serves
the same purpose that the Majumdar--Ghosh and AKLT points have long
served: not as a model of a particular compound, but as the fixed
reference from which the surrounding phase diagram can be understood.

The construction generalizes in more than one direction.
The recipe has three inputs: a motif of sites, a set of colored
product states on it, and the symmetry group under whose orbit the
target set of states is completed before projecting onto its
complement.
Nothing ties these inputs to four colors, tetrahedra, or the full
spin-rotation group.  Keeping the motif and colors but shrinking the
symmetry to rotations about a single axis reproduces, by the same
steps, anisotropic coloring models of the kagome
three-coloring type \cite{Changlani2018,Changlani2019}---which shows
that the known XXZ solvable points and the present SU(2)-symmetric
ones are members of one family, distinguished only by how much
symmetry the constraint is asked to respect.  Changing the local
motif instead changes the constrained objects themselves: projectors
on triangular faces of the same corner-sharing lattices produce
fully packed valence-bond \emph{loop} manifolds as exact spin-1
kernels \cite{Hari2026}, so colorings, loops, and the chiral
constraint studied here are all members of one family of exactly
enforceable local rules.  In the other direction, three-color chiral
constraints on triangles, larger simplices, and SU($N$) analogs can
be pursued with the same tools.
Spin ice demonstrated that a local flux rule generates emergent
gauge fields and fractionalization.  The models constructed here
demonstrate that a local handedness rule is equally consistent, and
that its consequences---residual entropy beyond that of ice, a
constraint geometry outside the linear Coulomb-phase
classification, and exactly protected
subspaces of chiral states---can be worked out in full.  What grows
from this point, under perturbations, at finite temperature, and in
other symmetry classes, is now a definite program rather than a
speculation.

\acknowledgments
We thank G.~Baskaran, Imre Hagym\'asi and Paula Mellado for valuable
discussions.
This work was supported by the Hungarian National Research, Development and Innovation Office (NKFIH) through OTKA Grant No.~K~142652.
The project supported by the Doctoral Excellence Fellowship Program (DCEP) is funded by the National Research Development and Innovation Fund of the Ministry of Culture and Innovation and the Budapest University of Technology and Economics.
The work of Y.I.\ and K.P.\ was performed, in part, at the Aspen
Center for Physics, which is supported by a grant from the Simons
Foundation (1161654, Troyer).
This research was also supported in part by grant NSF PHY-2309135 to
the Kavli Institute for Theoretical Physics.
Y.I.\ acknowledges support from the Abdus Salam International Centre
for Theoretical Physics through the Associates Programme, from the
Simons Foundation through Grant No.~284558FY19, from IIT Madras
through the Institute of Eminence program for establishing QuCenDiEM
(Project No.~SP22231244CPETWOQCDHOC), and the International Centre
for Theoretical Sciences for participation in the Discussion
Meeting---Fractionalized Quantum Matter (code: ICTS/DMFQM2025/07).
K.P.\ and P.K.\ acknowledge support from IIT Madras through the
Visiting Researcher program, during which this project was
initiated.

\vspace{\baselineskip}
\noindent\emph{Data availability.}---The numerical material
supporting the conclusions of this article---exact-diagonalization
multiplet tables, four-coloring enumerations, matching-count
calculations, chirality spectra, and the
symbolic-verification and mode-counting scripts---is available from the authors upon reasonable request.

\appendix

\section{Scalar chirality}
\label{app:chirality}

This appendix collects operator identities relating scalar chirality
to permutation and exchange operators, and used to rewrite the
spin-$1/2$ parent Hamiltonian in terms of chirality operators.

\subsection{Three spin-$1/2$ sites}

For three spin-$1/2$ sites, the scalar chirality of
Eq.~(\ref{eq:SpinScalarChirality}) can be expressed through cyclic
permutations as
\begin{equation}
  \hat\chi_{ijk}
  =
  \frac{i}{4}
  \left(
    \mathcal P_{ijk}-\mathcal P_{ikj}
  \right),
  \qquad
  \mathcal P_{ikj}=\mathcal P_{ijk}^{-1},
  \label{eq:chi_permutation}
\end{equation}
as noted in Ref.~\cite{WWZ1989}. Its eigenvalues are
$\pm\sqrt{3}/4$ on the two total-spin-$1/2$ doublets and zero
on the total-spin-$3/2$ quartet. Consistently,
\cite{Baskaran1989}
\begin{align}
  \hat\chi_{ijk}^{\,2}
  &=
  \frac{3}{32}
  -
  \frac{1}{8}
  \left(
    \hat{\mathbf S}_i\cdot\hat{\mathbf S}_j
    +
    \hat{\mathbf S}_j\cdot\hat{\mathbf S}_k
    +
    \hat{\mathbf S}_k\cdot\hat{\mathbf S}_i
  \right)
  \nonumber\\
  &=
  \frac{15}{64}
  -
  \frac{1}{16}
  \left(
    \hat{\mathbf S}_i+
    \hat{\mathbf S}_j+
    \hat{\mathbf S}_k
  \right)^2 .
  \label{eq:chi_squared}
\end{align}

\subsection{Four spin-$1/2$ sites}

On a tetrahedron, the consistently oriented sum of the four face
chiralities defines the tetrahedral chirality
$\hat\chi_{\mathrm{tet}}$ of Eq.~(\ref{eq:tetrahedronchirality}).
For four spin-$1/2$ sites, its eigenvalues are $\pm\sqrt3$ on the
two chiral singlets of Eq.~(\ref{eq:chistates}) and zero on the
remaining fourteen states.

Applying Eq.~(\ref{eq:chi_squared}) to the four faces gives
\begin{equation}
  \sum_{\langle ij\rangle}
  \hat{\mathbf S}_i\cdot\hat{\mathbf S}_j
  =
  \frac{3}{2}
  -
  4\sum_f\hat\chi_f^{\,2},
  \label{eq:bilinear_chi}
\end{equation}
where $f$ runs over the four faces. Similarly, using the spin-$1/2$
product rule
\(
S_i^\alpha S_i^\beta
=
\tfrac14\delta_{\alpha\beta}
+
\tfrac{i}{2}\epsilon_{\alpha\beta\gamma}S_i^\gamma
\),
the symmetric quadrilinear operator
\begin{equation}
  \mathcal Q
  =
  (\hat{\mathbf S}_1\cdot\hat{\mathbf S}_2)
  (\hat{\mathbf S}_3\cdot\hat{\mathbf S}_4)
  +
  (\hat{\mathbf S}_1\cdot\hat{\mathbf S}_3)
  (\hat{\mathbf S}_2\cdot\hat{\mathbf S}_4)
  +
  (\hat{\mathbf S}_1\cdot\hat{\mathbf S}_4)
  (\hat{\mathbf S}_2\cdot\hat{\mathbf S}_3)
\end{equation}
can be written as
\begin{equation}
  \mathcal Q
  =
  \frac{3}{16}
  -
  \sum_f\hat\chi_f^{\,2}
  +
  \frac12\hat\chi_{\mathrm{tet}}^{\,2}.
  \label{eq:Q_compact}
\end{equation}

These operators also have a simple representation in terms of the
total spin
\(
\hat{\mathbf S}_{\mathrm{tet}}=\sum_{i=1}^4\hat{\mathbf S}_i
\).
In particular,
\begin{equation}
  \mathcal Q
  =
  \frac18
  \left(\hat{\mathbf S}_{\mathrm{tet}}^{\,2}\right)^2
  -
  \frac78\hat{\mathbf S}_{\mathrm{tet}}^{\,2}
  +
  \frac{15}{16},
  \label{eq:Q_Stot}
\end{equation}
and
\begin{equation}
  \hat\chi_{\mathrm{tet}}^{\,2}
  =
  \frac14
  \left(
    \hat{\mathbf S}_{\mathrm{tet}}^{\,2}-2
  \right)
  \left(
    \hat{\mathbf S}_{\mathrm{tet}}^{\,2}-6
  \right)
  =
  3\,\mathcal P_{S_{\mathrm{tet}}=0} \,.
  \label{eq:chitet_squared_Stot}
\end{equation}
Thus $\hat\chi_{\mathrm{tet}}^{\,2}$ is three times the projector onto
the two-dimensional total-spin singlet sector. The chirality itself
is not a function of $\hat{\mathbf S}_{\mathrm{tet}}^{\,2}$, since it
distinguishes the two singlets by their eigenvalues $\pm\sqrt3$.

\subsection{Scalar chirality in coherent-state coordinates}
\label{app:chirality_coherent}

For a product of spin-coherent states
(\ref{eq:coherent_z}) with stereographic coordinates $z_i$, the
$S$-normalized scalar chirality of an oriented triangle is
\begin{equation}
  \chi_{ijk}
  =
  \frac{
    4\,\operatorname{Im}
    \left[
      (1+z_i^*z_j)
      (1+z_j^*z_k)
      (1+z_k^*z_i)
    \right]
  }{
    (1+|z_i|^2)
    (1+|z_j|^2)
    (1+|z_k|^2)
  } .
  \label{eq:chi_z_compact}
\end{equation}
This representation is particularly
convenient for evaluating chirality on the coherent zero-mode variety
parametrized by the M\"obius completion rule
(\ref{eq:Mobius}).

\section{Derivation of the spin-\texorpdfstring{$S$}{S} Hamiltonian}
\label{app:HamS}

Here we derive Eq.~(\ref{eq:spin_S_hamiltonian}) directly from the
spin-$1/2$ parent Hamiltonian.  Let $n=2S$ and consider the balanced
complete four-partite, or Tur\'an, graph $T(4n,4)=T(8S,4)$, whose
vertex set is partitioned into four independent subsets
$V_1,\ldots,V_4$, each containing $n$ spin-$1/2$ degrees of freedom.
Selecting one vertex from each subset defines one of the $n^4$
four-site cliques.  We place the spin-$1/2$ CCI parent term on every
such clique and sum,
\begin{equation}
  \widetilde{\mathcal H}
  =
  \sum_{v_1\in V_1}\cdots\sum_{v_4\in V_4}
  \mathcal H_{\mathrm{CCI}}^{(1/2)}
  (v_1,v_2,v_3,v_4).
  \label{eq:turan_sum}
\end{equation}

The different terms in
Eq.~(\ref{eq:spin_1_2_hamiltonian}) acquire simple multiplicity
factors in this sum.  The constant term occurs $n^4$ times.  For a
fixed pair of subsets $V_i,V_j$, a given bilinear term is accompanied
by arbitrary choices of the vertices in the other two subsets and
therefore acquires a factor $n^2$.  Similarly, a scalar-chirality
term involving three subsets acquires a factor $n$, while a product
of two bilinears involving all four subsets acquires no additional
multiplicity.

Introduce the collective spins
\begin{equation}
  \hat{\mathbf S}_i
  =
  \sum_{v\in V_i}
  \hat{\mathbf S}_v^{(1/2)},
  \qquad i=1,\ldots,4 .
  \label{eq:collective_spin_turan}
\end{equation}
Using the factorization of the sums over the independent subsets,
Eq.~(\ref{eq:turan_sum}) becomes
\begin{multline}
  \widetilde{\mathcal H}
  =
  \frac{n^4}{16}
  +
  \frac{n}{2\sqrt3}
  \sum_{\langle ijk\rangle}
  \hat{\mathbf S}_i\cdot
  \left(
    \hat{\mathbf S}_j\times\hat{\mathbf S}_k
  \right)
  \\
  -\frac{n^2}{12}
  \sum_{\langle ij\rangle}
  \hat{\mathbf S}_i\cdot\hat{\mathbf S}_j
  +
  \frac13
  \sum_{\langle ijkl\rangle}
  \left(
    \hat{\mathbf S}_i\cdot\hat{\mathbf S}_j
  \right)
  \left(
    \hat{\mathbf S}_k\cdot\hat{\mathbf S}_l
  \right).
  \label{eq:turan_collective}
\end{multline}
With $n=2S$, this is
\begin{multline}
  \widetilde{\mathcal H}
  =
  S^4
  +
  \frac{S}{\sqrt3}
  \sum_{\langle ijk\rangle}
  \hat{\mathbf S}_i\cdot
  \left(
    \hat{\mathbf S}_j\times\hat{\mathbf S}_k
  \right)
  \\
  -\frac{S^2}{3}
  \sum_{\langle ij\rangle}
  \hat{\mathbf S}_i\cdot\hat{\mathbf S}_j
  +
  \frac13
  \sum_{\langle ijkl\rangle}
  \left(
    \hat{\mathbf S}_i\cdot\hat{\mathbf S}_j
  \right)
  \left(
    \hat{\mathbf S}_k\cdot\hat{\mathbf S}_l
  \right).
  \label{eq:turan_collective_S}
\end{multline}
The six bilinear terms can be grouped according to the three
pairings of opposite bonds, so that
\begin{multline}
  S^4
  -\frac{S^2}{3}
  \sum_{\langle ij\rangle}
  \hat{\mathbf S}_i\cdot\hat{\mathbf S}_j
  +
  \frac13
  \sum_{\langle ijkl\rangle}
  \left(
    \hat{\mathbf S}_i\cdot\hat{\mathbf S}_j
  \right)
  \left(
    \hat{\mathbf S}_k\cdot\hat{\mathbf S}_l
  \right)
 \\ 
 =
  \frac13
  \sum_{\langle ijkl\rangle}
  \left(
    S^2-\hat{\mathbf S}_i\cdot\hat{\mathbf S}_j
  \right)
  \left(
    S^2-\hat{\mathbf S}_k\cdot\hat{\mathbf S}_l
  \right).
\end{multline}
Restricting each subset $V_i$ to its fully symmetric maximal-spin
sector,
\begin{equation}
  \hat{\mathbf S}_i^2=S(S+1),
\end{equation}
identifies $\hat{\mathbf S}_i$ with a physical spin-$S$ operator.
Equation~(\ref{eq:turan_collective_S}) then becomes precisely
Eq.~(\ref{eq:spin_S_hamiltonian}).  This is the usual
symmetrization construction underlying higher-spin AKLT states
\cite{Parameswaran2009}.

Positive semidefiniteness follows directly from the construction:
Eq.~(\ref{eq:turan_sum}) is a sum of positive-semidefinite
spin-$1/2$ parent terms, and its restriction to the maximal-spin
sector remains positive semidefinite.  The zero modes are inherited
just as directly.  A spin-$S$ coherent state can be represented as
the fully symmetric product of $2S$ identical spin-$1/2$ coherent
states,
\begin{equation}
  \ket{\mathbf n;S}
  \simeq
  \bigotimes_{v=1}^{2S}
  \ket{\mathbf n;\tfrac12}_v ,
\end{equation}
where the product already lies in the maximal-spin sector.  Hence a
spin-$S$ CCI product state corresponds on the decorated graph to
placing the same CCI direction on every spin-$1/2$ constituent of a
given subset $V_i$.  Every four-site clique in
Eq.~(\ref{eq:turan_sum}) then carries a spin-$1/2$ CCI configuration
and is annihilated by its local parent term.  The resulting spin-$S$
CCI coherent-product state is therefore an exact zero-energy state
of Eq.~(\ref{eq:spin_S_hamiltonian}).

\section{Geometry of the coherent zero-mode variety}
\label{app:coherent_geometry}

Here we collect the geometric results used in
Sec.~\ref{sec:mobius}: the classical-vector form of the coherent-state
energy, the exceptional locus of the M\"obius completion rule, and the
bounds on the tetrahedral chirality.

\subsection{Classical-vector form}
\label{app:classical}

For a product of spin-coherent states, the expectation value of
Eq.~(\ref{eq:spin_S_hamiltonian}) is
\begin{equation}
  \bra{\{\mathbf n_i\}}
  \mathcal H_{\mathrm{CCI}}^{(S)}
  \ket{\{\mathbf n_i\}}
  =
  S^4 H_{\rm CCI},
\end{equation}
where
\begin{multline}
  H_{\rm CCI}
  =
  \frac{1}{\sqrt3}
  \sum_{\langle ijk\rangle}
  \mathbf n_i\cdot
  (\mathbf n_j\times\mathbf n_k)
  \\
  +
  \frac13
  \sum_{\langle ijkl\rangle}
  (1-\mathbf n_i\cdot\mathbf n_j)
  (1-\mathbf n_k\cdot\mathbf n_l).
  \label{eq:HCCIclassical}
\end{multline}
Here $|\mathbf n_i|=1$. This is the Cartesian form of
$|B|^2/S^4$ and is therefore nonnegative. Its zero-energy
configurations are equivalently described by the stereographic
constraint Eq.~(\ref{eq:Bsceq0}).

Away from the three-coincident locus, the fourth spin can be written
directly in terms of the first three as
\begin{equation}
 \mathbf n_4=\frac{\mathbf B_{123}}{\|\mathbf B_{123}\|},
 \label{eq:nB123}
\end{equation}
where
\begin{multline}
 \mathbf B_{123}
 = (1-\mathbf n_2\cdot\mathbf n_3)\mathbf n_1
 +(1-\mathbf n_1\cdot\mathbf n_3)\mathbf n_2
 +(1-\mathbf n_1\cdot\mathbf n_2)\mathbf n_3
 \\
 +\sqrt3\bigl(
 \mathbf n_1\times\mathbf n_2
 +\mathbf n_2\times\mathbf n_3
 +\mathbf n_3\times\mathbf n_1\bigr).
 \label{eq:B123}
\end{multline}
Equation~(\ref{eq:nB123}) is the unit-vector form of the M\"obius
completion rule Eq.~(\ref{eq:Mobius}).  A generic local zero mode
therefore has six real parameters, the orientations of three spins.
After quotienting by three global spin rotations, the space of
inequivalent generic solutions is three dimensional.  For comparison,
the Heisenberg-tetrahedron constraint $\sum_i\mathbf n_i=0$ leaves a
two-dimensional manifold modulo global rotations
\cite{Moessner1998,Khatua2018}.

\subsection{Exceptional locus of the completion rule}
\label{app:specialloci}

The only configurations for which three spins do not uniquely determine
the fourth are those containing at least three coincident spin
directions. To see this, suppose for example that $z_1=z_2=z$. Then
\begin{equation}
  p_-(z,z,z_3,z_4)
  =
  \frac{1}{\sqrt6}(z-z_3)(z-z_4).
  \label{eq:B_pair_equal}
\end{equation}
The analogous factorizations for the other coincident pairs differ only
by site permutations and phase factors. Thus a zero mode containing a
coincident pair necessarily contains a third coincident spin, and the
exceptional branch is
\begin{equation}
  (z_1,z_2,z_3,z_4)=(z,z,z,w),
\end{equation}
together with its site permutations, with arbitrary
$w\in\hat{\mathbb C}$.

Equivalently, a matrix representative of the completion map
Eq.~(\ref{eq:Tmap}) is
\begin{equation}
  M(z_1,z_2)=
  \begin{pmatrix}
    -(\omega z_1+\omega^2z_2) & -z_1z_2\\
    1 & \omega z_2+\omega^2z_1
  \end{pmatrix},
\end{equation}
with $\det M=-(z_1-z_2)^2$. Hence the completion map is invertible for $z_1\ne z_2$. When
$z_1=z_2=z$, it gives $z_4=z$ unless also $z_3=z$, in which case
$z_4$ is unconstrained. 
A vanishing denominator in Eq.~(\ref{eq:Tmap}) corresponds to
$z_4=\infty$, and the M\"obius map extends holomorphically to this
point on the Riemann sphere.

\subsection{Equilateral representation and chirality sign}
\label{app:equilateral}

For a solution with four distinct spin directions, perform a global
rotation that sends $\mathbf n_3$ to the south pole, so that
$z_3=\infty$. Equation~(\ref{eq:Bsceq0}) then reduces to
\begin{equation}
  z_4+\omega z_1+\omega^2z_2=0,
  \label{eq:equilateral_constraint}
\end{equation}
whose general solution can be written as
\begin{equation}
  z_1=c+\rho,
  \quad
  z_2=c+\rho\omega,
  \quad
  z_4=c+\rho\omega^2,
 \label{eq:equilateral_param}
\end{equation}
where $c,\rho\in\mathbb C $.
Thus the three finite stereographic coordinates form an equilateral
triangle centered at $c$ with a fixed orientation.

Substituting Eq.~(\ref{eq:equilateral_param}) directly into
Eq.~(\ref{eq:tetrahedronchirality}) gives 
\begin{equation}
  \chi_{\mathrm{tet}}
  =
  -\frac{12\sqrt3\,|\rho|^2}
  {\bigl(1+|c+\rho|^2\bigr)
   \bigl(1+|c+\omega\rho|^2\bigr)
   \bigl(1+|c+\omega^2\rho|^2\bigr)} .
  \label{eq:chi_rho}
\end{equation}
The chirality therefore takes values in the interval
\begin{equation}
  -\frac{16}{3\sqrt3}
  \leq
  \chi_{\mathrm{tet}}
  \leq 0.
\end{equation}
$c=0$ and $|\rho|^2=1/2$ realizes the lower bound, which corresponds to the regular tetrahedral configuration, while $\chi_{\mathrm{tet}}=0$ is reached for $\rho=0$ on the three-coincident branch.

\subsection{Geometric meaning of the tetrahedral chirality}
\label{app:chirality_geometry}

The tetrahedral chirality has a simple geometric representation.
Equation~(\ref{eq:tetrahedronchirality}) can be written as
\begin{equation}
  \chi_{\mathrm{tet}}
  =
  -\det
  \begin{pmatrix}
    1 & \mathbf n_1^{\mathsf T} \\
    1 & \mathbf n_2^{\mathsf T} \\
    1 & \mathbf n_3^{\mathsf T} \\
    1 & \mathbf n_4^{\mathsf T}
  \end{pmatrix}.
  \label{eq:chi_homogeneous_det}
\end{equation}
Subtracting the first row from the remaining three gives
\begin{equation}
  \chi_{\mathrm{tet}}
  =
  -\det\!\left(
    \mathbf n_2-\mathbf n_1,
    \mathbf n_3-\mathbf n_1,
    \mathbf n_4-\mathbf n_1
  \right).
  \label{eq:chi_volume}
\end{equation}
Thus $\chi_{\mathrm{tet}}$ equals minus six times the oriented volume of
the tetrahedron whose vertices are the four spin directions
$\mathbf n_i$. Several properties follow immediately. The chirality
changes sign under an odd permutation of the four sites and is
invariant under an even permutation. It vanishes if and only if the
four spin directions are coplanar as points in spin space. Global
proper rotations leave $\chi_{\mathrm{tet}}$ invariant, whereas reversing
all spins changes its sign. Maximizing $|\chi_{\mathrm{tet}}|$ for four
unit vectors is therefore equivalent to maximizing the volume of a
tetrahedron whose vertices lie on the unit sphere. The maximum is
attained precisely by a regular tetrahedron, as proved below.

\subsection{Universal bound on the tetrahedral chirality}
\label{app:chirality_bounds}

The magnitude bound does not require the zero-mode condition
$p_-=0$; it holds for arbitrary four unit vectors. Define
\begin{equation}
 \bar{\mathbf n}=\frac14\sum_{i=1}^4\mathbf n_i,
 \qquad
 \mathbf u_i=\mathbf n_i-\bar{\mathbf n},
 \qquad
 Q=\sum_{i=1}^4\mathbf u_i\mathbf u_i^{\mathsf T}.
\end{equation}
Starting from Eq.~(\ref{eq:chi_homogeneous_det}), subtract
$\bar{\mathbf n}$ times the first column from the three
spin-coordinate columns. This leaves the determinant unchanged and
replaces each row $(1,\mathbf n_i^{\mathsf T})$ by
$(1,\mathbf u_i^{\mathsf T})$. Since
$\sum_i\mathbf u_i=0$, squaring the determinant and using the
corresponding Gram matrix gives
\begin{equation}
 \chi_{\mathrm{tet}}^2
 =
 \det
 \begin{pmatrix}
   4 & 0\\
   0 & Q
 \end{pmatrix}
 =
 4\det Q .
 \label{eq:chi_covariance}
\end{equation}
Moreover,
\begin{equation}
 \operatorname{tr}Q
 =
 \sum_{i=1}^4|\mathbf n_i-\bar{\mathbf n}|^2
 =
 4\bigl(1-|\bar{\mathbf n}|^2\bigr)
 \leq4.
\end{equation}
If $\lambda_{1,2,3}\geq0$ are the eigenvalues of the
positive-semidefinite matrix $Q$, the arithmetic--geometric mean
inequality gives
\begin{equation}
 \chi_{\mathrm{tet}}^2
 =
 4\lambda_1\lambda_2\lambda_3
 \leq
 4\left(
   \frac{\lambda_1+\lambda_2+\lambda_3}{3}
 \right)^3
 \leq
 \frac{256}{27},
\end{equation}
and hence
\begin{equation}
 |\chi_{\mathrm{tet}}|
 \leq
 \frac{16}{3\sqrt3}.
 \label{eq:chi_universal_bound}
\end{equation}

Equality requires saturation of both inequalities. The trace bound
is saturated only when $\bar{\mathbf n}=0$, while equality in the
arithmetic--geometric mean requires
$\lambda_1=\lambda_2=\lambda_3=4/3$. Thus
\begin{equation}
 Q=\frac43\,\mathbb I .
\end{equation}
To identify the corresponding spin configuration, introduce the
$3\times4$ matrix
\begin{equation}
 U=
 \left(
   \mathbf u_1\ 
   \mathbf u_2\ 
   \mathbf u_3\ 
   \mathbf u_4
 \right).
\end{equation}
Then $UU^{\mathsf T}=Q=(4/3)\mathbb I$, so
$U^{\mathsf T}U$ has nonzero eigenvalues
$\{4/3,4/3,4/3\}$. Moreover,
$U\mathbf 1=0$, and since $U$ has rank three,
$\mathbf 1$ spans the kernel of $U^{\mathsf T}U$. Therefore
\begin{equation}
 U^{\mathsf T}U
 =
 \frac43
 \left(
   \mathbb I_4-\frac14\mathbf 1\mathbf 1^{\mathsf T}
 \right).
\end{equation}
At equality $\bar{\mathbf n}=0$, so $\mathbf u_i=\mathbf n_i$.
The off-diagonal matrix elements therefore give
\begin{equation}
 \mathbf n_i\cdot\mathbf n_j=-\frac13,
 \qquad i\neq j.
\end{equation}
Thus equality is attained only by a regular tetrahedral frame, with
the sign of $\chi_{\mathrm{tet}}$ fixed by its orientation. Combining
Eq.~(\ref{eq:chi_universal_bound}) with Eq.~(\ref{eq:chi_rho}) proves
the window Eq.~(\ref{eq:chi_window}) and identifies the chiral
colorings as its unique lower-bound configurations, up to global
rotations and even site permutations.

\subsection{Apollonius circles and the six-cycle}
\label{app:apollonius}

For fixed $z_1\ne z_2$, introduce the projective coordinate
\begin{equation}
  \varsigma(z)=\frac{z-z_1}{z-z_2}.
\end{equation}
Equation~(\ref{eq:crossratio_id}) then takes the simple form
\begin{equation}
  \varsigma\bigl(T_{z_1,z_2}(z)\bigr)
  =
  -\omega\,\varsigma(z)
  =
  e^{-i\pi/3}\varsigma(z).
\end{equation}
Thus each application of the completion map rotates
$\varsigma$ by $60^\circ$ while preserving its modulus.  Consequently,
the six-cycle generated by a point $z$ lies on the Apollonius circle
\begin{equation}
  \frac{|z-z_1|}{|z-z_2|}
  =
  \text{const}.
  \label{eq:apollonius_circle}
\end{equation}
The complementary family
$\arg[(z-z_1)/(z-z_2)]=\text{const}$ consists of generalized circles
through $z_1$ and $z_2$; the completion map sends each member of this
family to the one whose projective angle differs by $-\pi/3$.
The two families form orthogonal pencils of generalized circles.

Since $T_{z_1,z_2}$ fixes $z_1$ and $z_2$, the update
\begin{equation}
  (z_1,z_2,z_3,z_4)
  \mapsto
  (z_1,z_2,z_4,T_{z_1,z_2}(z_4))
\end{equation}
is equivalently the simultaneous M\"obius action of $T_{z_1,z_2}$ on
the constrained quadruple.  For a generic fixed pair this transformation
is not a physical rotation of the Bloch sphere.  It preserves generalized
circles, but not spherical distances, and therefore neither the
individual face chiralities nor the magnitude of the tetrahedral
chirality are preserved along the orbit.  The sign of
$\chi_{\mathrm{tet}}$, however, remains fixed by
Eq.~(\ref{eq:chi_sign_main}).  For the six-cycle through the reference
coloring shown in Fig.~\ref{fig:Riemann_sphere}, one finds
\begin{equation}
  \chi_{\mathrm{tet}}
  =
  -\frac{16}{3\sqrt3}
  \left(
    1,\,
    \frac12,\,
    \frac16,\,
    \frac19,\,
    \frac16,\,
    \frac12
  \right).
  \label{eq:orbit_chi_values}
\end{equation}
Thus the M\"obius propagation preserves the handedness selected by the
constraint, but not the magnitude of the chirality.

A special simplification occurs when the two fixed spins are antipodal,
\begin{equation}
  z_2=-\frac{1}{z_1^*}.
\end{equation}
In this case $T_{z_1,z_2}\in PSU(2)$ and the M\"obius transformation
becomes a physical rotation of the Bloch sphere.  Rotating the fixed
pair to the poles, $z_1=0$ and $z_2=\infty$, gives
\begin{equation}
  z_4
  =
  T_{0,\infty}(z_3)
  =
  -\omega z_3
  =
  e^{-i\pi/3}z_3,
  \label{eq:antipodal_rotation_main}
\end{equation}
which is a rotation through $-\pi/3$ about the axis defined by the
fixed antipodal spins.  Thus the generic six-cycle is a genuinely
M\"obius orbit, becoming an ordinary rigid sixfold rotation only in
the antipodal case.

\section{Counting the zero-energy states}
\label{app:surjectivity}

\subsection{States annihilated by $\mathcal B$}

A convenient way to characterize the states annihilated by
$\mathcal B$ is to use the polynomial representation of spin states
\cite{Barnett2006}. At each site we identify
\begin{equation}
  (a_i^\dagger)^{2S-n}(b_i^\dagger)^n\ket{0}
  \quad\longleftrightarrow\quad
  z_i^n ,
  \qquad
  0\leq n\leq2S .
  \label{eq:spin_polynomial}
\end{equation}
A four-site spin-$S$ state is therefore represented by a polynomial
$f(z_1,z_2,z_3,z_4)$ whose degree in each variable is at most $2S$.

Consider now
\begin{equation}
  \mathcal B^\dagger:
  \mathscr H^{S-\frac12}
  \longrightarrow
  \mathscr H^S .
\end{equation}
A state in $\mathscr H^{S-\frac12}$ has degree at most $2S-1$ in
each variable. 
In the polynomial representation,
$\mathcal B^\dagger$ acts by multiplication with the polynomial
$p_+$ defined in Eq.~(\ref{eq:p_p}),
\begin{equation}
  \mathcal B^\dagger f=p_+f .
\end{equation}
Since the polynomial ring has no zero divisors and $p_+$ is not the
zero polynomial, $p_+f=0$ is possible only for $f=0$. Hence
multiplication by $p_+$, and with it $\mathcal B^\dagger$, is
injective on the full space $\mathscr H^{S-\frac12}$.

It follows that $\mathcal B$ is surjective onto
$\mathscr H^{S-\frac12}$: in finite dimensions $\mathcal B$ and
$\mathcal B^\dagger$ have equal rank, and injectivity gives
$\operatorname{rank}\mathcal B^\dagger=\dim\mathscr H^{S-\frac12}$.
Surjectivity holds separately in every total-spin sector as well.
Let $\mathcal P_{S_{\mathrm{tet}}}$ denote the projector onto the
sector with total spin $S_{\mathrm{tet}}$, which commutes with the
SU(2) scalar $\mathcal B$. Given $|w\rangle$ in that sector of
$\mathscr H^{S-\frac12}$, full-space surjectivity provides
$|v\rangle$ with $\mathcal B|v\rangle=|w\rangle$, and the projected
state satisfies
$\mathcal B\,\mathcal P_{S_{\mathrm{tet}}}|v\rangle
=\mathcal P_{S_{\mathrm{tet}}}\mathcal B|v\rangle=|w\rangle$. This
sector-resolved surjectivity is what enters the dimension count and
proves Eq.~(\ref{eq:dim_recursion}).

The same argument applies to $\mathcal B^{*}$. Applying
first $\mathcal B$ and then $\mathcal B^{*}$ therefore shows that
\begin{equation}
  \mathcal C=\mathcal B^{*}\mathcal B:
  \mathscr H^S
  \longrightarrow
  \mathscr H^{S-1}
\end{equation}
reaches the whole target space for $S\geq1$. This gives
Eq.~(\ref{eq:dim_recursion_C}).

\subsection{Why there are no additional zero-energy states}
\label{app:kernel_span}

We now prove that the states annihilated by $\mathcal B$ are exactly
those in $\mathscr V^S$. Equivalently, there are no additional states
in the local zero-energy kernel beyond the linear span of the globally
rotated CCI states.

Let $|f\rangle$ be orthogonal to $\mathscr V^S$, and denote its
polynomial representation by $f(z_1,z_2,z_3,z_4)$. With the
normalization of Eq.~(\ref{eq:spin_polynomial}), the overlap with a
product of four unnormalized coherent states is
\begin{equation}
  \langle w_1w_2w_3w_4|f\rangle
  =
  \bigl[(2S)!\bigr]^4
  f(\bar w_1,\bar w_2,\bar w_3,\bar w_4).
  \label{eq:coherent_polynomial_overlap}
\end{equation}
Thus $|f\rangle\perp\mathscr V^S$ implies that the polynomial $f$
vanishes on the complex conjugates of all globally rotated CCI
configurations.

The space $\mathscr V^S$ is a complex vector space invariant under
global SU(2) rotations. Invariance under the su(2) generators
therefore extends to their complex linear span $sl(2,\mathbb C)$,
and hence to the connected group SL(2,$\mathbb C$)
\cite{HallLieGroups}. The SL(2,$\mathbb C$) orbit of the reference
state consequently remains inside $\mathscr V^S$. In stereographic
coordinates this complexified action is the usual
PSL(2,$\mathbb C$) M\"obius action. Using
Eq.~(\ref{eq:coherent_polynomial_overlap}), and noting that complex
conjugation maps SL(2,$\mathbb C$) onto itself, we conclude that $f$
vanishes on the full PSL(2,$\mathbb C$) orbit of the conjugated
reference coloring.

To identify this orbit, consider first the polynomial $p_+$ associated
with $\mathcal B^\dagger$. Up to its normalization, it is the
quadratic form
\begin{equation}
  p_+
  =
  \frac{1}{2\sqrt6}
  \bm z^{\mathsf T}C_+\bm z,
  \qquad
  \bm z=(z_1,z_2,z_3,z_4)^{\mathsf T},
\end{equation}
with
\begin{equation}
  C_+
  =
  \begin{pmatrix}
    0&1&\omega^2&\omega\\
    1&0&\omega&\omega^2\\
    \omega^2&\omega&0&1\\
    \omega&\omega^2&1&0
  \end{pmatrix}.
  \label{eq:Cplus}
\end{equation}
Because $1+\omega+\omega^2=0$, every row of $C_+$ sums to zero, so
\begin{equation}
  C_+(1,1,1,1)^{\mathsf T}=0 .
\end{equation}
On the other hand,
\begin{equation}
  \det
  \begin{pmatrix}
    0&1&\omega^2\\
    1&0&\omega\\
    \omega^2&\omega&0
  \end{pmatrix}
  =2 ,
\end{equation}
and therefore
\begin{equation}
  \operatorname{rank}C_+=3 .
\end{equation}
If a homogeneous quadratic polynomial factors into two linear forms,
its quadratic-form matrix has rank at most two. Hence $p_+$ is
irreducible over $\mathbb C$.

The conjugated reference coloring lies on the hypersurface $p_+=0$.
For four distinct coordinates this condition has a simple projective
interpretation. Direct expansion gives
\begin{multline}
  (z_1-z_3)(z_2-z_4)
  -
  e^{-i\pi/3}(z_1-z_4)(z_2-z_3)
  \\
  =
  \sqrt6\,e^{i\pi/3}
  p_+(z_1,z_2,z_3,z_4),
  \label{eq:pplus_crossratio}
\end{multline}
so that $p_+=0$ is equivalent, for four distinct points, to the
equianharmonic cross-ratio condition
\begin{equation}
  \frac{(z_1-z_3)(z_2-z_4)}
       {(z_1-z_4)(z_2-z_3)}
  =
  e^{-i\pi/3}.
  \label{eq:equianharmonic_cr}
\end{equation}
For $p_-$ the corresponding value is the complex conjugate
$e^{+i\pi/3}$.

A M\"obius transformation is uniquely determined by the images of
three distinct points and preserves their cross ratio. Consequently,
every ordered quadruple of distinct points satisfying
Eq.~(\ref{eq:equianharmonic_cr}) is obtained from the conjugated
reference coloring by a global PSL(2,$\mathbb C$) transformation.
The complexified CCI orbit therefore coincides with the
distinct-point part of the hypersurface
\begin{equation}
  V(p_+)
  =
  \left\{
    (z_1,z_2,z_3,z_4)\in\mathbb C^4:
    p_+(z_1,z_2,z_3,z_4)=0
  \right\}.
  \label{eq:Vpplus}
\end{equation}

Since $p_+$ is irreducible, $V(p_+)$ is an irreducible hypersurface.
Its distinct-point locus is nonempty and Zariski open, and is
therefore dense in $V(p_+)$. Since $f$ vanishes on the complexified
CCI orbit, it follows that $f$ vanishes on a dense subset of
$V(p_+)$ and hence on the entire hypersurface. We now use the
standard algebraic-geometric fact that, for an irreducible polynomial
over $\mathbb C$, every polynomial vanishing identically on its zero
set is divisible by that polynomial \cite{CoxLittleOShea}. Therefore
\begin{equation}
  f=p_+q .
  \label{eq:f_pplus_q}
\end{equation}

Since $p_+$ has degree exactly one in every coordinate and the
polynomial ring has no zero divisors,
\begin{equation}
  \deg_{z_i}q
  =
  \deg_{z_i}f-1
  \leq2S-1
\end{equation}
whenever $q\neq0$. Thus $q$ represents a state
$|q\rangle\in\mathscr H^{S-\frac12}$. Moreover, as shown above,
multiplication by $p_+$ is precisely the polynomial representation
of $\mathcal B^\dagger$, and therefore
\begin{equation}
  |f\rangle
  =
  \mathcal B^\dagger|q\rangle .
  \label{eq:Vperp_imBdagger}
\end{equation}
Thus every state orthogonal to $\mathscr V^S$ belongs to the image
of $\mathcal B^\dagger$.

Now let $|\psi\rangle$ satisfy
\begin{equation}
  \mathcal B|\psi\rangle=0 .
\end{equation}
For every $|f\rangle\perp\mathscr V^S$,
Eq.~(\ref{eq:Vperp_imBdagger}) gives
\begin{equation}
  \langle f|\psi\rangle
  =
  \langle q|\mathcal B|\psi\rangle
  =
  0.
\end{equation}
Hence $|\psi\rangle$ is orthogonal to the entire orthogonal complement
of $\mathscr V^S$ and must itself belong to $\mathscr V^S$. Together
with the inclusion
$\mathscr V^S\subseteq\ker\mathcal B$ established in
Eq.~(\ref{eq:V_in_kerB}), this proves
\begin{equation}
  \mathscr V^S=\ker\mathcal B .
  \label{eq:V_equals_kerB_appendix}
\end{equation}

The same reasoning applies to the full color-ice space
$\mathscr W^S$. If $|f\rangle\perp\mathscr W^S$, its polynomial
vanishes on the complexified orbits of both chiralities. It is
therefore divisible by both $p_+$ and $p_-$. The two polynomials are
irreducible by the same rank argument and are not proportional, so
they are relatively prime. Hence
\begin{equation}
  f=p_+p_-q .
\end{equation}
For $S\geq1$, the polynomial $q$ has degree at most $2S-2$ in each
coordinate and therefore represents a state
$|q\rangle\in\mathscr H^{S-1}$. Since multiplication by $p_+p_-$ is
the polynomial representation of $\mathcal C^\dagger$, we obtain
\begin{equation}
  |f\rangle
  =
  \mathcal C^\dagger|q\rangle .
\end{equation}
Repeating the orthogonality argument then gives
\begin{equation}
  \mathscr W^S=\ker\mathcal C.
\end{equation}

\section{Exact gap of the local parent Hamiltonian}
\label{app:local_gap}

Here we prove Eq.~(\ref{eq:local_gap}) for the exact gap of a single
tetrahedral parent term. As discussed in
Sec.~\ref{sec:spectral_pairing},
$\mathcal B^\dagger\mathcal B$ on $\mathscr H^S$ and
$\mathcal B\mathcal B^\dagger$ on
$\mathscr H^{S-\frac12}$ have identical nonzero spectra.
It is therefore sufficient to determine the lowest eigenvalue of
$\mathcal B\mathcal B^\dagger$ on
$\mathscr H^{S-\frac12}$.

On this Hilbert space the Schwinger-boson number at every site is fixed,
\begin{equation}
  N_i
  =
  a_i^\dagger a_i+b_i^\dagger b_i
  =
  2S-1
  \equiv m .
  \label{eq:m_def}
\end{equation}
Since $\mathcal B$ contains one annihilation operator from each site,
it is linear in the boson operators at any given site. We define
\begin{equation}
  D_{i,a}=[\mathcal B,a_i^\dagger],
  \qquad
  D_{i,b}=[\mathcal B,b_i^\dagger].
  \label{eq:DiaDib}
\end{equation}

Normal ordering $\mathcal B\mathcal B^\dagger$ gives contributions
with zero through four contractions. Using
$a_i a_i^\dagger=1+a_i^\dagger a_i$ and
$b_i b_i^\dagger=1+b_i^\dagger b_i$, one obtains
\begin{align}
  \mathcal B\mathcal B^\dagger
  ={}&
  \mathcal B^\dagger\mathcal B
  +
  \sum_i
  \left(
    D_{i,a}^\dagger D_{i,a}
    +
    D_{i,b}^\dagger D_{i,b}
  \right)
  \nonumber\\
  &+
  \frac16\sum_{i<j}
  \left(
    N_iN_j+F_{ij}^\dagger F_{ij}
  \right)
  +
  \frac12\sum_i N_i
  +
  \mathbb I .
  \label{eq:BBdag_contractions}
\end{align}
Equation~(\ref{eq:BBdag_contractions}) is an operator identity in
the two-boson algebra of the four sites, obtained by normal ordering
all contractions of $\mathcal B\mathcal B^\dagger$. We have
additionally verified it exactly by evaluating all Fock-space matrix
elements with occupations up to three bosons per site; this
evaluation is exhaustive because every term in the identity is of
degree at most two in the mode operators of each site.
Restricting this identity to $\mathscr H^{S-\frac12}$, where
$N_i=m$, gives
\begin{align}
  \mathcal B\mathcal B^\dagger
  ={}&
  (m+1)^2\,\mathbb I
  +
  \mathcal B^\dagger\mathcal B
  +
  \frac16\sum_{i<j}F_{ij}^\dagger F_{ij}
  \nonumber\\
  &+
  \sum_i
  \left(
    D_{i,a}^\dagger D_{i,a}
    +
    D_{i,b}^\dagger D_{i,b}
  \right).
  \label{eq:BBdag_normal}
\end{align}
All terms following the first are positive semidefinite. Hence
\begin{equation}
  \mathcal B\mathcal B^\dagger
  \succeq
  (m+1)^2\,\mathbb I
  =
  4S^2\,\mathbb I .
  \label{eq:BBdag_bound}
\end{equation}

We now show that the bound is saturated. Let
\begin{equation}
  \ket{F_m}
  =
  \prod_{i=1}^4
  \frac{(a_i^\dagger)^m}{\sqrt{m!}}\ket{0}
\end{equation}
be the fully polarized state in $\mathscr H^{S-\frac12}$. From
Eqs.~(\ref{eq:Fij}) and (\ref{eq:B_in_F}), each of
$F_{ij}$, $\mathcal B$, $D_{i,a}$, and $D_{i,b}$ contains at least
one $b$-boson annihilation operator. Therefore
\begin{equation}
  F_{ij}\ket{F_m}
  =
  \mathcal B\ket{F_m}
  =
  D_{i,a}\ket{F_m}
  =
  D_{i,b}\ket{F_m}
  =
  0 .
\end{equation}
Equation~(\ref{eq:BBdag_normal}) then yields
\begin{equation}
  \mathcal B\mathcal B^\dagger\ket{F_m}
  =
  (m+1)^2\ket{F_m}
  =
  4S^2\ket{F_m}.
\end{equation}
Together with Eq.~(\ref{eq:BBdag_bound}), this proves
\begin{equation}
  \lambda_{\min}
  \left(
    \mathcal B\mathcal B^\dagger
    \big|_{\mathscr H^{S-\frac12}}
  \right)
  =
  4S^2 .
  \label{eq:BBdag_min}
\end{equation}

The state $\ket{F_m}$ is the highest-weight state of the unique
maximal-total-spin multiplet in $\mathscr H^{S-\frac12}$, with
$S_{\mathrm{tet}}=4S-2$. Since $\mathcal B\mathcal B^\dagger$ is an
SU(2) scalar,
the entire multiplet has eigenvalue $4S^2$. Together with the spectral
pairing between $\mathcal B\mathcal B^\dagger$ and
$\mathcal B^\dagger\mathcal B$, this establishes the exact local gap
\begin{equation}
  \Delta_{\rm loc}(S)=4S^2,
\end{equation}
and proves Eq.~(\ref{eq:local_gap}).

\section{Loop zero modes: reduction and counting}
\label{app:loops}

Here we collect the technical details of the loop analysis: the
polarized-loop reduction and transfer-matrix count quoted in
Sec.~\ref{sec:loops_uniform}, together with the geometric and
quantum-counting arguments for four-coloring backgrounds used in
Sec.~\ref{sec:coloring_loops}.

\subsection{Uniform-color backgrounds.}

Consider a loop of length $L$ whose sites are shared by $L$
tetrahedra, each containing two consecutive loop sites and two
environment sites polarized along the quantization axis.  Each term
of Eq.~(\ref{eq:Bop}) contains a pair of $b$ annihilation operators.
Every term with a $b$ operator on an environment site vanishes, and
the single surviving contribution has both $b$ operators on the two
loop sites.  Up to a nonzero scalar factor and an irrelevant phase,
the local zero-mode condition is therefore Eq.~(\ref{eq:bibj}).

Grouping the $2S+1$ states on each loop site into the fully polarized
state $U$ and the $2S$ deviated states $\bar U$, the allowed cyclic
sequences are counted by
\begin{equation}
  N_L(S)=\operatorname{Tr}T^L
  =\lambda_+^L+\lambda_-^L ,
    \label{eq:NLS}
\end{equation}
where
\begin{equation}
  \lambda_\pm=\frac{1\pm\sqrt{1+8S}}{2}.
\end{equation}
are the eigenvaues of the 
\begin{equation}
  T=
  \begin{pmatrix}
    1 & 2S\\
    1 & 0
  \end{pmatrix}
\end{equation}
transfer matrix.
Expanding this expression for $L=4$ and $6$ gives
Eq.~(\ref{eq:N4N6}).  For $S=\tfrac12$ the same count is the
monomer--dimer partition function of the cycle graph,
$N_4=1+4+2=7$ and $N_6=1+6+9+2=18$.

\subsection{Four-coloring backgrounds}

For the four-sublattice pyrochlore coloring, an elementary hexagon
contains three distinct colors, each appearing twice. Up to relabeling,
cyclic permutation, and reversal, its color sequence is therefore
$XYZXYZ$. To determine its holonomy explicitly, we label the three
colors on the loop as $A,B,C$, and denote by $D$ the fourth color,
which is absent from the loop. The sequence may then be written as
\begin{equation}
  ABCABC .
\end{equation}

\subsubsection{Identity holonomy for the pyrochlore hexagon}
\label{app:identity_holonomy}

The first three completion maps around the hexagon are
\begin{equation}
  T_{C,D},\qquad
  T_{A,D},\qquad
  T_{B,D},
\end{equation}
and on the reference coloring they propagate
\begin{equation}
  A
  \xrightarrow{T_{C,D}}
  B
  \xrightarrow{T_{A,D}}
  C
  \xrightarrow{T_{B,D}}
  A .
\end{equation}
Their product
\begin{equation}
  P
  =
  T_{B,D}\circ T_{A,D}\circ T_{C,D}
\end{equation}
therefore fixes $A$. Moreover, $D$ is a fixed point of each of the
three completion maps, and hence
\begin{equation}
  P(A)=A,
  \qquad
  P(D)=D .
\end{equation}

Using Eq.~(\ref{eq:crossratio_id}), the completion rule may
equivalently be written as
\begin{equation}
  \frac{T_{z_1,z_2}(z)-z_2}
       {T_{z_1,z_2}(z)-z_1}
  =
  (-\omega)^{-1}
  \frac{z-z_2}{z-z_1}.
  \label{eq:crossratio_id_inverse}
\end{equation}
Thus each of the three maps has multiplier $(-\omega)^{-1}$ at their
common fixed point $D$. The multiplier of their product is therefore
\begin{equation}
  P'(D)
  =
  (-\omega)^{-3}
  =
  -1 .
\end{equation}
A M\"obius transformation with two distinct fixed points and
multiplier $-1$ is an involution, so
\begin{equation}
  P^2=\mathrm{id}.
\end{equation}
The second half of the hexagon contains the same three completion
maps. Consequently, the full six-step holonomy is
\begin{equation}
  \mathcal M_6
  =
  P^2
  =
  \mathrm{id},
\end{equation}
which proves Eq.~(\ref{eq:pyrochlore_hexagon_identity}).

This argument depends only on the color sequence and on the local
completion rule. In particular, the four color directions need not
form the regular tetrahedral frame. Any four distinct directions
$z_A,z_B,z_C,z_D$ satisfying
$p_-(z_A,z_B,z_C,z_D)=0$ give the same identity holonomy, provided
the corresponding local completion maps are nondegenerate.

\subsection{Quantum kernel of an identity-holonomy loop}
\label{app:identity_holonomy_kernel}

We now show that identity holonomy has a stronger quantum
consequence. Consider an embedded loop of length $L$, with the
environment spins fixed in coherent states, and assume that every
local completion rule is a well-defined invertible M\"obius
transformation.

For a tetrahedron containing two neighboring loop sites $i$ and
$i+1$, fixing the other two spins reduces the four-site annihilation
operator $\mathcal B$ to a bilinear annihilation constraint on the
two loop spins. At the coherent-state level, its zero-energy
condition is precisely
\begin{equation}
  z_{i+1}=T_i(z_i).
  \label{eq:app_local_completion}
\end{equation}

The key observation is that these local M\"obius twists can be
removed by a site-dependent choice of stereographic coordinates.
Introduce
\begin{equation}
  z_i'=R_i(z_i),
\end{equation}
with $R_1=\mathrm{id}$, and choose the remaining transformations
recursively as
\begin{equation}
  R_{i+1}=R_i\circ T_i^{-1}.
  \label{eq:app_untwist_recursion}
\end{equation}
Equation~(\ref{eq:app_local_completion}) then becomes simply
\begin{equation}
  z_{i+1}'=z_i'.
  \label{eq:app_equal_z}
\end{equation}

For an open chain this change of coordinates can always be carried
out. Around a closed loop, after one circuit the transported
coordinate system is
\begin{equation}
  R_{L+1}=\mathcal M_L^{-1}.
\end{equation}
It therefore returns to the original coordinate system precisely
when
\begin{equation}
  \mathcal M_L=\mathrm{id}.
\end{equation}
Identity holonomy thus means that all local M\"obius twists can be
removed simultaneously.

Each M\"obius change of coordinate can be represented by an
invertible linear change of the two Schwinger-boson components at the
corresponding site. Under these changes the reduced two-site
annihilation constraint remains bilinear, while its coherent-state
zero condition becomes $z_i'=z_{i+1}'$.

A general bilinear constraint produces, on a pair of coherent states,
a factor of the form
\begin{equation}
  c_0+c_1z_i'+c_2z_{i+1}'+c_3z_i'z_{i+1}'.
\end{equation}
Requiring this expression to vanish for every
$z_i'=z_{i+1}'=z$ gives
$c_0=c_3=0$ and $c_2=-c_1$. Hence, up to a nonzero overall factor,
the only such bilinear constraint is proportional to
$z_{i+1}'-z_i'$. This is precisely the coherent-state zero condition
of the singlet pair-annihilation operator $F_{i,i+1}$ introduced in
Eq.~(\ref{eq:Fij}). Thus the loop zero-energy conditions reduce to
\begin{equation}
  F_{i,i+1}\ket{\psi}=0,
  \qquad
  i=1,\ldots,L,
  \qquad
  L+1\equiv1.
  \label{eq:app_F_constraints}
\end{equation}

The local transformations used above are generally nonunitary, so
this change of variables is not a physical site-dependent spin
rotation. It is an invertible change of basis in the local spin
Hilbert spaces and therefore gives a one-to-one correspondence
between the original loop kernel and the common kernel of
Eq.~(\ref{eq:app_F_constraints}).

The latter has a simple physical interpretation. On two spin-$S$
sites,
\begin{equation}
  F_{ij}^{\dagger}F_{ij}
  =
  2\left(
    S^2-\hat{\mathbf S}_i\cdot\hat{\mathbf S}_j
  \right),
\end{equation}
and in the pair-spin-$J$ sector its eigenvalue is
\begin{equation}
  \lambda_J
  =
  2S(2S+1)-J(J+1).
  \label{eq:app_F_spectrum}
\end{equation}
This vanishes only for $J=2S$. Thus every neighboring pair on the
untwisted loop must lie in its maximal-spin sector.

To determine the simultaneous kernel, one may regard each spin $S$
as $2S$ symmetrized spin-$1/2$ constituents. Maximal spin on the
bond $(i,i+1)$ requires the constituents belonging to these two sites
to be fully symmetric. Since neighboring bonds share sites and the
loop is connected, these conditions propagate around the entire
loop. All $2SL$ spin-$1/2$ constituents are therefore fully
symmetric, so the untwisted kernel is the single maximal-spin
multiplet
\begin{equation}
  J_{\rm tot}=LS.
\end{equation}
Its dimension is $2SL+1$. Since the change of basis is invertible,
the original embedded-loop kernel has the same dimension,
\begin{equation}
  \dim\mathscr K_{\rm loop}=2SL+1.
  \label{eq:app_loop_kernel_exact}
\end{equation}

It remains to identify the coherent states within this kernel. In the
untwisted variables all sites have the same stereographic coordinate,
so the propagated family becomes
\begin{equation}
  \ket{\Omega(z)}
  =
  \bigotimes_{i=1}^{L}\ket{z;S}_i .
  \label{eq:app_uniform_coherent}
\end{equation}
These are precisely the coherent states of the maximal-spin
$J_{\rm tot}=LS$ multiplet: they are obtained by applying the same
global spin rotation to every site. Since the spin-$LS$
representation is irreducible, the linear span of this
coherent-state family is the entire multiplet.

Transforming back to the original coordinates gives
$z_i=R_i^{-1}(z)$. Equation~(\ref{eq:app_untwist_recursion}) then
implies
\begin{equation}
  z_{i+1}=T_i(z_i),
\end{equation}
so these are precisely the coherent states obtained by propagating
an arbitrary initial spin direction around the original loop.
Consequently,
\begin{equation}
  \mathscr K_{\rm loop}
  =
  \operatorname{span}
  \left\{
    \ket{\widetilde\psi(z)}
    \,\middle|\,
    z\in\hat{\mathbb C}
  \right\},
  \qquad
  \dim\mathscr K_{\rm loop}=2SL+1.
  \label{eq:identity_holonomy_theorem}
\end{equation}

\section{Linearized coherent-state constraints and harmonic zero modes}
\label{app:rigidity}

We now characterize the infinitesimal coherent deformations of the
four-sublattice zero-energy state. The analysis is analogous to the
Maxwell counting of local constraints in classical frustrated magnets
\cite{Moessner1998} and to the constraint-matrix formulation of
classical spin liquids \cite{Yan2024formalism,Davier2023}. In momentum
space it is also closely related to the analysis of zero-energy
manifolds in a Luttinger--Tisza spectrum. The important distinction is
that here $\bm k$ labels fluctuations about a fixed four-sublattice
ground state, rather than candidate ordering wavevectors of the
original spins.

For a coherent product state the energy can be written as a sum of
positive local contributions,
\begin{equation}
  E
  =
  \sum_t W_t(\{z_i,\bar z_i\})\,|p_t(\{z_i\})|^2,
  \label{eq:coherent_energy_constraints}
\end{equation}
where $W_t>0$ and every coherent zero-energy state satisfies
$p_t=0$ on each tetrahedron $t$. Consider a small deformation of the
four-sublattice solution,
\begin{equation}
  z_i=z_i^{(0)}+\delta z_i .
\end{equation}
Since $p_t$ is holomorphic in the stereographic coordinates,
\begin{equation}
  p_t
  =
  \sum_i
  \mathsf R_{ti}\,\delta z_i
  +O(\delta z^2),
  \qquad
  \mathsf R_{ti}
  =
  \left.
  \frac{\partial p_t}{\partial z_i}
  \right|_{\{z^{(0)}\}},
  \label{eq:linear_constraint_matrix}
\end{equation}
where $\mathsf R$ is the linearized constraint matrix. Because
$p_t=0$ on the reference state, variations of the positive factors
$W_t$ do not contribute at quadratic order. The harmonic change in
the coherent-state energy is therefore
\begin{equation}
  E^{(2)}
  =
  \delta\bm z^\dagger
  \mathsf R^\dagger
  W^{(0)}
  \mathsf R
  \delta\bm z ,
  \label{eq:harmonic_R}
\end{equation}
where $W^{(0)}$ is diagonal in the local constraints and has positive
entries. Hence
\begin{equation}
  E^{(2)}=0
  \qquad\Longleftrightarrow\qquad
  \mathsf R\,\delta\bm z=0 .
  \label{eq:harmonic_kernel}
\end{equation}
The kernel of $\mathsf R$ therefore gives the infinitesimal coherent
zero modes of the four-sublattice state.

For a periodic background, $\mathsf R$ separates into Bloch matrices
$\mathsf R(\bm k)$. The dimension
\begin{equation}
  \nu(\bm k)
  =
  \dim\ker\mathsf R(\bm k)
\end{equation}
counts the independent complex zero modes at momentum $\bm k$. The
geometry of the set on which $\nu(\bm k)>0$ provides a useful
classification of the coherent flexibility of the ordered state:
a zero mode throughout the Brillouin zone gives an extensive number
of infinitesimal deformations, zero modes restricted to lines give a
subextensive number proportional to the linear system size, while
zero modes only at isolated momenta give an $O(1)$ tangent space.

This reciprocal-space picture resembles the codimension
classification of degenerate minima in Luttinger--Tisza analyses,
but its meaning is different. Here the zero set describes soft
deformations around one particular coherent ground-state component.
Moreover, a zero mode of $\mathsf R$ is only guaranteed to satisfy
the local constraints to first order; it need not extend to a finite
continuous family of exact zero modes. This distinction will be
important for the checkerboard lattice.

\subsection{Linearizing the local constraint}

Consider first a single tetrahedron in the four-sublattice state, with
the sites ordered according to their colors $A,B,C,D$.  Its coherent
zero-energy condition is
\begin{equation}
  p_-(z_A,z_B,z_C,z_D)=0,
\end{equation}
with $p_-$ given in Eq.~(\ref{eq:p_m}).  We now make an infinitesimal
deformation
\begin{equation}
  z_X\longrightarrow z^X+\delta z_X,
  \qquad
  X=A,B,C,D,
\end{equation}
about the tetrahedral solution of
Eq.~(\ref{eq:z_color_directions}).  To first order,
\begin{equation}
  \delta p_-
  =
  g_A\delta z_A
  +g_B\delta z_B
  +g_C\delta z_C
  +g_D\delta z_D,
  \label{eq:local_linearized_constraint}
\end{equation}
where
\begin{align}
  g_A
  &=
  \left.
  \frac{\partial p_-}{\partial z_A}
  \right|_{\rm tet}
  =
  \frac{1}{\sqrt6}
  \left(
    z^B+\omega z^C+\omega^2z^D
  \right),
  \nonumber\\
  g_B
  &=
  \frac{1}{\sqrt6}
  \left(
    z^A+\omega z^D+\omega^2z^C
  \right),
  \nonumber\\
  g_C
  &=
  \frac{1}{\sqrt6}
  \left(
    z^D+\omega z^A+\omega^2z^B
  \right),
  \nonumber\\
  g_D
  &=
  \frac{1}{\sqrt6}
  \left(
    z^C+\omega z^B+\omega^2z^A
  \right).
\end{align}
Using $z^B=-z^A$ and $z^D=-z^C$, these coefficients satisfy
\begin{equation}
  g_B=-g_A,
  \qquad
  g_D=-g_C,
  \qquad
  \frac{g_C}{g_A}
  =
  i(2-\sqrt3).
  \label{eq:rigidity_gradients}
\end{equation}
It is therefore convenient to define
\begin{equation}
  \gamma
  \equiv
  \frac{g_C}{g_A}
  =
  i(2-\sqrt3),
\end{equation}
and divide Eq.~(\ref{eq:local_linearized_constraint}) by $g_A$.
The linearized constraint on every tetrahedron then takes the simple
form
\begin{equation}
  \delta z_A-\delta z_B
  +
  \gamma
  \left(
    \delta z_C-\delta z_D
  \right)
  =
  0.
  \label{eq:local_linearized_simple}
\end{equation}

Equation~(\ref{eq:local_linearized_simple}) is the basic equation used
below.  The lattice enters only through the way in which the four
colors belonging to different tetrahedra are shared between magnetic
unit cells.

For the full lattice, collect all infinitesimal deformations into a
vector $\delta\bm z$.  The linearized constraints can then be written
as
\begin{equation}
  \mathsf R\,\delta\bm z=0,
  \label{eq:R_real_space}
\end{equation}
where $\mathsf R$ has one row for each local tetrahedral constraint
and one column for each spin.  Thus
\begin{equation}
  \ker\mathsf R
\end{equation}
is precisely the space of coherent deformations that satisfy all
local constraints to first order.

This is only a linear statement.  If
$\mathsf R\,\delta\bm z=0$, then
\begin{equation}
  p_-
  \left(
    z^{(0)}+\epsilon\delta z
  \right)
  =
  O(\epsilon^2),
\end{equation}
but the infinitesimal deformation need not extend to an exact
finite-amplitude family.  The checkerboard lattice below provides an
important example of such an obstruction.

\subsection{Bloch form}

Because the four-sublattice background is periodic, the linearized
problem separates into momentum sectors.  Let $\alpha$ label the
sites of the magnetic unit cell and write
\begin{equation}
  \delta z_\alpha(\bm R)
  =
  u_\alpha(\bm k)
  e^{i\bm k\cdot\bm R}.
  \label{eq:Bloch_deformation}
\end{equation}
Equation~(\ref{eq:R_real_space}) then becomes
\begin{equation}
  \mathsf R(\bm k)\,
  \bm u(\bm k)
  =
  0.
  \label{eq:Rk_equation}
\end{equation}
If a magnetic cell contains $n_s$ sites, the number of independent
complex infinitesimal deformations at momentum $\bm k$ is
\begin{equation}
  \nu(\bm k)
  =
  n_s-\operatorname{rank}\mathsf R(\bm k).
  \label{eq:nullity_k}
\end{equation}
For a finite periodic system the total number of complex tangent
directions is therefore
\begin{equation}
  N_{\rm lin}
  =
  \sum_{\bm k}\nu(\bm k).
  \label{eq:Nlin_sum}
\end{equation}

The Bloch matrices are not unique: changing the origin of a site or
a local constraint multiplies a column or row by a nonzero Bloch
phase, while changing the ordering permutes rows or columns.
None of these operations changes the rank.  We therefore use below
the conventions that make the matrices simplest.

There are always three zero modes at $\bm k=0$.  Indeed, a common
M\"obius transformation preserves the coherent zero-mode condition.
Infinitesimally it acts as
\begin{equation}
  \delta z_i
  =
  a+bz_i+cz_i^2,
  \label{eq:global_Mobius_tangent}
\end{equation}
with three complex parameters $a,b,c$.  These are spatially uniform
and therefore occur at $\Gamma$.  Physical global spin rotations form
a three-real-dimensional subset of these global M\"obius
deformations.

\subsection{Corner-sharing lattices}

\paragraph*{Checkerboard lattice.}

Choose a $2\times2$ magnetic cell with sites
\begin{equation}
  A=(0,0),
  \qquad
  B=(1,0),
  \qquad
  C=(0,1),
  \qquad
  D=(1,1),
\end{equation}
and magnetic translations $(2,0)$ and $(0,2)$.  Define the Bloch
phases
\begin{equation}
  X=e^{i\bm k\cdot\bm A_1},
  \qquad
  Y=e^{i\bm k\cdot\bm A_2}.
\end{equation}
There are two crossed plaquettes per magnetic cell.  In the basis
$(u_A,u_B,u_C,u_D)$ their linearized constraints are represented by
\begin{equation}
  \mathsf R_{\rm cb}(\bm k)
  =
  \begin{pmatrix}
    1  & -1 & \gamma   & -\gamma\\
    XY & -Y & \gamma X & -\gamma
  \end{pmatrix}.
  \label{eq:R_checkerboard}
\end{equation}
Since this is a $2\times4$ matrix,
\begin{equation}
  \nu(\bm k)\geq2
\end{equation}
at every momentum.  The two rows are independent for
$\bm k\neq0$, so
\begin{equation}
  \nu(\bm k)=2,
  \qquad
  \bm k\neq0,
\end{equation}
whereas at $\Gamma$ the two rows coincide and
$\nu(0)=3$.  On an $L\times L$ torus of magnetic cells,
\begin{equation}
  N_{\rm lin}
  =
  2(L^2-1)+3
  =
  2L^2+1
  =
  \frac{N}{2}+1.
  \label{eq:Nlin_checkerboard}
\end{equation}

Thus the checkerboard lattice has an extensive tangent space around
the four-sublattice state.

\paragraph*{Pyrochlore lattice.}

The same counting becomes equally transparent on the pyrochlore
lattice.  Use the usual four-site primitive cell with an ``up''
tetrahedron inside the cell and a ``down'' tetrahedron connecting
neighboring cells.  With primitive-cell Bloch phases $X,Y,Z$, a
convenient form is
\begin{equation}
  \mathsf R_{\rm pyr}(\bm k)
  =
  \begin{pmatrix}
    1 & -1 & \gamma   & -\gamma\\
    1 & -X & \gamma Y & -\gamma Z
  \end{pmatrix}.
  \label{eq:R_pyrochlore}
\end{equation}
Again the rank is two everywhere except at
$X=Y=Z=1$, where it is one.  Hence
\begin{equation}
  \nu(\bm k)=2
  \quad
  (\bm k\neq0),
  \qquad
  \nu(0)=3,
\end{equation}
and on an $L^3$ torus
\begin{equation}
  N_{\rm lin}
  =
  2(L^3-1)+3
  =
  \frac{N}{2}+1.
  \label{eq:Nlin_pyrochlore}
\end{equation}

The linear counts are therefore identical for the checkerboard and
pyrochlore lattices.  Their nonlinear behavior is nevertheless
different.  The compact empty-square modes of the checkerboard
lattice close only to first order: their exact M\"obius holonomy is
parabolic rather than the identity.  The pyrochlore hexagons, by
contrast, have identity holonomy
(Appendix~\ref{app:identity_holonomy}), so the corresponding
infinitesimal modes extend to exact continuous weathervane families.

\subsection{fcc lattice}

For the fcc lattice we use the cubic magnetic cell of
Fig.~\ref{fig:fcc_lattice}, with translations
\begin{equation}
  \bm A_1=(2,0,0),
  \qquad
  \bm A_2=(0,2,0),
  \qquad
  \bm A_3=(0,0,2),
\end{equation}
and basis sites
\begin{equation}
  A=(0,0,0),
  \quad
  B=(0,1,1),
  \quad
  C=(1,0,1),
  \quad
  D=(1,1,0).
\end{equation}
Let
\begin{equation}
  X=e^{i\bm k\cdot\bm A_1},
  \qquad
  Y=e^{i\bm k\cdot\bm A_2},
  \qquad
  Z=e^{i\bm k\cdot\bm A_3}.
\end{equation}
The eight elementary tetrahedra in the cubic magnetic cell give
\begin{equation}
\mathsf R_{\rm fcc}(\bm k)
=
\begin{pmatrix}
  1   & -1 & \gamma   & -\gamma\\
  Z   & -1 & \gamma   & -\gamma Z\\
  Y   & -1 & \gamma Y & -\gamma\\
  YZ  & -1 & \gamma Y & -\gamma Z\\
  X   & -X & \gamma   & -\gamma\\
  XZ  & -X & \gamma   & -\gamma Z\\
  XY  & -X & \gamma Y & -\gamma\\
  XYZ & -X & \gamma Y & -\gamma Z
\end{pmatrix}.
\label{eq:R_fcc}
\end{equation}

The momentum dependence of its rank follows analytically from simple
$4\times4$ minors.  For example,
\begin{subequations}
\begin{align}
  \det\mathsf R_{\{1,2,3,4\}}
  &=
  \gamma^2
  (Y-1)^2(Z-1)^2,
  \\
  \det\mathsf R_{\{1,2,5,6\}}
  &=
  -\gamma^2
  (X-1)^2(Z-1)^2,
  \\
  \det\mathsf R_{\{1,3,5,7\}}
  &=
  \gamma^2
  (X-1)^2(Y-1)^2,
\end{align}
\label{eq:fcc_minors}
\end{subequations}
where the subscripts specify the four selected rows of
Eq.~(\ref{eq:R_fcc}).  At least one of these minors is nonzero unless
two of the three Bloch phases are equal to unity.  Thus a nontrivial
kernel can occur only on
\begin{equation}
  Y=Z=1,
  \qquad
  X=Z=1,
  \qquad
  X=Y=1,
\end{equation}
which are the three $\langle100\rangle$ axes through $\Gamma$.

Direct substitution into Eq.~(\ref{eq:R_fcc}) gives rank two on each
axis away from $\Gamma$, and rank one at $\Gamma$.  Consequently,
\begin{equation}
  \nu(\bm k)=2
\end{equation}
on the three axes away from their intersection.  On an $L^3$ torus
each axis contains $L-1$ nonzero momenta, and therefore
\begin{equation}
  N_{\rm lin}
  =
  3\times2(L-1)+3
  =
  6L-3.
  \label{eq:Nlin_fcc}
\end{equation}

A Fourier superposition of modes along, for example, the $X$ axis
can be localized in the $x$ direction while remaining uniform in the
transverse directions.  It therefore produces a deformation
supported on a single $(100)$ plane.  The three momentum-space axes
are thus precisely the Bloch representation of the planar zero modes
constructed in Fig.~\ref{fig:fcc_lattice}(b).

\subsection{Square and honeycomb lattices}

\paragraph*{Square lattice.}

The square-lattice model contains one local tetrahedral constraint on
every elementary square.  Using the same $2\times2$ four-color cell
as for the checkerboard lattice gives
\begin{equation}
  \mathsf R_{\rm sq}(\bm k)
  =
  \begin{pmatrix}
    1  & -1 & \gamma   & -\gamma\\
    X  & -1 & \gamma X & -\gamma\\
    Y  & -Y & \gamma   & -\gamma\\
    XY & -Y & \gamma X & -\gamma
  \end{pmatrix}.
  \label{eq:R_square}
\end{equation}
Its determinant factorizes completely:
\begin{equation}
  \det\mathsf R_{\rm sq}(\bm k)
  =
  \gamma^2
  (X-1)^2(Y-1)^2.
  \label{eq:det_R_square}
\end{equation}
The matrix therefore has a kernel only on the two lines
\begin{equation}
  X=1
  \qquad\text{or}\qquad
  Y=1.
\end{equation}
Away from $\Gamma$ its rank on either line is two, giving
$\nu=2$, while at $\Gamma$ the rank is one and $\nu=3$.
Hence
\begin{equation}
  N_{\rm lin}
  =
  2\times2(L-1)+3
  =
  4L-1.
  \label{eq:Nlin_square}
\end{equation}
These two null lines are the momentum-space counterparts of the two
families of real-space line defects in
Fig.~\ref{fig:linear-defect}(a).

\paragraph*{Honeycomb lattice.}

For the honeycomb lattice let $\bm a_1,\bm a_2$ be primitive
translations of the two-site honeycomb cell, with geometric
sublattices $u$ and $v$.  We choose
\begin{equation}
  u_{\bm n}
  \sim
  \left\{
    v_{\bm n},
    v_{\bm n-\bm a_1},
    v_{\bm n-\bm a_2}
  \right\}
\end{equation}
for the three nearest neighbors of $u_{\bm n}$.  The four-color
state has a $2\times2$ enlargement of this cell.  A convenient color
assignment is
\begin{align}
  &(u_{00},u_{01},u_{10},u_{11})
  =
  (A,B,C,D),
  \nonumber\\
  &(v_{00},v_{01},v_{10},v_{11})
  =
  (D,C,B,A).
\end{align}
With magnetic translations
$2\bm a_1,2\bm a_2$ and Bloch phases $X,Y$, order the amplitudes as
\begin{equation}
  \bm u
  =
  (
    u_{00},u_{01},u_{10},u_{11},
    v_{00},v_{01},v_{10},v_{11}
  )^T.
\end{equation}
The linearized matrix can be written in block form,
\begin{equation}
  \mathsf R_{\rm hc}(\bm k)
  =
  \begin{pmatrix}
    D_u & B_{uv}\\
    B_{vu} & D_v
  \end{pmatrix},
  \label{eq:R_honeycomb_block}
\end{equation}
where
\begin{align}
  D_u
  & =
  \operatorname{diag}(1,-1,\gamma,-\gamma),
\\
  D_v
 & =
  \operatorname{diag}(-\gamma,\gamma,-1,1),
\end{align}
and
\begin{equation}
  B_{uv}
  =
  \begin{pmatrix}
    -\gamma & \gamma Y^{-1} & -X^{-1} & 0\\
    -\gamma & \gamma        & 0       & X^{-1}\\
    -\gamma & 0             & -1      & Y^{-1}\\
    0       & \gamma        & -1      & 1
  \end{pmatrix},
\end{equation}
\begin{equation}
  B_{vu}
  =
  \begin{pmatrix}
    1 & -1 & \gamma   & 0\\
    Y & -1 & 0        & -\gamma\\
    X & 0  & \gamma   & -\gamma\\
    0 & -X & \gamma Y & -\gamma
  \end{pmatrix}.
\end{equation}
Its determinant is
\begin{equation}
  \det\mathsf R_{\rm hc}(\bm k)
  =
  \frac{\gamma^4}{X^2Y^2}
  (X-1)^2
  (Y-1)^2
  (X-Y)^2.
  \label{eq:det_R_honeycomb}
\end{equation}
Thus the null set consists of the three symmetry-related lines
\begin{equation}
  X=1,
  \qquad
  Y=1,
  \qquad
  X=Y.
\end{equation}
On each line away from $\Gamma$ the rank is six, so the nullity is
two.  At $\Gamma$ the rank is five, leaving the three global
M\"obius modes.  Therefore
\begin{equation}
  N_{\rm lin}
  =
  3\times2(L-1)+3
  =
  6L-3.
  \label{eq:Nlin_honeycomb}
\end{equation}
The three null lines are the reciprocal-space representation of the
three $C_6$-related families of line deformations shown in
Fig.~\ref{fig:linear-defect}(b).

\subsection{Triangular lattice: rhombus covering}

For the triangular lattice let $\bm a_1,\bm a_2$ be primitive
translations and define
\begin{equation}
  \bm d_1=\bm a_1,
  \qquad
  \bm d_2=\bm a_2,
  \qquad
  \bm d_3=\bm a_2-\bm a_1.
\end{equation}
The four-sublattice state is represented by the parities of the
coordinates along $\bm a_1,\bm a_2$,
\begin{equation}
  A=(0,0),
  \qquad
  B=(1,0),
  \qquad
  C=(0,1),
  \qquad
  D=(1,1),
\end{equation}
with magnetic translations $2\bm a_1,2\bm a_2$.

The three rhombus orientations are generated by the pairs
$(\bm d_1,\bm d_2)$,
$(\bm d_2,\bm d_3)$, and
$(\bm d_3,\bm d_1)$.  The full $12\times4$ constraint matrix can be
written as
\begin{equation}
  \mathsf R_{\rm rh}(\bm k)
  =
  \begin{pmatrix}
    \mathsf R_{12}\\
    \mathsf R_{23}\\
    \mathsf R_{31}
  \end{pmatrix},
  \label{eq:R_triangular_rhombi}
\end{equation}
with
\begin{equation}
  \mathsf R_{12}
  =
  \begin{pmatrix}
    1  & -1 & \gamma   & -\gamma\\
    X  & -1 & \gamma X & -\gamma\\
    Y  & -Y & \gamma   & -\gamma\\
    XY & -Y & \gamma X & -\gamma
  \end{pmatrix},
\end{equation}
\begin{equation}
  \mathsf R_{23}
  =
  \begin{pmatrix}
    1 & -Y/X & \gamma   & -\gamma/X\\
    Y & -1   & \gamma   & -\gamma\\
    Y & -Y/X & \gamma   & -\gamma Y/X\\
    Y & -Y   & \gamma Y & -\gamma
  \end{pmatrix},
\end{equation}
and
\begin{equation}
  \mathsf R_{31}
  =
  \begin{pmatrix}
    1 & -1   & \gamma   & -\gamma/X\\
    X & -1   & \gamma   & -\gamma\\
    Y & -Y/X & \gamma   & -\gamma\\
    Y & -Y   & \gamma X & -\gamma
  \end{pmatrix}.
\end{equation}

The determinants of the three $4\times4$ blocks are
\begin{subequations}
\begin{align}
  \det\mathsf R_{12}
  &=
  \gamma^2
  (X-1)^2(Y-1)^2,
  \\
  \det\mathsf R_{23}
  &=
  \frac{\gamma^2}{X^2}
  (X-Y)^2(Y-1)^2,
  \\
  \det\mathsf R_{31}
  &=
  \frac{\gamma^2}{X^2}
  (X-1)^2(X-Y)^2.
\end{align}
\label{eq:triangular_rhombus_minors}
\end{subequations}
For any $(X,Y)\neq(1,1)$ at least one of these determinants is
nonzero.  Hence
\begin{equation}
  \operatorname{rank}
  \mathsf R_{\rm rh}(\bm k)
  =
  4,
  \qquad
  \bm k\neq0,
\end{equation}
and there are no nonuniform infinitesimal coherent deformations.
At $\Gamma$ all local constraints reduce to the same color equation,
the rank drops to one, and the kernel consists of the three global
M\"obius modes:
\begin{equation}
  N_{\rm lin}=3.
  \label{eq:Nlin_triangular_rhombi}
\end{equation}

\subsection{Triangular lattice: claw covering}

The alternative triangular-lattice construction contains two
four-site motifs centered on every site.  In the notation above, the
three outer sites of the two motifs are displaced from the center by
\begin{align}
  \mathcal C_+
  &=
  \{
    \bm d_1,\,
    \bm d_3,\,
    -\bm d_2
  \},
  \nonumber\\
  \mathcal C_-
  &=
  \{
    \bm d_2,\,
    -\bm d_1,\,
    -\bm d_3
  \}.
\end{align}
These are the two alternating sets of nearest neighbors shown in
Fig.~\ref{fig:triangular_lattice_alt}.

Ordering first the four $\mathcal C_+$ constraints and then the four
$\mathcal C_-$ constraints gives
\begin{equation}
\mathsf R_{\rm claw}(\bm k)
=
\begin{pmatrix}
  1 & -1     & \gamma/Y   & -\gamma/X\\
  X & -1     & \gamma     & -\gamma/Y\\
  1 & -Y/X   & \gamma     & -\gamma\\
  Y & -1     & \gamma X   & -\gamma\\[1mm]
  1 & -1/X   & \gamma     & -\gamma/Y\\
  1 & -1     & \gamma X/Y & -\gamma\\
  Y & -1     & \gamma     & -\gamma/X\\
  X & -Y     & \gamma     & -\gamma
\end{pmatrix}.
\label{eq:R_triangular_claw}
\end{equation}

The absence of nonzero-momentum zero modes follows from three
$4\times4$ minors.  With the row numbering of
Eq.~(\ref{eq:R_triangular_claw}),
\begin{subequations}
\begin{align}
  \det\mathsf R_{\{1,4,6,7\}}
  &=
  -\frac{\gamma^2}{XY}
  (X-1)^2(Y-1)^2,
  \\
  \det\mathsf R_{\{2,4,6,8\}}
  &=
  \frac{\gamma^2}{Y^2}
  (X^2-Y)(Y-1)^3,
  \\
  \det\mathsf R_{\{3,4,7,8\}}
  &=
  \frac{\gamma^2}{X^2}
  (X-1)^3(X-Y^2).
\end{align}
\label{eq:triangular_claw_minors}
\end{subequations}
If $X\neq1$ and $Y\neq1$, the first minor is nonzero.  If
$X=1$ but $Y\neq1$, the second becomes
$-\gamma^2(Y-1)^4/Y^2$, while if $Y=1$ but $X\neq1$, the third becomes
$\gamma^2(X-1)^4/X^2$.  Thus
\begin{equation}
  \operatorname{rank}
  \mathsf R_{\rm claw}(\bm k)
  =
  4
  \qquad
  \text{for every }\bm k\neq0.
\end{equation}
At $\Gamma$ the rank is one and only the three global M\"obius modes
remain:
\begin{equation}
  N_{\rm lin}=3.
  \label{eq:Nlin_triangular_claw}
\end{equation}

It is noteworthy that the rhombus and claw coverings have the same
linearized coherent zero-mode space around the four-sublattice
state, even though their quantum zero-energy kernels are very
different.  The number of infinitesimal coherent deformations and
the dimension of the full quantum kernel are therefore independent
characteristics of the model.

\subsection{Summary}

The results are summarized in Table~\ref{tab:rigidity}.  The
dimension and geometry of the null set of $\mathsf R(\bm k)$ provide
a direct reciprocal-space measure of the coherent flexibility of the
four-sublattice state.

\begin{table*}[tb]
\caption{\label{tab:rigidity}
Linearized coherent deformations around the four-sublattice ordered
state.  The column ``cell'' gives the number of spin variables and
local constraints in the magnetic unit cell.  The ``null set''
specifies the momenta for which the linearized constraint matrix
$\mathsf R(\bm k)$ has a nontrivial kernel, and ``nullity'' gives its
dimension away from $\Gamma$.  The last column gives the total number
of complex infinitesimal deformations on a periodic system of $L^d$
magnetic cells.  All lattices have three global M\"obius modes at
$\Gamma$.}
\begin{ruledtabular}
\begin{tabular}{llllll}
lattice & sharing & cell & null set & nullity & linear modes\\
\hline
checkerboard
& corner
& $4/2$
& full zone
& $2$
& $N/2+1$
\\
pyrochlore
& corner
& $4/2$
& full zone
& $2$
& $N/2+1$
\\
fcc
& edge
& $4/8$
& three $\langle100\rangle$ axes
& $2$
& $6L-3$
\\
square
& edge
& $4/4$
& two axes
& $2$
& $4L-1$
\\
honeycomb
& edge
& $8/8$
& three $C_6$-related lines
& $2$
& $6L-3$
\\
triangular, rhombi
& face
& $4/12$
& $\Gamma$ only
& ---
& $3$
\\
triangular, claws
& edge
& $4/8$
& $\Gamma$ only
& ---
& $3$
\end{tabular}
\end{ruledtabular}
\end{table*}

The hierarchy is therefore particularly transparent.  On the
corner-sharing checkerboard and pyrochlore lattices the kernel exists
throughout the Brillouin zone, producing an extensive number of
infinitesimal deformations.  On the fcc, square, and honeycomb
lattices it survives only on one-dimensional momentum-space sets,
giving $O(L)$ modes and corresponding real-space plane or line
deformations.  For both triangular-lattice constructions, only the
three global M\"obius modes remain.

The linearized analysis only describes the tangent space around a
given coherent state; it does not determine the full nonlinear
zero-mode variety or the quantum kernel.  This distinction is already
visible on the checkerboard lattice, where the empty-square modes
satisfy the constraints to first order but are obstructed at higher
orders.  On the pyrochlore lattice, by contrast, the identity
holonomy of a hexagon allows the corresponding modes to develop into
exact continuous zero-mode families.  The triangular models provide
the opposite example: although the four-sublattice coherent state has
no nonuniform infinitesimal deformations, the quantum kernel still
contains additional polarized and entangled zero-energy states.
Thus the constraint matrix characterizes the infinitesimal coherent
degrees of freedom around a chosen ordered state, not the full
ground-state degeneracy.

\section{Matching bounds for the polarized sector}
\label{app:matching}

This appendix derives the bounds quoted in
Sec.~\ref{sec:matching} by applying known results from graph theory to
the dual graph $G$. Under the correspondence established there, the
polarized product zero modes are counted by the monomer--dimer
partition function at unit activity,
\begin{equation}
  \mathcal D_{\rm match}(G)
  =
  Z_G(1)
  =
  \sum_{k\geq0}m_k(G),
\end{equation}
where $m_k(G)$ is the number of matchings containing $k$ edges.
For the checkerboard and pyrochlore lattices, $G$ is a $4$-regular
bipartite graph with $N_t=N/2$ vertices. We can therefore directly
apply rigorous lower and upper bounds known for the matching
partition function of regular bipartite graphs.

The lower-matching bounds of Friedland--Gurvits and Csikv\'ari
\cite{FriedlandGurvits2008,Csikvari2017} give
\begin{equation}
  Z_G(1)
  \geq
  \left(
    \frac{19+13\sqrt{13}}{34}
  \right)^{N_t}
  \simeq
  1.3919^N .
  \label{eq:matching_lower_app}
\end{equation}
Conversely, among $4$-regular graphs the matching count is maximized
by disjoint unions of $K_{4,4}$ \cite{DaviesJenssenPerkinsRoberts2017}.
Since
\begin{equation}
  Z_{K_{4,4}}(1)
  =
  \sum_{k=0}^{4}\binom{4}{k}^{\!2}k!
  =
  1+16+72+96+24
  =
  209,
  \label{eq:K44}
\end{equation}
and each $K_{4,4}$ block contains eight dual-graph vertices,
\begin{equation}
  Z_G(1)
  \leq
  209^{N_t/8}
  =
  209^{N/16}
  \simeq
  1.3964^N .
  \label{eq:matching_upper_app}
\end{equation}
Together, Eqs.~(\ref{eq:matching_lower_app}) and
(\ref{eq:matching_upper_app}) give Eq.~(\ref{eq:match_bounds}).

For the periodic $16$-site checkerboard cluster, the dual graph is
precisely $K_{4,4}$. The five terms
$1+16+72+96+24$ in Eq.~(\ref{eq:K44}) are the numbers $m_k$ of
matchings with $k=0,\ldots,4$ occupied edges and therefore count
polarized product states with $k$ down spins. These are fixed-$J^z$
states rather than total-spin multiplets. For example, the $16$
one-down-spin states consist of the $J=8$ descendant of the
fully polarized multiplet and $15$ highest-weight states with
$J=7$, in agreement with the $15$ zero-energy
$J=7$ multiplets in Table~\ref{tab:planar16}.

\section{High-polarization counting of zero modes}
\label{app:highspin_count}

Here we explain why different chirality assignments have identical
zero-mode counts close to full polarization, as observed in
Table~\ref{tab:planar16}. We consider spin $1/2$ and work in a fixed
quantization axis. Let $k$ be the number of down spins, so that
\begin{equation}
  J^z=\frac{N}{2}-k .
  \label{eq:app_Sz_k}
\end{equation}
The dimension of this fixed-$J^z$ sector is $\binom Nk$.

For spin $1/2$, each local CCI parent term projects onto a single
forbidden chiral singlet $\ket{\phi_t}$ on tetrahedron $t$. Since a
four-spin singlet has $J=0$, $\ket{\phi_t}$ lies entirely in the
sector with two up and two down spins. Its contribution to the global
$k$-down-spin sector therefore generates constraint vectors of the
form
\begin{equation}
  \ket{\phi_t}\otimes\ket{\eta},
  \label{eq:app_constraint_vector}
\end{equation}
where $\ket{\eta}$ contains the remaining $k-2$ down spins on the
$N-4$ sites outside tetrahedron $t$.

For each tetrahedron there are
$\binom{N-4}{k-2}$ such vectors. Let $\mathscr R_k$ denote their
linear span over all tetrahedra and all choices of $\ket{\eta}$.
Since the Hamiltonian is a sum of local projectors, the zero-energy
subspace in the $k$-down-spin sector is the orthogonal complement of
$\mathscr R_k$. Hence
\begin{equation}
  \mathcal N_0(k)
  =
  \binom Nk-\dim\mathscr R_k .
  \label{eq:highspin_rank}
\end{equation}
The number of vectors generating $\mathscr R_k$ is
\begin{equation}
  M_k
  =
  N_{\rm tet}\binom{N-4}{k-2}.
\end{equation}
Therefore
\begin{equation}
  \mathcal N_0(k)
  \geq
  \binom Nk
  -
  N_{\rm tet}\binom{N-4}{k-2},
  \label{eq:highspin_bound}
\end{equation}
with equality whenever the constraint vectors are linearly
independent:
\begin{equation}
  \mathcal N_0(k)
  =
  \binom Nk
  -
  N_{\rm tet}\binom{N-4}{k-2}.
  \label{eq:highspin_count}
\end{equation}

This immediately explains the insensitivity to the chirality pattern
in the independent-constraint regime. Changing the local chirality
changes the individual forbidden singlets $\ket{\phi_t}$, but not the
number of independent constraints. The chirality assignment can affect
the zero-mode count only once linear relations develop among constraint
vectors belonging to different tetrahedra.

Table~\ref{tab:planar16} is organized in total-spin multiplets rather
than fixed-$J^z$ sectors. If $n_J$ denotes the number of zero-energy
multiplets with total spin $J$, then, for the high-polarization sectors
considered here,
\begin{equation}
  \mathcal N_0(k)
  =
  \sum_{J\geq N/2-k} n_J ,
  \label{eq:Sz_from_multiplets}
\end{equation}
because each spin-$J$ multiplet contributes exactly one state at a
given $J^z$ whenever $J\geq J^z$.

For the periodic $N=16$ clusters, the constraints are independent in
the two-down-spin sector. Using
\begin{equation}
  N_{\rm tet}
  =
  8,\ 16,\ 16,\ 48
\end{equation}
for the checkerboard, square, honeycomb, and triangular clusters,
respectively, Eq.~(\ref{eq:highspin_count}) gives
\begin{equation}
  \mathcal N_0(2)
  =
  112,\ 104,\ 104,\ 72 .
  \label{eq:k2_counts}
\end{equation}
These values agree with the sums of the $J\geq6$ multiplet counts in
Table~\ref{tab:planar16} and are independent of the chirality
assignment.

For $k=3$, the constraints remain independent for the checkerboard,
square, and honeycomb clusters, giving
\begin{equation}
  \mathcal N_0(3)
  =
  464,\ 368,\ 368 ,
  \label{eq:k3_counts}
\end{equation}
in agreement with the sums over $J\geq5$. For the triangular cluster,
by contrast,
\begin{equation}
  N_{\rm tet}\binom{N-4}{k-2}
  =
  48\times12
  >
  \binom{16}{3},
\end{equation}
so the constraint vectors must already be linearly dependent.

As the polarization is lowered further, nontrivial linear relations
among constraints become increasingly important. Their rank can then
depend on the relative chiralities of the local parent terms, and the
uniform and alternating spectra begin to separate. For the $16$-site
square cluster this first occurs at $J=4$, where the alternating and
uniform Hamiltonians have $453$ and $452$ zero-energy multiplets,
respectively. For the honeycomb cluster the spectra remain identical
through $J=4$ and first differ at $J=3$, with $478$ and $476$
multiplets. Thus the chirality dependence of the low-spin spectrum can
be understood as a rank-deficiency effect among overlapping local
constraints.

This counting should be distinguished from the matching construction
of Sec.~\ref{sec:matching}. A matching imposes the stronger condition
that no tetrahedron contain two down spins and therefore produces
individual product zero modes. The rank counting above instead
implements the full quantum constraint: configurations with two down
spins on the same tetrahedron are allowed as long as their local wave
function is orthogonal to the forbidden chiral singlet. The matching
states therefore span only a subset of the high-polarization
zero-energy sector. For example, on the $16$-site checkerboard cluster,
the matching construction gives $m_2=72$ two-down-spin product states,
whereas the complete $k=2$ zero-energy sector has dimension
$\mathcal N_0(2)=112$.

\section{Anderson towers for tetrahedral magnetic order}
\label{app:tower}

We determine the symmetry content of the Anderson tower associated
with tetrahedral magnetic order, following the finite-size symmetry
analysis of
Refs.~\cite{Bernu_PRL.69.2590_1992,Bernu_PRB.50.10048_1994}.
A finite-spin realization contains one additional ingredient: the
quantum reference state may transform with a one-dimensional phase
under the combined lattice and spin rotations that leave the
classical tetrahedral configuration invariant. This phase does not
change the tower multiplicities, but it can shift their discrete
lattice-symmetry labels.

\subsection{Single-chirality $A_4$ tower}
\label{app:tower_A4}

A chirality-selecting interaction leaves the proper tetrahedral group
$T\simeq A_4$ of the $12$ even permutations as the relevant discrete
symmetry of a single tetrahedral orientation. Its conjugacy classes
are
\[
  \{e\},\qquad
  4C_3^+,\qquad
  4C_3^-,\qquad
  3C_2 ,
\]
where the two sets of threefold rotations are distinct conjugacy
classes in $A_4$. We choose $C_3^+$ to contain the cycle
$\hat P_{(123)}$ used in Eq.~(\ref{eq:P_on_matchings}). The
irreducible representations are the singlets
$\mathsf A,\mathsf E_\pm$, with $\mathsf E_\pm$ a
complex-conjugate pair, and the triplet $\mathsf T$. Their characters
are listed in Table~\ref{tab:T_character_table}(a).

\begin{table}[tp]
\caption{
Character tables relevant to tetrahedral order.
(a) The proper tetrahedral group $T\simeq A_4$, with
$\omega=e^{2\pi i/3}$ and $C_3^+$ chosen to contain the cycle
$(123)$.
(b) The full permutation group $S_4\simeq T_d$. In passing from
$A_4$ to $S_4$, the classes $4C_3^\pm$ merge into the single class
of eight three-cycles; the last two columns contain the odd
permutations, which exchange the two chiralities. 
}
\label{tab:T_character_table}
\begin{ruledtabular}
\begin{tabular}{c cccc}
\multicolumn{5}{c}{(a) $A_4\simeq T$}\\[1mm]
& $e$ & $4C_3^+$ & $4C_3^-$ & $3C_2$ \\
\hline
$\mathsf A$   & $1$ & $1$        & $1$        & $1$  \\
$\mathsf E_+$ & $1$ & $\omega$   & $\omega^2$ & $1$  \\
$\mathsf E_-$ & $1$ & $\omega^2$ & $\omega$   & $1$  \\
$\mathsf T$   & $3$ & $0$        & $0$        & $-1$
\end{tabular}

\vspace{2mm}

\begin{tabular}{c ccccc}
\multicolumn{6}{c}{(b) $S_4\simeq T_d$}\\[1mm]
& $e$
& $8(123)$
& $3(12)(34)$
& $6(1234)$
& $6(12)$ \\
\hline
$\mathsf A_1$ & $1$ & $1$  & $1$  & $1$  & $1$  \\
$\mathsf A_2$ & $1$ & $1$  & $1$  & $-1$ & $-1$ \\
$\mathsf E$   & $2$ & $-1$ & $2$  & $0$  & $0$  \\
$\mathsf T_1$ & $3$ & $0$  & $-1$ & $1$  & $-1$ \\
$\mathsf T_2$ & $3$ & $0$  & $-1$ & $-1$ & $1$
\end{tabular}
\end{ruledtabular}
\end{table}

Let $\{\bm S_a^{(0)}\}_{a=1}^4$ be a reference tetrahedral
configuration. For every $g\in T$, the permutation of the four
sublattices can be compensated by a global spin rotation $R_g$.
The combined operation therefore leaves the classical ordered
configuration invariant.

For the corresponding quantum coherent reference state, however, the
same operation need only return the state to the same ray,
\begin{equation}
  \hat U(R_g)\hat P_g\ket{\Psi}
  =
  \eta(g)\ket{\Psi},
  \qquad
  g\in T ,
  \label{eq:tower_stabilizer_character_app}
\end{equation}
where $\eta$ is a one-dimensional representation of $A_4$,
\begin{equation}
  \eta
  \in
  \left\{
    \mathsf A,\mathsf E_+,\mathsf E_-
  \right\}.
\end{equation}

The origin of this phase is purely quantum mechanical. The operation
$(R_g,g)$ returns every classical spin direction to itself, but the
quantum state can acquire a phase.
Equivalently, $\eta$ specifies a twisted boundary condition for the
collective rotor under the discrete tetrahedral identifications of
the classical order-parameter manifold. 

The collective-coordinate Hilbert space carries the induced
representation
\begin{equation}
  \mathcal H_{\rm tower}^{(\eta)}
  =
  \operatorname{Ind}_{T_{\rm diag}}^{
    SO(3)_{\rm spin}\times T_{\rm lattice}}
  \eta ,
  \label{eq:tower_induced_eta}
\end{equation}
where
\begin{equation}
  T_{\rm diag}
  =
  \left\{
    (R_g,g)\mid g\in T
  \right\}.
\end{equation}
At fixed integer total spin $J$, Peter--Weyl decomposition, or
equivalently Frobenius reciprocity, gives
\begin{equation}
  \mathcal H_J^{(\eta)}
  =
  \mathcal V_J
  \otimes
  \mathcal M_J^{(\eta)},
\end{equation}
where $\mathcal V_J$ is the $(2J+1)$-dimensional spin-$J$
representation and
\begin{equation}
  \mathcal M_J^{(\eta)}
  \simeq
  \left(D^J\!\downarrow_T\right)\otimes\eta .
  \label{eq:tower_at_J_eta}
\end{equation}
Here we used the equivalence of the integer-spin representation
$D^J$ and its dual. A nontrivial stabilizer character therefore only
tensors the tetrahedral lattice representation by a one-dimensional
irrep; it cannot change the number of spin-$J$ multiplets.

It is useful first to consider the untwisted case
$\eta=\mathsf A$. Using the standard SO(3) character
\begin{equation}
  \chi_J(\theta)
  =
  \frac{\sin[(J+\tfrac12)\theta]}{\sin(\theta/2)},
\end{equation}
and evaluating it at the tetrahedral rotation angles
$\theta=0,\,2\pi/3,$ and $\pi$, one obtains
\begin{subequations}
\label{eq:tower_class_characters}
\begin{align}
  \chi_J(e)&=2J+1, \\
  \chi_J(C_3^\pm)&=u_J, \\
  \chi_J(C_2)&=(-1)^J ,
\end{align}
\end{subequations}
where
\begin{equation}
  u_J
  =
  \frac{\sin[(2J+1)\pi/3]}{\sin(\pi/3)}
  =
  \begin{cases}
    1,  & J=0\!\!\pmod3,\\
    0,  & J=1\!\!\pmod3,\\
    -1, & J=2\!\!\pmod3.
  \end{cases}
  \label{eq:uJ_periodic}
\end{equation}
The character projection
\begin{equation}
  \nu_\Gamma(J)
  =
  \frac{1}{12}
  \sum_{\mathcal C}
  |\mathcal C|\,
  \chi_\Gamma^*(\mathcal C)
  \chi_J(\mathcal C)
\end{equation}
then yields
\begin{align}
  \nu_A(J)
  &=
  \frac{2J+1+8u_J+3(-1)^J}{12},
  \label{eq:tower_nuA}\\
  \nu_{E_+}(J)
  =
  \nu_{E_-}(J)
  &=
  \frac{2J+1-4u_J+3(-1)^J}{12},
  \label{eq:tower_nuE}\\
  \nu_T(J)
  &=
  \frac{2J+1-(-1)^J}{4}.
  \label{eq:tower_nuT}
\end{align}
The dimension identity
\begin{equation}
  \nu_A(J)
  +
  \nu_{E_+}(J)
  +
  \nu_{E_-}(J)
  +
  3\nu_T(J)
  =
  2J+1
  \label{eq:tower_dimension_identity}
\end{equation}
shows that the tower contains $2J+1$ spin-$J$ multiplets, or
$(2J+1)^2$ states at fixed $J$, as required for a rigid rotor on
SO(3).

The untwisted decomposition is listed in
Table~\ref{tab:tetrahedral_tower}. For a general stabilizer character
$\eta$, every irrep in the untwisted tower is simply tensored by
$\eta$. Equivalently,
\begin{equation}
  \nu_\Gamma^{(\eta)}(J)
  =
  \nu_{\Gamma\otimes\eta^*}(J).
  \label{eq:tower_twisted_multiplicity}
\end{equation}
The relevant tensor products are
\begin{equation}
  \mathsf E_+\otimes\mathsf E_+
  =
  \mathsf E_-,
  \quad
  \mathsf E_+\otimes\mathsf E_-
  =
  \mathsf A,
  \quad
  \mathsf E_\pm\otimes\mathsf T
  =
  \mathsf T .
  \label{eq:A4_tensor_rules}
\end{equation}
Thus a twist cyclically permutes
$\mathsf A,\mathsf E_+,\mathsf E_-$ while leaving the tetrahedral
triplet unchanged.

The class characters are periodic in $J$ with period six, giving
\begin{equation}
  D^{J+6}\!\downarrow_T
  =
  D^J\!\downarrow_T
  \oplus
  \mathsf A
  \oplus
  \mathsf E_+
  \oplus
  \mathsf E_-
  \oplus
  3\mathsf T ,
\end{equation}
where the added term is the regular representation of $T$.

\begin{table}[bp]
\caption{
Untwisted ($\eta=\mathsf A$) multiplicities of the $A_4$
irreducible representations in the single-chirality tetrahedral
Anderson tower. For a nontrivial stabilizer character $\eta$, the
irreps in each row are tensored by $\eta$ according to
Eq.~(\ref{eq:tower_at_J_eta}). The last column gives
$\dim\mathcal M_J=2J+1$, the number of spin-$J$ multiplets.
}
\label{tab:tetrahedral_tower}
\begin{ruledtabular}
\begin{tabular}{c cccc c}
$J$
& $\nu_A$
& $\nu_{E_+}$
& $\nu_{E_-}$
& $\nu_T$
& $\dim\mathcal M_J$
\\
\hline
$0$ & $1$ & $0$ & $0$ & $0$ & $1$  \\
$1$ & $0$ & $0$ & $0$ & $1$ & $3$  \\
$2$ & $0$ & $1$ & $1$ & $1$ & $5$  \\
$3$ & $1$ & $0$ & $0$ & $2$ & $7$  \\
$4$ & $1$ & $1$ & $1$ & $2$ & $9$  \\
$5$ & $0$ & $1$ & $1$ & $3$ & $11$ \\
$6$ & $2$ & $1$ & $1$ & $3$ & $13$ \\
$7$ & $1$ & $1$ & $1$ & $4$ & $15$ \\
$8$ & $1$ & $2$ & $2$ & $4$ & $17$
\end{tabular}
\end{ruledtabular}
\end{table}

\subsection{Tower of the full color-ice subspace and
\texorpdfstring{$S_4$}{S4} symmetry}
\label{app:tower_S4}

Without explicit chirality selection, both enantiomeric tetrahedral
configurations belong to the collective order-parameter manifold.
Odd permutations
of the sublattices reverse the orientation and exchange the two
chiral sectors. If they are symmetries, the discrete symmetry is
therefore enlarged from $A_4$ to $S_4\simeq T_d$, with
$A_4\triangleleft S_4$ and
$S_4/A_4\simeq\mathbb Z_2$.

The $A_4$ quantum numbers of a single chiral sector then have to be
combined into irreducible representations of $S_4$. Restricting an
$S_4$ irrep to the even permutations gives
\begin{equation}
  \mathsf A_{1,2}\!\downarrow_{A_4}
  =
  \mathsf A,
  \quad
  \mathsf E\!\downarrow_{A_4}
  =
  \mathsf E_+\oplus\mathsf E_-,
  \quad
  \mathsf T_{1,2}\!\downarrow_{A_4}
  =
  \mathsf T .
  \label{eq:S4_to_A4_branching}
\end{equation}
Thus $\mathsf A_1$ and $\mathsf A_2$ are indistinguishable if only
even permutations are retained, and similarly both
$\mathsf T_1$ and $\mathsf T_2$ reduce to the same tetrahedral triplet.
By contrast, the two complex-conjugate one-dimensional irreps
$\mathsf E_+$ and $\mathsf E_-$ combine into the two-dimensional
$\mathsf E$ irrep once the odd permutations relating the two
chiralities are restored.

Equivalently, starting from an $A_4$ irrep in one chiral sector and
restoring the odd permutations generates the corresponding $S_4$
representation,
\begin{subequations}
  \label{eq:A4_to_S4_induction}
\begin{align}
  \operatorname{Ind}_{A_4}^{S_4}\mathsf A
  &=
  \mathsf A_1\oplus\mathsf A_2,
\\
  \operatorname{Ind}_{A_4}^{S_4}\mathsf E_\pm
  &=
  \mathsf E,
\\  \operatorname{Ind}_{A_4}^{S_4}\mathsf T
  &=
  \mathsf T_1\oplus\mathsf T_2 .
\end{align}
\end{subequations}
The induction therefore amounts physically to adjoining the
opposite-chirality partner generated by an odd sublattice
permutation.

For a stabilizer character $\eta$, the lattice representation at
fixed $J$ is therefore
\begin{equation}
  \mathcal M_J^{(\pm,\eta)}
  =
  \operatorname{Ind}_{A_4}^{S_4}
  \mathcal M_J^{(\eta)}
  =
  \operatorname{Ind}_{A_4}^{S_4}
  \left[
    \left(D^J\!\downarrow_{A_4}\right)
    \otimes\eta
  \right].
  \label{eq:tower_S4_general}
\end{equation}
Writing the twisted $A_4$ multiplicities as
$\nu_\Gamma^{(\eta)}(J)$ gives
\begin{multline}
  \mathcal M_J^{(\pm,\eta)}
  =
  \nu_A^{(\eta)}(J)
  \left(
    \mathsf A_1\oplus\mathsf A_2
  \right)
  \\
  +
  \left[
    \nu_{E_+}^{(\eta)}(J)
    +
    \nu_{E_-}^{(\eta)}(J)
  \right]\mathsf E
  \\
  +
  \nu_T(J)
  \left(
    \mathsf T_1\oplus\mathsf T_2
  \right).
  \label{eq:tower_S4_decomposition}
\end{multline}
Its dimension is independent of the twist,
\begin{equation}
  \dim\mathcal M_J^{(\pm,\eta)}
  =
  2(2J+1),
  \label{eq:S4_tower_dimension}
\end{equation}
so the full two-chirality tower contains twice as many spin-$J$
multiplets as a single chiral component.

For the untwisted case $\eta=\mathsf A$, the first few sectors are
\begin{align}
  \mathcal M_0^{(\pm)}
  &=
  \mathsf A_1\oplus\mathsf A_2,
  \\
  \mathcal M_1^{(\pm)}
  &=
  \mathsf T_1\oplus\mathsf T_2,
  \\
  \mathcal M_2^{(\pm)}
  &=
  2\mathsf E
  \oplus
  \mathsf T_1
  \oplus
  \mathsf T_2,
  \\
  \mathcal M_3^{(\pm)}
  &=
  \mathsf A_1
  \oplus
  \mathsf A_2
  \oplus
  2\mathsf T_1
  \oplus
  2\mathsf T_2 .
\end{align}
For a nontrivial $\eta$, the corresponding $S_4$ content follows
directly from Eq.~(\ref{eq:tower_S4_general}); the total multiplicity
remains $2(2J+1)$.

\subsection{Relation to the finite-spin color-ice subspaces}
\label{app:tower_color_ice}

The color-ice subspaces introduced above provide finite-spin
realizations of the tetrahedral towers. The fixed-chirality subspace
$\mathscr V^S$ realizes the $A_4$ tower, while $\mathscr W^S$
contains both chiralities and realizes the corresponding $S_4$
structure.

For $S=1/2$, the CCI subspace $\mathscr V^{1/2}$ contains a unique
$J=0$ state, the chiral singlet $\ket{-}$. Under the three-cycle
$C_3^+=(123)$ it transforms as \( \hat P_{(123)}\ket{-} = \omega\ket{-} \),
and therefore carries the one-dimensional $A_4$ irrep
$\mathsf E_+$. Since a spin-$S$ coherent state is the symmetric
product of $2S$ identical spin-$1/2$ spinors, its stabilizer character is
\(  \eta_S=\mathsf E_+^{\otimes\,2S} \).
Thus
\begin{equation}
  \eta_S
  =
  \begin{cases}
    \mathsf A,   & 2S=0\pmod3,\\
    \mathsf E_+, & 2S=1\pmod3,\\
    \mathsf E_-, & 2S=2\pmod3.
  \end{cases}
  \label{eq:etaS_mod3}
\end{equation}
The opposite chiral sector carries the conjugate character
$\eta_S^*$.

Identifying $J=S_{\mathrm{tet}}$, the fixed-chirality subspace reproduces
the complete twisted rotor for
$0\le J\le2S$:
\begin{equation}
  \mathscr V^S_J
  \simeq
  \left(D^J\!\downarrow_{A_4}\right)
  \otimes\eta_S,
  \qquad
  \dim\mathscr V^S_J=2J+1 .
  \label{eq:VS_twisted_tower}
\end{equation}
Thus Table~\ref{tab:tetrahedral_tower} gives the rotor multiplicities
for every $S$, while the one-dimensional irreps
$\mathsf A,\mathsf E_+,\mathsf E_-$ are cyclically shifted according
to Eq.~(\ref{eq:etaS_mod3}).

The full color-ice subspace is the sum of the two chiral sectors,
\begin{equation}
  \mathscr W^S_J
  =
  \mathscr V^S_{+,J}
  +
  \mathscr V^S_{-,J}.
\end{equation}
For $0\le J<2S$ they are linearly independent, so that
\begin{equation}
  \mathscr W^S_J
  =
  \mathscr V^S_{+,J}
  \oplus
  \mathscr V^S_{-,J}
  \simeq
  \mathcal M_J^{(\pm,\eta_S)},
  \qquad
  \dim\mathscr W^S_J=2(2J+1).
  \label{eq:WS_twisted_tower}
\end{equation}
Equivalently, in this range the kernel-dimension relation becomes
\begin{equation}
  \dim\mathscr H^S_J-\dim\mathscr H^{S-1}_J
  =
  4J+2 .
  \label{eq:kernel_dimension_tower}
\end{equation}

At $J=2S$ the two finite-spin chiral subspaces begin to overlap, and
the $\mathscr W^S_J$ multiplicities fall below the rigid-rotor value.
The finite-spin spaces subsequently truncate and terminate in the
unique fully polarized multiplet at $J=4S$. These deviations are
finite-$S$ effects: for any fixed $J$, the rotor multiplicities and
symmetry content are recovered once $S$ is sufficiently large.

For the four-sublattice triangular clusters,
$2\widetilde S=N/4$, and therefore
\begin{equation}
  \eta_{\widetilde S}
  =
  \mathsf E_+^{\otimes N/4}.
  \label{eq:eta_cluster}
\end{equation}
Thus the $N=12$ cluster realizes the untwisted tower,
$\eta_{3/2}=\mathsf A$, whereas the $N=16$ cluster carries
$\eta_2=\mathsf E_+$. This produces the shifted lattice quantum
numbers discussed in Sec.~\ref{sec:triangular_lattice}.


\bibliography{citations}
\end{document}